\documentclass[useAMS,usenatbib]{mn2e} 

\usepackage{times}
\usepackage{graphicx}
\usepackage{amssymb}
\usepackage{bm}
\usepackage{url}

\newcommand{\ion}[2]{{#1~\small#2}}

\newcommand{\NHI}{$N\rm _{H\,{\sevensize I}}$}

\newcommand{\HI}{$\rm H\,{\sevensize I}$}

\newcommand{\sfr}{M$_{\odot}$ yr$^{-1}$}

\newcommand{\para}{\boldsymbol{\theta}}

\title[The physical properties of $z>2$ Lyman limit systems]{The physical properties of $z>2$ Lyman limit systems: new constraints for feedback and accretion models.}
\author[Fumagalli et al.]{Michele Fumagalli$^{1}$\thanks{E-mail: michele.fumagalli@durham.ac.uk},
John M. O'Meara$^{2}$, J. Xavier Prochaska$^{3,4}$ \\
  $^{1}$Institute for Computational Cosmology and Centre for Extragalactic Astronomy, 
  Department of Physics, Durham University, South Road, Durham, DH1 3LE, UK. \\
  $^{2}$Department of Chemistry and Physics, Saint Michael's College, One Winooski Park, 
     Colchester, VT 05439, USA \\
  $^{3}$Department of Astronomy and Astrophysics, University of California, 1156 High Street, 
     Santa Cruz, CA 95064 USA \\
  $^{4}$University of California Observatories, Lick Observatory 1156 High Street, Santa Cruz, 
     CA 95064 USA \\
}

\begin{document}


\pagerange{\pageref{firstpage}--\pageref{lastpage}} \pubyear{\the\year}

\maketitle

\label{firstpage}

\begin{abstract}
We study the physical properties of a homogeneous sample of 157 optically-thick absorption line systems 
at redshifts $\sim1.8-4.4$,  selected from a high-dispersion spectroscopic survey of Lyman 
limit systems (LLSs). By means of multiple ionisation models and Bayesian techniques, 
we derive the posterior probability distribution functions for the density, metallicity, 
temperature, and dust content of the absorbing gas. 
We find that $z>2$ LLSs are highly ionised with ionisation parameters
between $-3\lesssim\log U\lesssim -2$, depending on the \ion{H}{I} column density.
LLSs are characterised by low temperatures ($T<5\times10^4~$K) and reside in 
dust-poor environments. Between $z\sim2.5-3.5$, $\sim 80\%$ of the LLSs have physical densities between 
$n_{\rm H}\sim10^{-3.5}-10^{-2}~\rm cm^{-3}$ for the assumed UV background, but 
we caution that a degeneracy between the ionisation parameter and the intensity of the radiation field 
prevents robust inference on the density and sizes of LLSs. Conversely, metallicity estimates 
are less sensitive to the assumptions behind ionisation corrections.
LLSs at $z>2$ are characterised by a broad unimodal distribution over $>4$ orders of magnitude, 
with a peak at $\log Z/Z_\odot\sim-2$. LLSs are metal poor, significantly less enriched than DLAs,
with $\sim70\%$ of the metallicity PDF below $\log Z/Z_\odot\le-1.5$.
The median metallicity of super LLSs with $\log N_{\rm HI}\ge19$ rapidly 
evolves with redshift, with a ten-fold increase between $z\sim2.1-3.6$ ($\sim1.5$ Gyr). 
Based on this sample, we find that LLSs at $z=2.5-3.5$ account for $\sim 15\%$ of all the metals produced by 
UV-selected galaxies. The implications for theories of cold gas accretion and metal ejection 
from galaxies are also discussed.
\end{abstract}


\section{Introduction}\label{sec:intro}

A detailed census of the metal budget in the Universe as a function of redshift provides valuable 
integral constraints to some of the most fundamental astrophysical processes that regulate galaxy 
evolution, including star formation, galactic outflows, and gas accretion \citep[e.g.][]{fuk98,bou07,pee14,raf14}. 
Both at high redshift ($z\sim 2.5$) and in the local Universe, approximately 
$\sim 1/3$ of the metals can be found within galaxies \citep{bou07,pee14}, 
with the more diffuse intergalactic medium (IGM) and circumgalactic medium (CGM)
being a significant repository of cosmic metals. Indeed, 
metals are ubiquitously detected in absorption line studies, 
albeit with the notable exception of a few pristine gas pockets \citep{fum11pri,sim12}.
Thus, the study of metallicity in absorption provides valuable insight into the chemical 
enrichment of cosmic structures across a wide range of densities, from the modest 
overdensities traced by the Ly$\alpha$ forest in the IGM \citep{sch03,sim04,sim11}, up to the
largest overdensities traced by damped Ly$\alpha$ systems (DLAs) found against quasars or $\gamma-$ray 
bursts sightlines \citep{pet94,pro03,raf12,cuc15}.

Until recently, a bottleneck in compiling a full census of cosmic metals was the lack of a systematic
study of the metallicity in large samples of Lyman limit systems (LLSs), defined as optically-thick clouds with 
neutral hydrogen column densities between $17.2 \le \log N_{\rm HI} < 20.3$\footnote{Throughout this paper, 
logarithmic column densities are expressed in units of $\rm cm^{-2}$.}. 
Differently from DLAs and the IGM, which have been the subject 
of dedicated campaigns to characterise their metal content, LLSs have received far less attention in 
past years, with most previous studies focusing on small samples, often restricted to the 
subset of so-called super LLSs (SLLSs) with $19 \le \log N_{\rm HI} < 20.3$.
Part of this unbalance can be attributed to the difficulties in measuring \NHI\ in optically-thick 
absorbers between $17.2 \le \log N_{\rm HI} < 19$, for which saturated hydrogen Lyman series lines and 
the lack of damping wings prevent precise estimates of the neutral hydrogen column density.
Furthermore, differently from the study of metals in neutral DLAs, LLSs are significantly ionised and 
observers can only access tracers of the underlying metal content of LLSs via metal lines. 
Detailed ionisation modelling  is thus needed to infer the intrinsic 
metal content of LLSs (see Section \ref{sec:grid}).

One of the first studies of the LLS metallicity was presented in 
\citet{ste90}, who analysed 8 systems between $2.90 < z < 3.23$ with column densities
$17.0 < \log N_{\rm HI} \le 19.3$, finding with photoionisation calculations metallicity between
$-3.0 < \log Z/Z_\odot \le -1.5$.
Since this study, however, the attention has focused mostly on SLLSs. 
For instance, \citet{des03} studied 12 SLLSs between $z\sim 1.8-4.3$ and conducted photoionisation 
modelling to assess the importance of ionisation corrections (ICs), concluding that generally
ICs are $< 0.2~\rm dex$ for $\log N_{\rm HI} \ge 19.3$ and thus considered negligible.
\citet{per07} and \citet{per08} studied the abundance of 6 SLLSs at $z<1.5$ and 13 SLLSs at $z\ge 3$.
Again using photoionisation modelling, they concluded that ICs are small compared to observational 
uncertainties, typically below $\sim 0.2~\rm dex$ and not exceeding $\sim 0.35~\rm dex$.
Together with values from the literature, they reported that the metallicity of SLLSs evolves more rapidly 
with redshift than for DLAs, and that SLLSs tend to have higher metallicity than DLAs especially at 
$z <2$ \citep[cf][]{kul07}. Nevertheless, SLLSs do not appear to substantially contribute to the total 
metal budget of cosmic structures at $z\sim 2.5$ \citep[but see][]{pro06}.
More recently, \citet{som13} presented the analysis of the abundances of 5 SLLSs between $1.7 < z < 2.4$ 
with the aid of ionisation models, finding that a varying degree of ionisation correction was needed 
for individual systems. Together with a large compilation of data from the literature, they 
also concluded that SLLSs are on average more enriched than DLAs, and may evolve faster with redshift 
\citep[see also][]{mei09,bat12}, a result that could be partially explained by
different selections as a function of redshift \citep{des09}.

In recent years, there has been a
renewed interest in probing the metal content of LLSs at lower column densities
in the context of CGM studies and the postulated connection between 
LLSs and cold accretion \citep[e.g.][]{fau11,fum11sim,van12,fum14}. 
\citet{fum13} presented the analysis of a composite spectrum of 38 LLSs at 
$z\sim 2.8$ together with simple ionisation modelling, finding evidence that LLSs have typical 
metallicity below $\log Z/Z_\odot \sim -1.5$. \citet{leh13} studied 28 partial LLSs and LLSs with column densities 
$16.2 \lesssim \log N_{\rm HI} \lesssim 18.5$ at $z\lesssim 1$, uncovering a bi-modal distribution with two branches 
peaking at $\log Z/Z_\odot \sim -1.6$ and $\log Z/Z_\odot \sim -0.3$.  
Finally, \citet{coo15} presented the analysis of 17 LLSs at $z \sim 3.2-4.4$, selected 
based on the lack of visible metal lines in Sloan Digital Sky Survey (SDSS) spectra. Together with 
the study of a small but representative sample of LLSs, after ionisation corrections, 
they offered additional evidence that high-redshift LLSs span a range of metallicity between 
$-3.0 < \log Z/Z_\odot \le -1.5$.

In this work, we aim to provide a coherent analysis of the metal content of a large sample of LLSs
from the High Dispersion Lyman Limit System (HD-LLS) survey \citep{pro15}, which includes 157 LLSs with column densities 
$17.3 \le \log N_{\rm HI} < 20.3$ between $z \sim 1.76 - 4.39$. This study represents a ten-fold increase 
in the sample size for analyses of the metal abundances of LLSs derived from 
high dispersion data (Section \ref{sec:obser}). By means of multiple grids of ionisation models,
combined with a Bayesian formalism and Markov Chain Monte Carlo (MCMC) techniques 
(Section \ref{sec:parfit}), we derive posterior probability distribution functions (PDFs) for 
the metallicity and the physical density of LLSs, also assessing the robustness of ICs (Section \ref{sec:grid})
and exploring possible systematic effects arising from different model assumptions.
We also include a homogeneous re-analysis of data 
for systems with $17.2 \le \log N_{\rm HI} < 20.3$ from the literature, especially at $z<2$.
Finally, we discuss the physical properties of LLSs (Section \ref{sec:metal}), both in the context of the cosmic 
budget of metals (Section \ref{sec:budget}) and of the properties of the CGM (Section \ref{sec:cgm}).
Summary and conclusions follow in Section \ref{sec:summary}. 
The readers who are primarily interested in the astrophysical implications of our work may wish to focus 
mainly on Section \ref{sec:metal}.
Throughout this work, we assume solar abundances from \citet{asp09} for which $Z_\odot = 0.0134$, 
and the ``Planck 2013'' cosmology \citep{pla14} with Hubble constant 
$H_0 = (67.8 \pm 0.78)~\rm km~s^{-1}~Mpc^{-1}$ and matter density parameter $\Omega_{\rm m} = 0.308 \pm 0.010$. 

\begin{table*}
\caption{List of the LLSs included in this study.}\label{tab:sample} 
\centering
\begin{tabular}{c c c c c c c c}
\hline
\hline
System &  Redshift  &	$\log N_{\rm HI}^a$  &   Reference  & System &  Redshift  &	$\log N_{\rm HI}$  &   Reference \\
\hline
\multicolumn{8}{c}{\it HDLLS statistical sample} \\
\hline
J000345-232346z2.187 & 2.187 & $19.65\pm0.15 $ & [P15]& J102509+045246z3.130 & 3.130 & $18.10\pm0.50*$ & [P15] \\
J003454+163920z3.754 & 3.754 & $20.05\pm0.20 $ & [P15]& J102832-004607z2.824 & 2.824 & $18.00\pm0.30 $ & [P15] \\
J004049-402514z2.816 & 2.816 & $17.55\pm0.15 $ & [P15]& J103249+054118z2.761 & 2.761 & $17.60\pm0.20 $ & [P15] \\
J010355-300946z2.908 & 2.908 & $19.10\pm0.15 $ & [P15]& J103456+035859z2.849 & 2.849 & $19.60\pm0.20 $ & [P15] \\							
J010516-184642z2.927 & 2.927 & $20.00\pm0.20 $ & [P15]& J103456+035859z3.003 & 3.003 & $19.10\pm0.15 $ & [P15] \\
J010619+004823z3.286 & 3.286 & $19.05\pm0.25 $ & [P15]& J103456+035859z3.059 & 3.059 & $19.15\pm0.25 $ & [P15] \\							
J010619+004823z3.321 & 3.321 & $19.10\pm0.20 $ & [P15]& J103514+544040z2.457 & 2.457 & $19.65\pm0.25 $ & [P15] \\							
J010619+004823z4.172 & 4.172 & $19.05\pm0.20 $ & [P15]& J103514+544040z2.846 & 2.846 & $19.70\pm0.15 $ & [P15] \\
J012156+144823z2.662 & 2.662 & $19.25\pm0.20 $ & [P15]& J104018+572448z3.266 & 3.266 & $18.30\pm0.60*$ & [P15] \\
J012403+004432z3.078 & 3.078 & $20.20\pm0.20 $ & [P15]& J110325-264506z1.839 & 1.839 & $19.40\pm0.15 $ & [P15] \\
J012700-004559z2.944 & 2.944 & $19.80\pm0.20 $ & [P15]& J110708+043618z2.601 & 2.601 & $19.90\pm0.20 $ & [P15] \\
J013340+040059z3.995 & 3.995 & $20.10\pm0.30 $ & [P15]& J111008+024458z3.476 & 3.476 & $18.10\pm0.40*$ & [P15] \\							
J013340+040059z4.117 & 4.117 & $18.60\pm0.80*$ & [P15]& J111113-080402z3.481 & 3.481 & $20.00\pm0.20 $ & [P15] \\							
J013421+330756z4.279 & 4.279 & $17.70\pm0.15 $ & [P15]& J111113-080402z3.811 & 3.811 & $18.20\pm0.50*$ & [P15] \\
J014850-090712z2.995 & 2.995 & $17.55\pm0.15 $ & [P15]& J113130+604420z2.362 & 2.362 & $20.05\pm0.15 $ & [P15] \\
J015741-010629z2.631 & 2.631 & $19.45\pm0.20 $ & [P15]& J113418+574204z3.410 & 3.410 & $17.97\pm0.19 $ & [P15] \\
J015741-010629z3.385 & 3.385 & $18.35\pm0.75*$ & [P15]& J113621+005021z3.248 & 3.248 & $18.10\pm0.60*$ & [P15] \\
J020455+364918z1.955 & 1.955 & $20.10\pm0.20 $ & [P15]& J115659+551308z2.616 & 2.616 & $19.10\pm0.30 $ & [P15] \\
J020455+364918z2.690 & 2.690 & $18.15\pm0.65*$ & [P15]& J115906+133737z3.723 & 3.723 & $19.90\pm0.15 $ & [P15] \\
J020944+051717z3.988 & 3.988 & $18.00\pm0.30 $ & [P15]& J115940-003203z1.904 & 1.904 & $20.05\pm0.15 $ & [P15] \\							
J020950-000506z2.523 & 2.523 & $19.05\pm0.15 $ & [P15]& J120331+152254z2.708 & 2.708 & $18.30\pm0.70*$ & [P15] \\
J020951-000513z2.523 & 2.523 & $19.00\pm0.20 $ & [P15]& J120918+095427z2.363 & 2.363 & $20.25\pm0.20 $ & [P15] \\
J020951-000513z2.547 & 2.547 & $18.10\pm0.50*$ & [P15]& J120918+095427z3.023 & 3.023 & $19.20\pm0.30 $ & [P15] \\
J023903-003850z2.868 & 2.868 & $18.80\pm0.45*$ & [P15]& J121539+090608z2.523 & 2.523 & $20.20\pm0.20 $ & [P15] \\
J024122-363319z2.739 & 2.739 & $18.00\pm0.70*$ & [P15]& J124820+311043z4.075 & 4.075 & $19.95\pm0.15 $ & [P15] \\
J030341-002322z2.443 & 2.443 & $19.90\pm0.15 $ & [P15]& J124957-015928z3.530 & 3.530 & $18.10\pm0.40*$ & [P15] \\
J030341-002322z2.941 & 2.941 & $18.60\pm0.30*$ & [P15]& J125336-022808z3.603 & 3.603 & $19.35\pm0.20 $ & [P15] \\
J033854-000520z2.746 & 2.746 & $20.00\pm0.20 $ & [P15]& J125759-011130z2.918 & 2.918 & $19.95\pm0.20 $ & [P15] \\
J033900-013318z3.116 & 3.116 & $19.50\pm0.20 $ & [P15]& J130756+042215z2.250 & 2.250 & $20.00\pm0.15 $ & [P15] \\
J034024-051909z2.174 & 2.174 & $19.35\pm0.20 $ & [P15]& J130756+042215z2.749 & 2.749 & $18.20\pm0.60*$ & [P15] \\
J034227-260243z3.012 & 3.012 & $18.10\pm0.30 $ & [P15]& J132554+125546z3.767 & 3.767 & $19.60\pm0.20 $ & [P15] \\
J034402-065300z3.843 & 3.843 & $19.55\pm0.15 $ & [P15]& J132729+484500z3.058 & 3.058 & $19.35\pm0.25 $ & [P15] \\
J042610-220217z4.175 & 4.175 & $17.40\pm0.15 $ & [P15]& J133146+483826z2.910 & 2.910 & $19.65\pm0.35*$ & [P15] \\
J043906-504740z2.796 & 2.796 & $18.10\pm0.60*$ & [P15]& J133254+005251z3.421 & 3.421 & $19.20\pm0.20 $ & [P15] \\
J045142-132033z2.998 & 2.998 & $17.55\pm0.15 $ & [P15]& J133757+021820z3.270 & 3.270 & $19.95\pm0.15 $ & [P15] \\
J073149+285448z3.608 & 3.608 & $18.15\pm0.45*$ & [P15]& J133942+054822z2.952 & 2.952 & $17.65\pm0.25 $ & [P15] \\
J073621+651312z2.909 & 2.909 & $18.30\pm0.70*$ & [P15]& J134002+110630z2.508 & 2.508 & $20.15\pm0.15 $ & [P15] \\
J075155+451619z2.927 & 2.927 & $19.80\pm0.20 $ & [P15]& J134811+281802z2.448 & 2.448 & $19.85\pm0.15 $ & [P15] \\							
J081054+460358z3.472 & 3.472 & $19.90\pm0.30 $ & [P15]& J134816-013509z2.883 & 2.883 & $18.60\pm0.70*$ & [P15] \\							
J081435+502946z3.004 & 3.004 & $19.75\pm0.15 $ & [P15]& J134939+124230z3.158 & 3.158 & $19.60\pm0.30 $ & [P15] \\							
J081618+482328z3.343 & 3.343 & $18.30\pm0.50*$ & [P15]& J135706-174401z3.007 & 3.007 & $19.40\pm0.25 $ & [P15] \\							
J082619+314848z2.856 & 2.856 & $19.40\pm0.20 $ & [P15]& J140243+590958z2.986 & 2.986 & $19.30\pm0.30 $ & [P15] \\
J082849+085854z2.044 & 2.044 & $19.90\pm0.10 $ & [P15]& J140248+014634z3.456 & 3.456 & $19.20\pm0.30 $ & [P15] \\
J085959+020519z2.845 & 2.845 & $17.90\pm0.60*$ & [P15]& J140747+645419z2.935 & 2.935 & $20.20\pm0.20 $ & [P15] \\
J091210+054742z2.522 & 2.522 & $19.35\pm0.20 $ & [P15]& J142903-014518z3.427 & 3.427 & $18.00\pm0.40*$ & [P15] \\
J091546+054942z2.663 & 2.663 & $18.20\pm0.70*$ & [P15]& J145408+511443z3.231 & 3.231 & $20.05\pm0.15 $ & [P15] \\
J092459+095103z3.219 & 3.219 & $19.30\pm0.20 $ & [P15]& J145649-193852z2.170 & 2.170 & $19.75\pm0.20 $ & [P15] \\							
J092705+562114z1.775 & 1.775 & $19.00\pm0.10 $ & [P15]& J145649-193852z2.351 & 2.351 & $19.55\pm0.20 $ & [P15] \\
J093153-000051z2.927 & 2.927 & $19.25\pm0.25 $ & [P15]& J145807+120937z2.648 & 2.648 & $18.35\pm1.05*$ & [P15] \\
J094253-110425z2.917 & 2.917 & $17.50\pm0.15 $ & [P15]& J145907+002401z2.767 & 2.767 & $20.00\pm0.20 $ & [P15] \\
J094932+033531z3.311 & 3.311 & $19.85\pm0.15 $ & [P15]& J150654+522005z4.114 & 4.114 & $18.25\pm0.65*$ & [P15] \\
J095256+332939z3.144 & 3.144 & $19.95\pm0.20 $ & [P15]& J150932+111313z1.821 & 1.821 & $18.50\pm0.50*$ & [P15] \\							 
J095256+332939z3.211 & 3.211 & $19.90\pm0.20 $ & [P15]& J151047+083535z2.722 & 2.722 & $19.40\pm0.40*$ & [P15] \\							 
J095256+332939z3.262 & 3.262 & $20.00\pm0.30 $ & [P15]& J155036+053749z2.980 & 2.980 & $19.75\pm0.25 $ & [P15] \\							 
J095309+523029z1.768 & 1.768 & $20.10\pm0.10 $ & [P15]& J155103+090849z2.700 & 2.700 & $17.50\pm0.20 $ & [P15] \\
J095542+411655z2.812 & 2.812 & $19.90\pm0.15 $ & [P15]& J155556+480015z3.131 & 3.131 & $19.60\pm0.15 $ & [P15] \\						       
J100428+001825z2.746 & 2.746 & $19.80\pm0.20 $ & [P15]& J155738+232057z2.773 & 2.773 & $19.40\pm0.40*$ & [P15] \\							 
J101539+111815z2.870 & 2.870 & $18.20\pm0.70*$ & [P15]& J155810-003120z2.630 & 2.630 & $19.60\pm0.20 $ & [P15] \\
J101939+524627z1.834 & 1.834 & $19.10\pm0.30 $ & [P15]& J160843+071508z1.763 & 1.763 & $19.40\pm0.30 $ & [P15] \\
\hline    
\end{tabular}
\flushleft{$^a$ Asterisks mark uncertain column densities for which we assume a flat distribution centred on the listed value
and with half width defined by the quoted error.}
\end{table*}

\begin{table*}
\contcaption{List of the LLSs included in this study.} 
\centering
\begin{tabular}{c c c c c c c c}
\hline
\hline
System &  Redshift  &	$\log N_{\rm HI}$  &   Reference  & System &  Redshift  &	$\log N_{\rm HI}$  &   Reference \\
\hline
\multicolumn{8}{c}{\it HDLLS statistical sample (continued)} \\
\hline
J161545+060852z2.988 & 2.988 & $19.00\pm0.50*$ & [P15]&  J212916+003756z2.735 & 2.735 & $20.10\pm0.20 $ & [P15] \\
J162116-004250z3.105 & 3.105 & $19.80\pm0.20 $ & [P15]&  J212916+003756z2.917 & 2.917 & $18.10\pm0.40*$ & [P15] \\
J171227+575506z2.315 & 2.315 & $20.20\pm0.15 $ & [P15]&  J214144-384041z2.893 & 2.893 & $20.00\pm0.15 $ & [P15] \\
J171227+575506z2.849 & 2.849 & $18.10\pm0.50*$ & [P15]&  J220639-181846z2.698 & 2.698 & $20.00\pm0.15 $ & [P15] \\
J172323+224358z4.391 & 4.391 & $18.25\pm0.25 $ & [P15]&  J223408+000001z2.652 & 2.652 & $19.00\pm0.30*$ & [P15] \\
J173352+540030z2.779 & 2.779 & $19.70\pm0.20 $ & [P15]&  J223438+005730z2.604 & 2.604 & $19.50\pm0.25 $ & [P15] \\
J173352+540030z3.151 & 3.151 & $18.50\pm0.60*$ & [P15]&  J223819-092106z3.128 & 3.128 & $18.35\pm0.65*$ & [P15] \\
J183753-584809z2.729 & 2.729 & $18.10\pm0.60*$ & [P15]&  J224147+135203z3.654 & 3.654 & $20.20\pm0.20 $ & [P15] \\
J200324-325144z3.172 & 3.172 & $19.75\pm0.15 $ & [P15]&  J230301-093930z3.312 & 3.312 & $17.90\pm0.20 $ & [P15] \\
J200324-325144z3.188 & 3.188 & $19.88\pm0.13 $ & [P15]&  J231543+145606z2.943 & 2.943 & $18.80\pm0.30 $ & [P15] \\
J200324-325144z3.548 & 3.548 & $18.03\pm0.23 $ & [P15]&  J231543+145606z3.135 & 3.135 & $19.95\pm0.15 $ & [P15] \\
J205344-354652z2.332 & 2.332 & $19.00\pm0.25 $ & [P15]&  J231643-334912z2.386 & 2.386 & $19.00\pm0.20 $ & [P15] \\
J205344-354652z2.350 & 2.350 & $19.60\pm0.25 $ & [P15]&  J231934-104036z2.675 & 2.675 & $19.45\pm0.15 $ & [P15] \\
J205344-354652z2.989 & 2.989 & $20.10\pm0.15 $ & [P15]&  J232340+275800z3.267 & 3.267 & $19.20\pm0.60*$ & [P15] \\
J205344-354652z3.094 & 3.094 & $19.05\pm0.15 $ & [P15]&  J232340+275800z3.565 & 3.565 & $19.15\pm0.35*$ & [P15] \\
J205344-354652z3.172 & 3.172 & $18.25\pm0.55*$ & [P15]&  J233446-090812z3.226 & 3.226 & $17.70\pm0.30 $ & [P15] \\
J212329-005052z2.059 & 2.059 & $19.25\pm0.15 $ & [P15]&  J234855-144436z2.775 & 2.775 & $17.50\pm0.20 $ & [P15] \\
J212912-153841z2.638 & 2.638 & $19.25\pm0.15 $ & [P15]&  J235057-005209z2.930 & 2.930 & $18.15\pm0.75*$ & [P15] \\
J212912-153841z2.769 & 2.769 & $19.20\pm0.20 $ & [P15]&  J235833-544042z2.895 & 2.895 & $17.40\pm0.20 $ & [P15] \\
J212912-153841z2.968 & 2.968 & $17.30\pm0.20 $ & [P15]& 							\\
\hline
\multicolumn{8}{c}{\it Literature sample} \\
\hline
J000520+052410z0.851&  0.851 & $19.08\pm0.04$&  [M09]   &     J122037-004032z0.975&  0.975 & $20.20\pm0.07$&  [M08]   \\
J001210-012207z1.386&  1.386 & $20.26\pm0.05$&  [M09]   &     J122414+003709z1.266&  1.266 & $20.00\pm0.07$&  [M07]   \\
J002127+010420z1.326&  1.326 & $20.04\pm0.11$&  [M09]   &     J122607+173649z2.557&  2.556 & $19.32\pm0.15$&  [D03]   \\
J002133+004300z0.520&  0.520 & $19.54\pm0.05$&  [D09]   &     J122836+101841z0.938&  0.938 & $19.41\pm0.05$&  [M08]   \\
J002133+004300z0.942&  0.942 & $19.38\pm0.13$&  [D09]   &     J131119-012030z1.762&  1.762 & $20.00\pm0.08$&  [S13]   \\
J011800+032000z4.128&  4.128 & $20.02\pm0.15$&  [P07]   &     J131956+272808z0.661&  0.661 & $18.30\pm0.30$&  [K12]   \\
J012126+034707z2.976&  2.976 & $19.53\pm0.10$&  [P07]   &     J132323-002155z0.716&  0.716 & $20.21\pm0.20$&  [P06b]  \\
J012403+004432z2.988&  2.988 & $19.18\pm0.10$&  [P07]   &     J133007-205616z0.853&  0.853 & $19.40\pm0.05$&  [M08]   \\
J013340+040100z3.139&  3.139 & $19.01\pm0.10$&  [P07]   &     J141217+091625z2.668&  2.668 & $19.75\pm0.10$&  [D03]   \\
J013405+005109z0.842&  0.842 & $19.93\pm0.15$&  [P06a]  &     J143511+360437z0.203&  0.203 & $19.80\pm0.10$&  [B12]   \\
J013724-422417z3.101&  3.101 & $19.81\pm0.10$&  [P07]   &     J143645-005150z0.738&  0.738 & $20.08\pm0.11$&  [M08]   \\
J013724-422417z3.101&  3.665 & $19.11\pm0.10$&  [P07]   &     J144653+011355z2.087&  2.087 & $20.18\pm0.10$&  [D03]   \\
J015428+044818z0.160&  0.160 & $19.48\pm0.10$&  [S15]   &     J145418+121054z2.255&  2.255 & $20.30\pm0.15$&  [D03]   \\
J015733-004824z1.416&  1.416 & $19.90\pm0.07$&  [D09]   &     J145418+121054z3.171&  3.171 & $19.70\pm0.15$&  [D03]   \\
J021857+081728z1.769&  1.769 & $20.20\pm0.10$&  [D09]   &     J145508-004507z1.093&  1.093 & $20.08\pm0.06$&  [M08]   \\
J031155-765151z0.203&  0.203 & $18.22\pm0.20$&  [L09]   &     J151326+084850z2.088&  2.088 & $19.47\pm0.10$&  [D03]   \\
J035405-272425z1.405&  1.405 & $20.18\pm0.15$&  [M07]   &     J152510+002633z0.567&  0.567 & $19.78\pm0.08$&  [D09]   \\
J042707-130253z1.408&  1.408 & $19.04\pm0.05$&  [M09]   &     J155103+090849z2.320&  2.320 & $19.70\pm0.05$&  [S13]   \\
J044117-431343z0.101&  0.101 & $19.63\pm0.15$&  [S15]   &     J155304+354828z0.083&  0.083 & $19.55\pm0.15$&  [B12]   \\
J045608-215909z0.474&  0.474 & $19.45\pm0.05$&  [S15]   &     J163145+115602z0.900&  0.900 & $19.70\pm0.05$&  [M09]   \\
J082601-223026z0.911&  0.911 & $19.04\pm0.05$&  [M07]   &     J163428+703132z1.041&  1.041 & $17.23\pm0.15$&  [Z04]   \\
J092554+400414z0.248&  0.248 & $19.55\pm0.15$&  [B12]   &     J205145+195006z1.116&  1.116 & $20.00\pm0.15$&  [M09]   \\
J092837+602521z0.154&  0.154 & $19.35\pm0.15$&  [B12]   &     J210244-355306z2.507&  2.507 & $20.21\pm0.10$&  [D03]   \\
J100102+594414z0.303&  0.303 & $19.32\pm0.10$&  [B12]   &     J211927-353741z1.996&  1.996 & $20.06\pm0.10$&  [D03]   \\
J100902+071343z0.356&  0.356 & $18.40\pm0.41$&  [T11]   &     J213135-120705z0.430&  0.429 & $19.18\pm0.05$&  [S15]   \\
J100930-002619z0.843&  0.843 & $20.20\pm0.06$&  [M07]   &     J215145+213013z1.002&  1.002 & $19.30\pm0.10$&  [N08]   \\
J100930-002619z0.887&  0.887 & $19.48\pm0.05$&  [M07]   &     J215501-092224z0.081&  0.081 & $17.98\pm0.05$&  [J05]   \\
J101033-004724z1.327&  1.327 & $19.81\pm0.05$&  [M07]   &     J215502+135826z3.142&  3.142 & $19.94\pm0.10$&  [D03]   \\
J102837-010028z0.632&  0.632 & $19.95\pm0.07$&  [D09]   &     J215502+135826z3.565&  3.565 & $19.37\pm0.15$&  [D03]   \\
J102837-010028z0.709&  0.709 & $20.04\pm0.06$&  [D09]   &     J215502+135826z4.212&  4.212 & $19.61\pm0.10$&  [D03]   \\
J103744+002809z1.424&  1.424 & $20.04\pm0.12$&  [M08]   &     J221527-161133z3.656&  3.656 & $19.01\pm0.15$&  [P07]   \\
J103921-271916z2.139&  2.139 & $19.55\pm0.15$&  [S13]   &     J221527-161133z3.662&  3.662 & $20.05\pm0.15$&  [P07]   \\
J103921-271916z2.139&  2.139 & $19.60\pm0.10$&  [D09]   &     J221651-671443z3.368&  3.368 & $19.80\pm0.10$&  [P07]   \\
J105440-002048z0.830&  0.830 & $18.95\pm0.18$&  [M08]   &     J233121+003807z1.141&  1.141 & $20.00\pm0.05$&  [M07]   \\
J105440-002048z0.951&  0.951 & $19.28\pm0.05$&  [M08]   &     J234403+034226z3.882&  3.882 & $19.50\pm0.10$&  [D03]   \\
J110325-264515z1.838&  1.838 & $19.50\pm0.05$&  [D03]   &     J235253-002851z0.873&  0.873 & $19.18\pm0.09$&  [M09]   \\
J110736+000328z0.954&  0.954 & $20.26\pm0.14$&  [M06]   &     J235253-002851z1.032&  1.032 & $19.81\pm0.13$&  [M09]   \\
J121549-003432z1.554&  1.554 & $19.56\pm0.05$&  [M08]   &     J235253-002851z1.247&  1.247 & $19.60\pm0.24$&  [M09]   \\
J121920+063838z0.006&  0.006 & $19.32\pm0.03$&  [T05]   &     \\
\hline    
\end{tabular}
\flushleft{References: 
[D03] \citet{des03}; 
[Z04] \citet{zon04};
[T05] \citet{tri05}; 
[J05] \citet{jen05}; 
[P06a] \citet{per06a}; 
[P06b] \citet{per06b}; 
[P07] \citet{per07}; 
[M07] \citet{mei07}; 
[M08] \citet{mei08}; 
[N08] \citet{nes08};
[M09] \citet{mei09dat}; 
[D09] \citet{des09}; 
[L09] \citet{leh09};
[T11] \citet{tum11}; 
[K12] \citet{kac12}; 
[B12] \citet{bat12}; 
[S13] \citet{som13}; 
[S15] \citet{som15}; 
[P15] \citet{pro15}.}
\end{table*}

\begin{figure}
\centering
\includegraphics[scale=0.45]{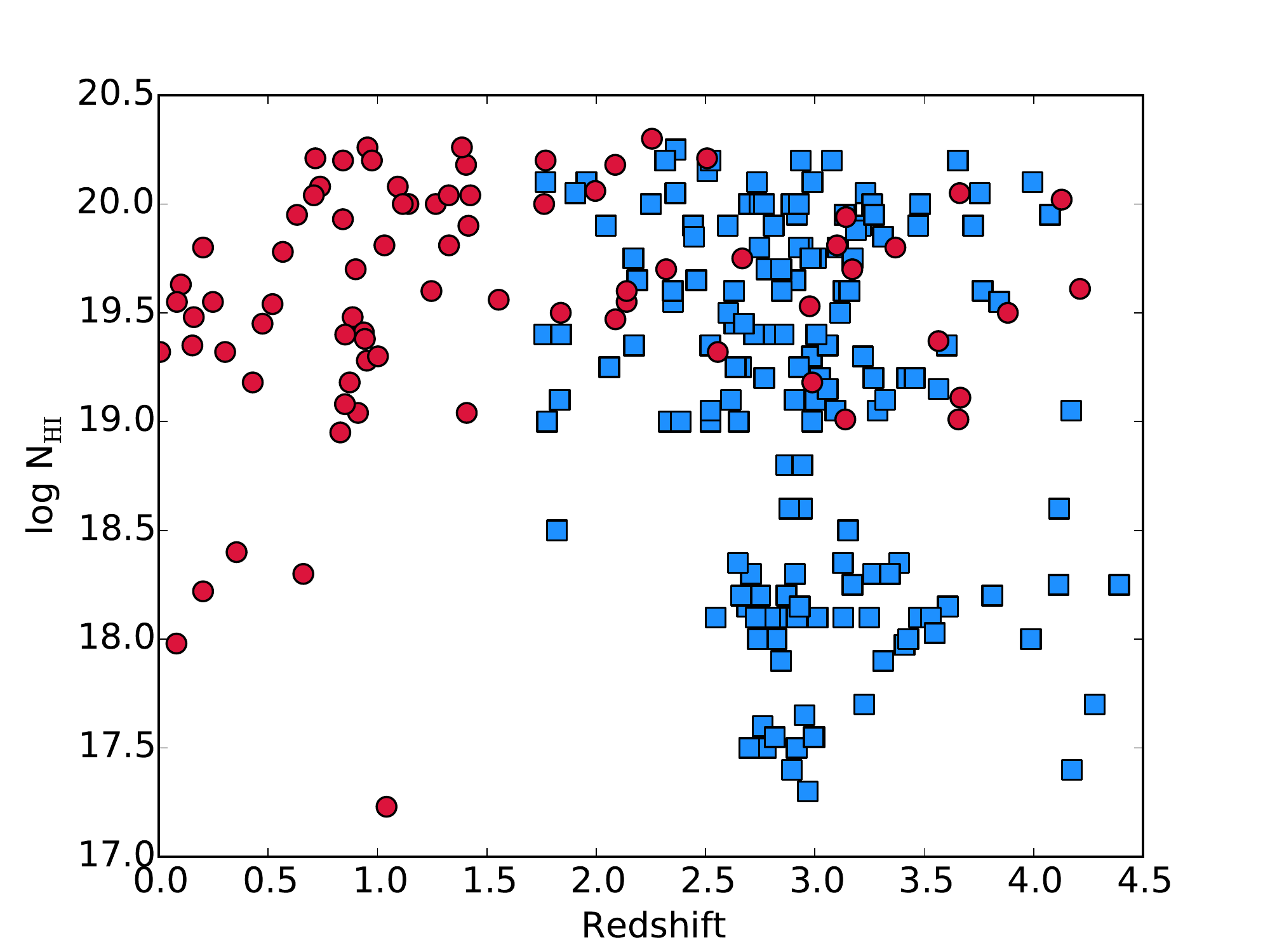}
\caption{Scatter plot of the redshift and hydrogen column density of the LLSs included in this study.
The HD-LLS sample is shown with blue squares, while the red circles mark data from the literature.}\label{fig:sample}
\end{figure}

\section{Observations}\label{sec:obser}

Our primary dataset includes 157 LLSs from the HD-LLS survey presented in \citet{pro15}, which is
composed by an \HI\ selected sample of optically-thick absorbers between $z = 1.76 - 4.39$.
Systems from this sample have been observed at high resolution with echelle or echellette 
spectrographs at the Keck and Magellan telescopes.  We refer the reader 
to the work by \citet{pro15} for additional details on the observations, data reduction and analysis,
including the measurement of column densities for hydrogen and metal ions. 
Relevant to this analysis, column densities have been measured using the apparent optical depth method
\citep{sav91} focusing on transitions outside the Ly$\alpha$ forest. This means that 
common transitions such as \ion{O}{VI}, \ion{C}{III}, \ion{Si}{III} are not included for the
majority of the systems analysed. The implications of this restriction will be discussed below.

In this study, we also collect 77 LLSs (strictly with $17.3 \le \log N_{\rm HI} < 20.3$) 
from the literature, when ion column densities are available (see Table \ref{tab:sample}), bringing
the total sample to 234 systems. Although we analyse literature data together with our own sample in 
a self-consistent manner, in the following we will primarily refer to the HD-LLS subset for a statistical analysis 
of the LLS population, thus avoiding possible selection biases from heterogeneous compilations of LLSs in the 
literature. A list of the LLSs included in this study and a summary of their redshifts and \HI\ column densities
is provided in Table \ref{tab:sample} and Figure \ref{fig:sample}. 
As highlighted in Figure \ref{fig:sample}, by leveraging the HD-LLS sample, our analysis significantly 
augments the samples used in previous studies, especially for $17.3 \le \log N_{\rm HI} < 19.0$.

\section{Ionisation modelling and parameter estimation}\label{sec:parfit}

Differently from the higher column-density DLAs, the bulk of the gas in SLLSs and LLSs is not fully
neutral, and therefore the observed ions are only tracers of the underlying chemical abundance 
of the gas. It is therefore necessary to apply ICs to translate the observed ion 
abundance into a measurement of the gas phase metallicity. The standard technique followed 
by many authors is to compute ICs relying on parametric one-dimensional radiative transfer calculations,
typically at equilibrium, in which radiation with specified spectral properties strikes the face of a
``cloud'' that is generally assumed to be a slab of homogeneous medium with specified physical characteristics
(e.g. density and metallicity). This problem is very tractable from a theoretical point of view, thanks to 
publicly-available radiative transfer codes. However, in real astrophysical environments, many of the parameters 
that specify the geometry of the problem, the incident radiation field, or the sources of opacity in the 
radiative transfer equation, are unconstrained. Thus, common practice is to generate large grids of models, and 
to use observables to constrain the unknown parameters. 

Once a parametric grid of models is at hand, one wishes to identify the ``best'' set of parameters ${\bf \theta}$
by comparing observables ${\bf N}$ 
with model predictions ${\bf M}$. In this context, ${\bf N}$ is a set of
column densities $N_{\rm x}$ and associated errors $\sigma_{\rm x}$, while ${\bf M}$ is a set
of column densities $\bar N_{\rm x}$ 
computed from parametric ionisation models. Here, the vector index for  ${\bf \theta}$
runs over the $D$ parameters that define the dimension of the problem, and the vector indexes for ${\bf M}$ and 
${\bf N}$ run over all the column densities of $i$-th elements in $j$-th ionisation stages. 
In the following, we will adopt a Bayesian formalism \citep[see also][]{cri15,coo15}, with which we 
can explore the posterior PDFs for parameters of interest (e.g. density and metallicity), 
after marginalising over additional nuisance parameters 
that describe the radiative transfer problem (e.g. dust, local sources, temperature). 

More specifically, the joint posterior PDFs of the parameters given the observables and the models is defined by
\begin{equation}\label{eq:bay}
p(\para | {\bf N,M}) = \frac{p(\para|{\bf M}) p({\bf N}|\para,{\bf M})}{p({\bf N}|{\bf M})}\:,
\end{equation}
with  $p(\para | {\bf M})$ the prior on the parameters given the models, $p({\bf N}|\para,{\bf M})$
is the likelihood of the data given the models, and $p({\bf N}|{\bf M})$ is the marginal distribution. 
In this work, the likelihood is defined as
product of Gaussian functions
\begin{equation}\label{eq:like}
p({\bf N}|\para,{\bf M}) = \prod_{x=1}^{i\times j} \frac{1}{\sqrt{2\pi}\sigma_x}\exp \left(-\frac{(N_{\rm x}-\bar N_{\rm x}(\para))^2}{2\sigma_{\rm x}^2}\right)\:.
\end{equation}
In presence of upper or lower limits for a given ion, the corresponding term in the product of the right-hand 
side of Equation \ref{eq:like} is replaced by a rescaled cumulative distribution function
or Q-function, respectively. Throughout this work, 
we will attribute equal probability to the values of unknown parameters (e.g. volume density, metallicity) 
by means of flat priors. For the redshift and the \ion{H}{I} column density, which are measured for each individual
system, we assume instead Gaussian priors centred at the observed values. The only exception is for  
LLSs with saturated Lyman series lines, for which we assume a top-hat function between the minimum and maximum values 
allowed by observations \citep[for more details see][and Table \ref{tab:sample}]{pro15}. 
To reconstruct the posterior PDFs for individual systems, we sample 
the full parameter space using {\sc emcee} \citep{for13}, an affine invariant MCMC ensemble 
sampler. In Appendix \ref{sec:valid}, we present results of the analysis of mock data to validate 
our procedure. 

We stress that an underlying assumption of this method is that we are comparing model predictions to 
observations after integrating the ionic column densities over the entire depth of the absorbing cloud. 
This means that, effectively, we are smoothing possible dishomogeneities in the metal 
distribution of individual LLSs \citep[e.g.][]{pro10}. Furthermore, the number of transitions 
included in the analysis of each system varies according to the wavelength range of the spectra, the 
data quality, and the redshift of the LLS. We refer the readers to \citet{pro15} for details on individual 
absorbers, noting that about 8 metal transitions are included for a typical LLS.

\section{Systematic uncertainties in the ionisation corrections}\label{sec:grid}

In this work, we use the {\sc Cloudy} \citep[c13.03;][]{fer13} radiative transfer code
to construct grids of ionisation models. Given the limited number of transitions accessible in spectra, 
it is generally not practical to constrain all the possible free parameters relevant to this radiative 
transfer problem. Thus, a few simplifying assumptions are often necessary, or at least routinely applied in the 
literature. Nevertheless, understanding the implications of the assumptions made in the derivation of 
physical quantities is of clear importance, despite being a computationally-intensive task. 
Subtle systematic errors or any unexplored degeneracy may in fact taint the final results. 

In this section, before turning our attention to the statistical analysis of the LLS metallicity 
and its astrophysical implications, we aim to discuss the robustness of the applied IC by comparing the 
inferred distributions of metallicity and density under varying model assumptions. 
Our goal is to provide a simple but quantitative assessment of the degeneracy related 
to ICs in this type of work. We will start by considering a ``minimal'' model, that is the model with 
the least number of free parameters. Next, we will compare the results from this simple but widely-used 
model with results from more complex models, in which additional degrees of freedom are introduced.
In practice, it may not always be possible to adopt heavily parametric models, but these different 
calculations should offer, to first order, an estimate of the systematic uncertainties of our results.  

\begin{figure*}
\centering
\includegraphics[scale=0.68]{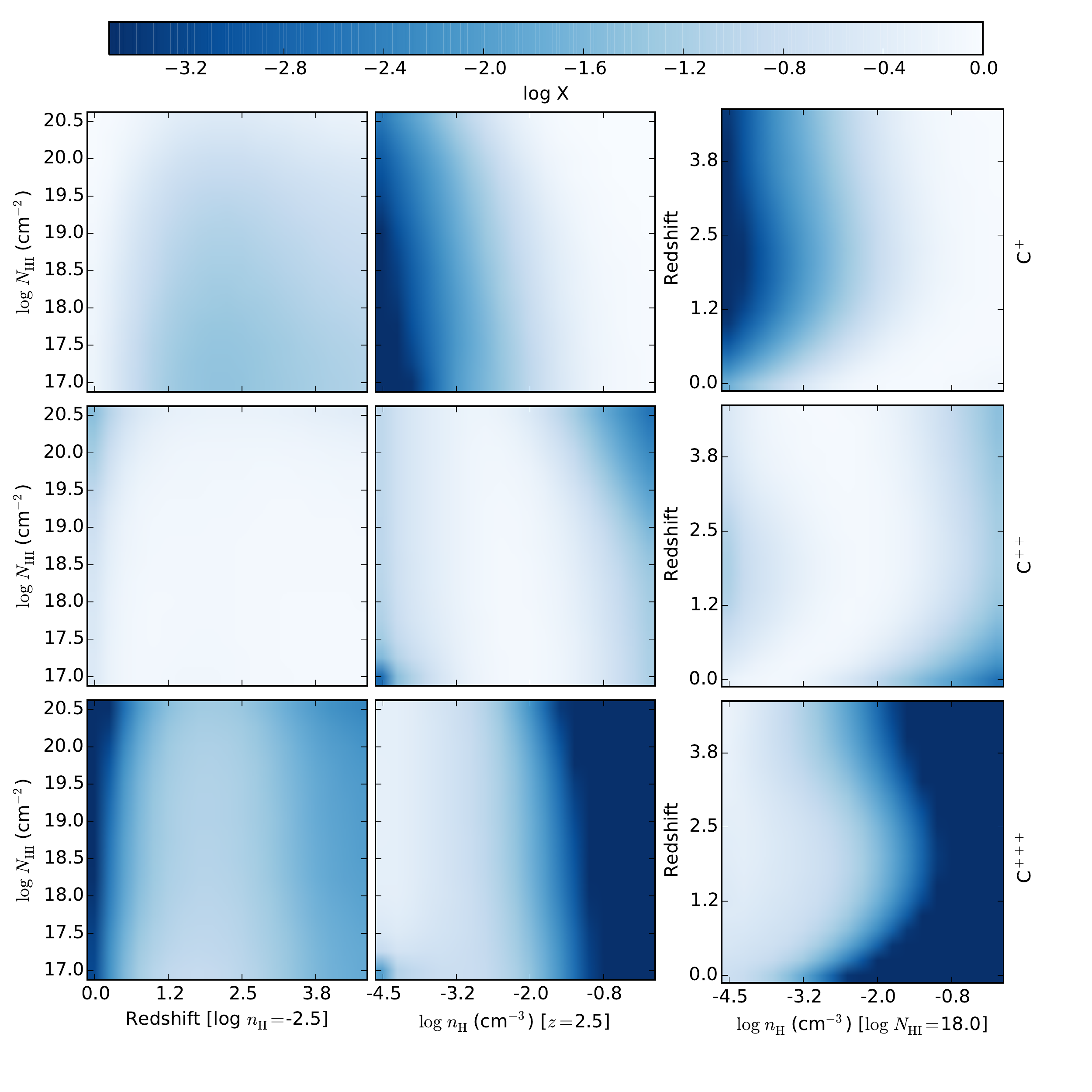}
\caption{Slices of the fraction of carbon in three ionisation stages  (C$^+$ top, 
C$^{++}$ middle, C$^{+++}$ bottom) as predicted by the minimal photoionisation model
as a function of density, redshift, and column density.
The ion fractions are shown as a function of redshift and \NHI\ 
for $n_{\rm H}=10^{-2.5}~\rm cm^{-3}$ (left), of \NHI\ 
and $n_{\rm H}$ for $z=2.5$ (centre), and of redshift and $n_{\rm H}$ for $\log N_{\rm HI} = 18$ 
(right).}\label{fig:carb}
\end{figure*}

\subsection{Minimal model}\label{sec:mingrid}

\begin{table}
\caption{Free paramaters in the minimal grid of models}\label{tab:minimal} 
\centering
\begin{tabular}{c c c c}
\hline
\hline
Parameter   		        &  Min.     &	  Max.  &   Step    \\
            		        &	    &		&	    \\
\hline  
$\log Z/Z_\odot$                &     -4.0  &	  1.0	&     0.25   \\ %
$z$      	                &      0.0  &	  4.5	&     0.25   \\ %
$\log N_{\rm HI}$ [cm$^{-2}$]   &     17.0  &	  20.5	&     0.25   \\ %
$\log n_{\rm H}$ [cm$^{-3}$] 	&     -4.5  &	   0.0	&     0.25   \\ %
\hline    
\end{tabular}
\flushleft{The columns of the table are: (1) the free parameter as described in the text; (2) the minimum allowed value; 
(3) the maximum allowed value; (4) the step adopted in the grid.}
\end{table}

Our minimal model consists of a static gas slab of constant density $n_{\rm H}$, and thus we restrict our analysis 
to a single phase medium. This gas slab is illuminated on one side by the redshift-dependent metagalactic
UV background radiation $J_{\rm uvb}(z)$ as specified by the \citet[][]{haa12} model.
The cosmic microwave background is also included. The geometry of the problem is specified by the 
neutral hydrogen column density $N_{\rm HI}$, which in turn defines the depth of the cloud $\Delta r$.
In the minimal model, we do not include the effects of grains, and we assume that all the 
metals are in the gas phase. The relative abundance of each element $A_{i}$ is assumed to be consistent with 
the solar neighbourhood, as compiled in \citet{asp09}. 
Each {\sc Cloudy} model is iterated until a converged solution is reached. 
Table \ref{tab:minimal} summarises the parameter space covered by this minimal grid of models.
The grid has dimension $D = 4$, but only two parameters (density and metallicity) are 
unconstrained, as both redshift and column density can be directly measured with varying degree of accuracy. 
Throughout this analysis, we focus on a set of ions that are commonly detected in our sample of 
LLSs \citep[see e.g.][]{pro15}.

Before diving into the derivation of the density and metallicity posterior PDFs, 
we briefly discuss how ICs vary with the free parameters under consideration\footnote{In this work, 
$X_{\rm E^j} \equiv \left(n_{i,j}/n_{i}\right)$ defines the fraction of the $i$-th element ``E''
that is found in the $j$-th ionisation stage. Thus, as an example, $X_{\rm H^0}$ is the neutral fraction of 
hydrogen.}. As an example, Figure \ref{fig:carb} shows the fraction of carbon in three ionisation stages as 
a function of density, redshift, and column density. Well known trends from photoionisation 
calculations can be seen in this figure. 

Considering low ions first (e.g. C$^+$), one can see that $X_{\rm C^+}$ increases towards high 
column and volume densities with curves of constant ionisation fraction that are approximately 
diagonal in the $\log n_{\rm H}$--$\log N_{\rm HI}$ plane. Also visible is the effect of the 
redshift-dependent photoionisation rate, $\Gamma_{\rm HI}$, which is encoded in the \citet{haa12} model.
Due to the evolution of the UVB, the lowest neutral fraction can be observed at $z\sim 2$ for a
fixed column density (see right panels), with a rapid increase towards $z\sim 0$ as $\Gamma_{\rm HI}$  plunges. 
Similar features are also commonly found in the variation of $X$  with redshift and density
for common low-ionisation species (e.g. for Si$^+$, O$^0$, H$^0$, Al$^+$, Fe$^+$, Mg$^+$).  
Figure \ref{fig:carb} also highlights how the fraction of progressively more 
ionised species shifts with density at constant redshift. At $z\sim 2.5$  
most of the carbon is singly ionised for $\log n_{\rm H} \gtrsim -1.5$, doubly ionised  
between $-3 \lesssim \log n_{\rm H} \lesssim -1.5$, and triply ionised for $\log n_{\rm H} \lesssim -3$. 
Thus, carbon (or silicon) are expected to be predominantly ionised within LLSs, yielding
strong \ion{C}{III} and \ion{Si}{III} absorption as seen in previous work \citep[e.g.][see also Sect.
\ref{sec:ioncor}]{pro10lls,rib11,fum13}. For reference,  the mean cosmic density at $z\sim 2.5$ is 
$\sim 8\times 10^{-6}~\rm cm^{-3}$.

With the minimal grid of models in hand, we apply the formalism described in Section \ref{sec:parfit} to infer the 
posterior PDFs for density and metallicity.
Specifically, we reconstruct the posterior 
PDFs for each of the 234 LLSs in our sample by running the MCMC code over each set of observations. 
In principle, we could then assign a unique value of density and metallicity to each systems 
(e.g. through percentiles of the reconstructed PDFs). However, in some cases, the posterior PDFs 
show broad and/or  bi-modal distributions, reflecting the degeneracy between parameters 
given a limited number of observables. In the following, we therefore prefer to study 
the statistical properties of the LLS population by exploiting the information contained in the
PDFs rather than in ``best fit'' values for individual systems.
Comparisons between medians of the posterior PDFs will be used to assess the level of convergence 
of the different physical parameters with respect to model assumptions and to evaluate  the metal mass density 
of LLSs (equation \ref{eq:omega}). 
To further assess how the sample variance influences the shape of the reconstructed PDFs,
we adopt bootstrap techniques. Specifically, we construct 1000 realisations for the 
PDF of quantity of interest by combining sets of 234 LLSs, drawn from the full sample but 
allowing for repetitions. 

\begin{figure*}
\centering
\includegraphics[scale=0.4]{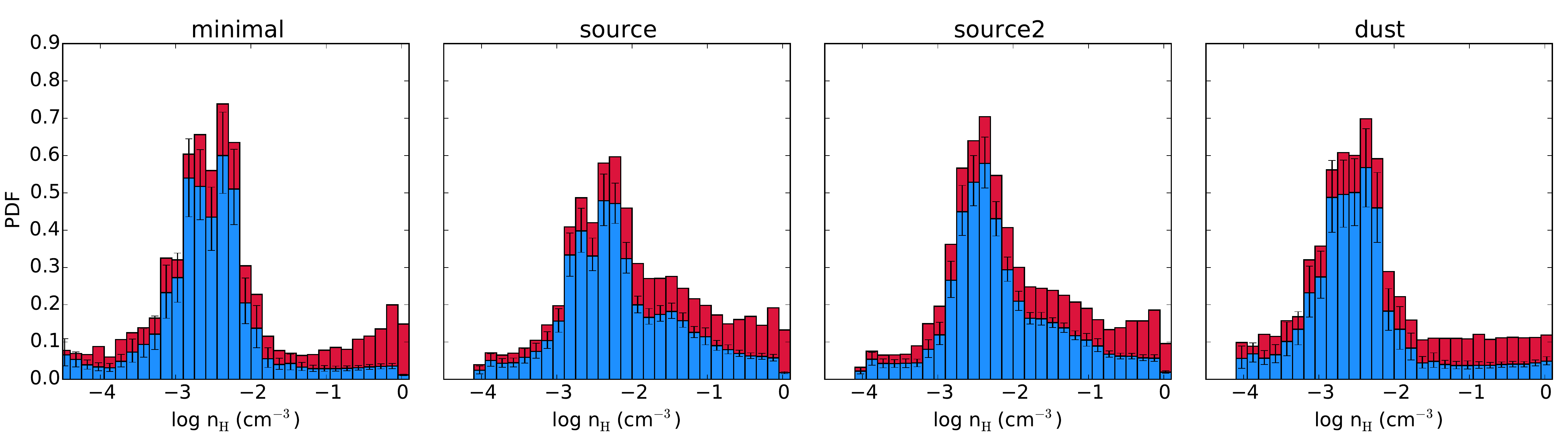}
\caption{Posterior PDFs for the density of the entire LLS sample (red, in the background) 
and the HD-LLS subset (blue, in the foreground), both of which are normalised 
to the total sample size. Each panel shows the PDF obtained by combining the marginalised posterior PDFs for individual systems under the assumption of different photoionisation models, as labelled.
Error bars show the 10th and 90th confidence intervals from bootstrapping.  
Despite non negligible differences in the shape of these PDFs, typical densities for these LLSs lie in the range 
$-3.5 \lesssim \log n_{\rm H} \lesssim -2$ with broader tails.}\label{fig:pdfden}
\end{figure*}

\begin{figure*}
\centering
\includegraphics[scale=0.4]{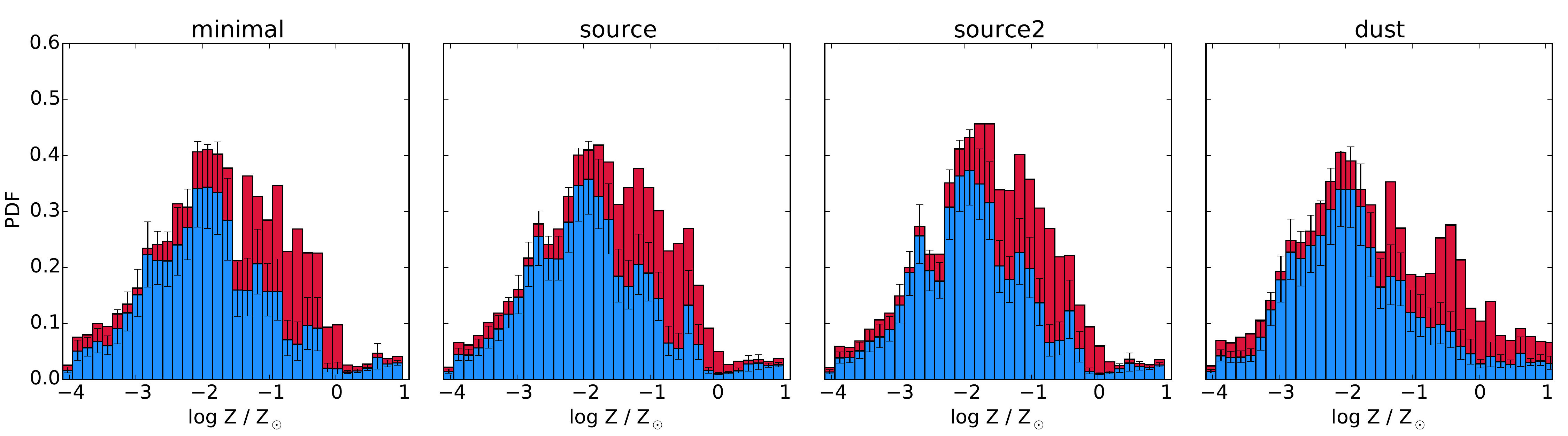}
\caption{Same as Figure \ref{fig:pdfden}, but for the posterior PDFs of the metallicity.
The metallicity distributions for this sample are characterised by a peak around $\log Z/Z_\odot \sim -2$, 
with broader tails and possibly a secondary peak close to $\log Z/Z_\odot\sim -1$.
Appreciable differences among different models can be seen.}\label{fig:pdfmet}
\end{figure*}

\begin{figure}
\centering
\includegraphics[scale=0.48]{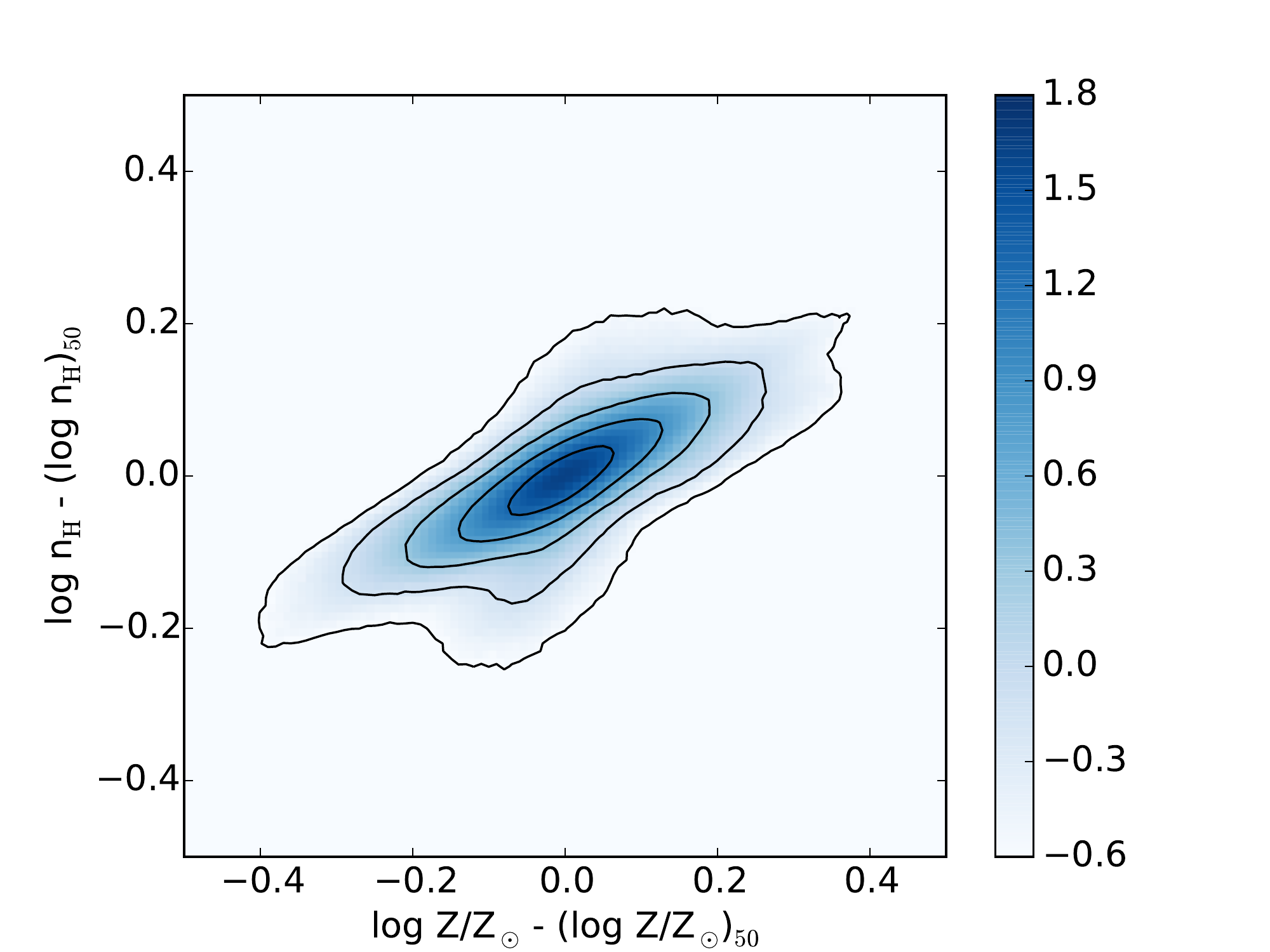}
\caption{Joint posterior PDFs for the metallicity and density in the HD-LLS sub-sample, 
obtained by combining the PDFs for individual systems relative to the respective medians 
and under the assumption of a minimal photoionisation model. Darker colours (with logarithmic scaling) mark 
regions of higher probability density in this two-dimensional space.}\label{fig:jointmin}
\end{figure}

The marginalised PDFs for the density and metallicity are shown in left-most panels of 
Figure \ref{fig:pdfden} and \ref{fig:pdfmet}, both for the entire sample (red) and the 
HD-LLS sub-sample (blue). From these figures, we infer that, under the assumption of a minimal photoionisation model,
the density distribution for the HD-LLS sub-sample is characterised by a well-defined peak between
$-3 \lesssim \log n_{\rm H} \lesssim -2$, while the PDF of the full sample exhibits also a tail at
higher densities, attributable to the lower redshift and higher column density of the LLSs
from the literature (Figure \ref{fig:sample}). Similarly, the metallicity distribution for the HD-LLS dataset
is peaked around $\log Z/Z_\odot \sim -2$, with broader tails both at high and low metallicity.
Again, due to the different nature of the systems from the literature, the entire sample shows a prominent 
tail towards higher metallicity, possibly with a hint of a second peak around $\log Z/Z_\odot \sim -1$.

\begin{figure*}
\centering
\includegraphics[scale=0.65]{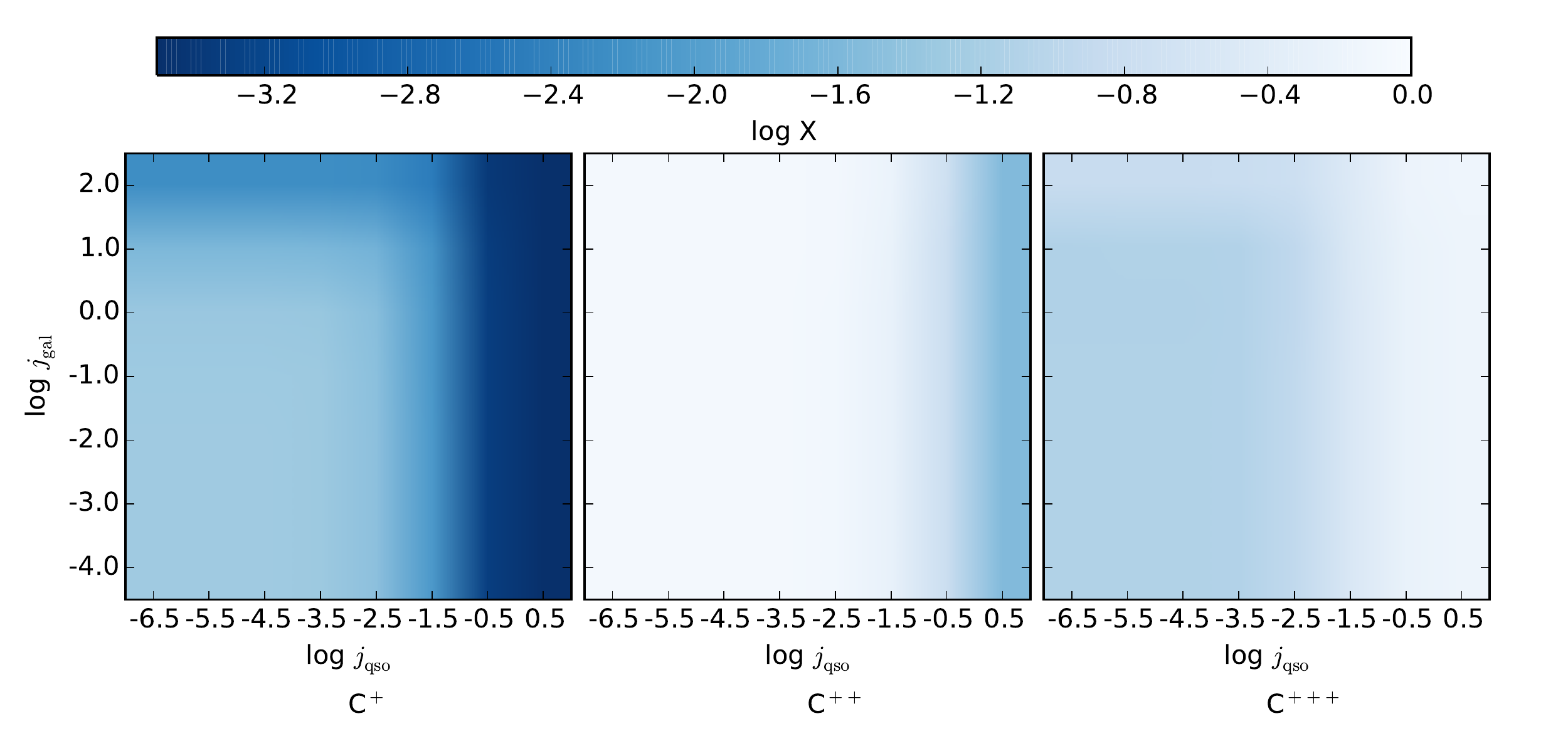}
\caption{From left to right, the variation in the fraction of singly, doubly, and triply ionised carbon
as a function of additional contribution from galaxies and quasars to the UVB radiation field (see text 
for details). Progressively higher contributions from local sources reduce the fraction of 
carbon in low ionisation stages, with the harder SEDs of quasars contributing the most to the ionisation of 
C$^{+++}$. Each panel shows a slice in the source grid of models at $z = 2.5$, $\log N_{\rm HI} = 18$,
$\log n_{\rm H} = -2.5$, and $\log Z/Z_\odot=-2.5$.}\label{fig:carbjj}
\end{figure*}

In an attempt to quantify whether this minimal model provides an acceptable description of the data, we 
examine the average residuals by comparing the observed ions to the model predictions given the median 
of the PDFs for the free parameters in the model. We also compute the residuals for the third most-deviant ion in 
each system, to provide further insight into the ability of the model to capture the column density for 
a wide range of ions.
With this exercise, we find that $<15\%$ of the systems have either the mean residual or the residual of the 
third most-deviant ion in excess of 3 times the observed error. 
We also examine the joint PDF of density and metallicity to assess how well we can constrain 
individual parameters with this model. In Figure \ref{fig:jointmin}, we show the 
mean joint PDF for the HD-LLS sub-sample, which we construct by combining the joint PDFs for 
individual systems after normalising them to the respective medians.
This figure shows that, on average, most of the probability is contained within a single region, 
centred around zero. This means that, on average, both densities and metallicity are characterised by well defined 
unimodal PDFs. Furthermore, despite some degree of correlation \citep[see also][]{coo15}, the posterior PDFs are on 
average well contained within $\pm 0.2$ dex from the median, indicating that most of the probability is 
found close to the peak of the distribution. A similar shape is found for the joint PDF for the entire sample (not shown), 
although the heterogeneous quality of the data from the literature (with some systems having only a couple of 
measured transitions) broadens the width of the joint PDF.
In turn, this means that the large spread observed in Figure 
\ref{fig:pdfden} and Figure \ref{fig:pdfmet} arises from intrinsic 
scatter in the physical properties of the LLS population.

Altogether, we conclude that a very simple photoionisation model is able to capture the ion distribution 
for most of the systems under analysis,
although with a few significant outliers. Moreover, high ions such as N$^{+4}$ or O$^{+5}$, included 
in our analysis when measured in the observations, are known to be under-produced in standard photoionisation models
\citep[e.g.][]{leh14}, and our models are no exception. Despite this good agreement, as we will show in the following 
via series of model expansions, the reconstructed PDFs are subject to systematic uncertainties arising from the 
adopted ionisation models.

\begin{table}
\caption{Free paramaters in the ``source'' and ``source2'' grids of models}\label{tab:source} 
\centering
\begin{tabular}{c c c c}
\hline
\hline
Parameter   		        &  Min.     &	  Max.  &   Step    \\
            		        &	    &		&	    \\
\hline  
$\log Z/Z_\odot$                &     -4.0  &	  1.0	&     0.5   \\ %
$z$      	                &      0.0  &	  4.5	&     0.5   \\ %
$\log N_{\rm HI}$ [cm$^{-2}$]   &     17.0  &	  20.5	&     0.5   \\ %
$\log n_{\rm H}$ [cm$^{-3}$] 	&     -4.0  &	   0.0	&     0.5   \\ %
$\log j_{\rm gal}$        	&     -4.0  &	   2.0	&     1.0   \\ %
$\log j_{\rm qso}$        	&     -6.5  &	   0.5	&     1.0   \\ %
\hline    
\end{tabular}
\flushleft{The columns of the table are: (1) the free parameter as described in the text; (2) the minimum allowed value; 
(3) the maximum allowed value; (4) the step adopted in the grid.}
\end{table}

\subsection{Uncertainties in the UVB model and proximity to local sources}

An important assumption of the minimal model is a fixed source of ionising radiation, which 
introduces a one-to-one mapping between density and ionisation parameter. However, as discussed at 
length in the literature, both the intrinsic uncertainty in the UVB model 
\citep[e.g.][]{fau09,haa12,bec13} and the possible contribution of a local radiation field 
\citep[e.g.][]{dod01,des03,sch06,mei09,nag10,fum11sim} 
make calculations that rely on a specific UVB prone to substantial uncertainties \citep[e.g.][]{aga05,fec11,sim11,cri15}. 
For these reasons, parametric models have been devised to allow for variations in the source of radiation.

One such parametrisation can be found in the recent work by \citet{cri15}, who introduce 
a free parameter $\alpha_{\rm UV}$ to tune the hardness of the UVB from a hard (AGN dominated) to 
a soft (galaxy dominated) spectrum. Here we follow a similar procedure, which we however 
generalise to allow for a free parametrisation in the amplitude of the
radiation field, suitable for the treatment of local sources. In the model of \citet{cri15}, in fact, 
the UVB spectrum is renormalised to satisfy independent measurements of the hydrogen and helium  
photoionisation rates within a $\pm 0.3$ dex interval. More substantial variations in the 
normalisation are, however, especially relevant for LLSs, as many pieces of evidence consistently place 
optically-thick absorbers in proximity to galaxies or even quasars \citep{fau11,fum11sim,van12,fum13,pro13}. However, as we will show below, by allowing for a varying amplitude in the radiation field, we 
break the one-to-one relation between the physical density and the ionisation parameter,
thus introducing a degeneracy in the model. 

To obtain a more generalise form for the radiation field, 
we construct a source term by combining three contributions: the intensity from the UVB,
$J_{\rm uvb}(\nu)$, the intensity from local galaxies, $J_{\rm gal}(\nu)$, and the intensity from a local AGN, 
$J_{\rm qso}(\nu)$. The combined input spectral energy distribution (SED) therefore becomes
$J (\nu)  = J_{\rm uvb}(\nu) + j_{\rm gal} J_{\rm gal}(\nu) + j_{\rm qso} J_{\rm qso}(\nu)$, where $j_{\rm gal}$ and 
$j_{\rm qso}$ are free parameters (constrained to be positive). 
Clearly, many parameters regulate the amplitude of the local radiation field for quasars and galaxies. 
To minimise the number of free parameters, and given that we are not trying to constrain the astrophysical origin 
of a radiation field in excess of the UVB, we combine the intrinsic properties of local sources and the escape fraction
of ionising radiation in two ``phenomenological'' parameters ($j_{\rm gal}$ and $j_{\rm qso}$) that globally describe 
the effects of local sources, as detailed below.

 For $J_{\rm gal}$, we create a {\sc starburst99} model \citep[v7.0.1;][]{lei99} 
using default input parameters, a continuous 
star formation rate (SFR) $\dot\psi =  1$~\sfr, and the Geneva (2012) stellar tracks with no rotation and solar 
metallicity \citep{eks12}. The normalisation coefficient $j_{\rm gal}$ is then used to account, altogether, for the 
intrinsic SFR of the local source, the escape fraction of ionising radiation $f_{\rm esc}$, the distance 
between the cloud and the source $d_{\rm cs}$, and a dust extinction $\kappa$ that does not 
depend on wavelength. Thus, in our model
\begin{equation}
j_{\rm gal} = \frac{\dot\psi}{1\rm M_\odot~yr^{-1}}\left(\frac{100~\rm kpc}{d_{\rm cs}}\right)^2 f_{\rm esc} (1-\kappa)\:.
\end{equation}
As an example, $\log j_{\rm gal} \le -4$ yields the \citet{haa12} UVB within $\sim 1\%$ at all frequencies, 
while  $\log j_{\rm gal} \sim 2$ describes a cloud at $\sim 30$ kpc from a galaxy with $\dot\psi = 100~\rm M_\odot~yr^{-1}$
and $f_{\rm esc} = 0.1$.
For $J_{\rm qso}(\nu)$, instead, we use the mean quasar SED from \citet{ric06}, which we combine
with the other sources with weight $j_{\rm qso}$ to account for varying degrees of intrinsic luminosity
$L_{\rm qso,bol}$, proximity to the source, and constant dust extinction:
\begin{equation}
j_{\rm qso} = \frac{L_{\rm qso,bol}}{4.6\times 10^{47}~\rm erg~s^{-1}}\left(\frac{100~\rm kpc}{d_{\rm cs}}\right)^2 (1-\kappa)\:.
\end{equation}
With this parametrisation, the UVB is recovered within $1\%$ at all frequency for   $\log j_{\rm qso} \le -6$, 
while $\log j_{\rm qso} \sim 0.5$ describes a cloud that lies at $\sim 50$ kpc from an average type 1 quasar. 
The full parameter space occupied by this grid of models, dubbed ``source'', is described in Table \ref{tab:source}.
Finally, to assess the systematic difference arising from different choices of the UVB, we re-run a second grid of
source models, labelled ``source2'', by using the \citet{fau09} UVB instead of the \citet{haa12} model.

\begin{figure}
\centering
\includegraphics[scale=0.58]{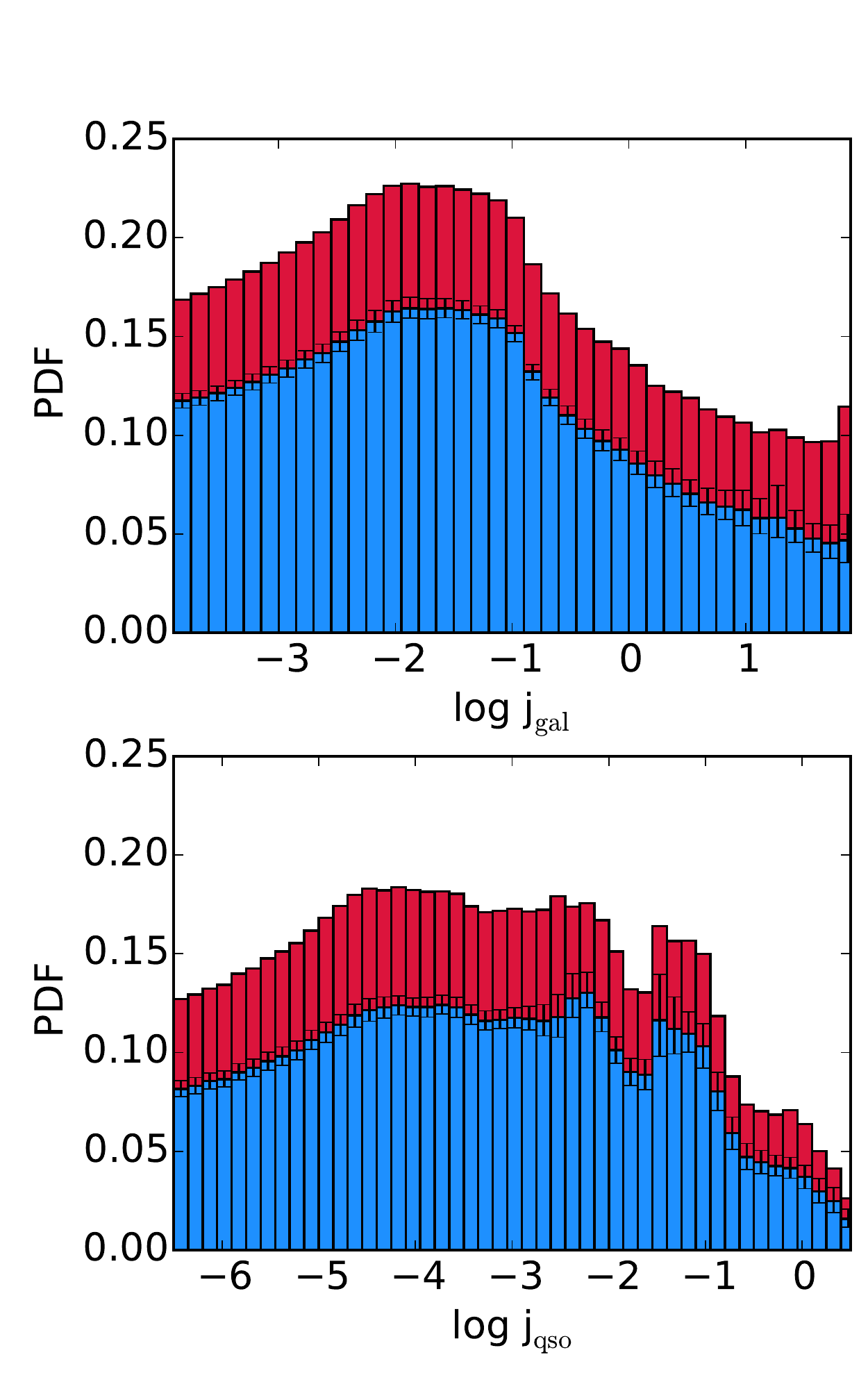}
\caption{Same as Figure \ref{fig:pdfden}, but for the posterior PDFs for the 
additional contribution of galaxies (top) and quasars (bottom) to the UVB 
in the source model. The broad PDFs for the population, together with very broad
PDFs for individual systems (Figure \ref{fig:jointsource}), suggest that these data do not 
constrain the spectral shape and amplitude of the radiation field.}\label{fig:jqsogal}
\end{figure}

\begin{figure}
\centering
\begin{tabular}{c}
\includegraphics[scale=0.44]{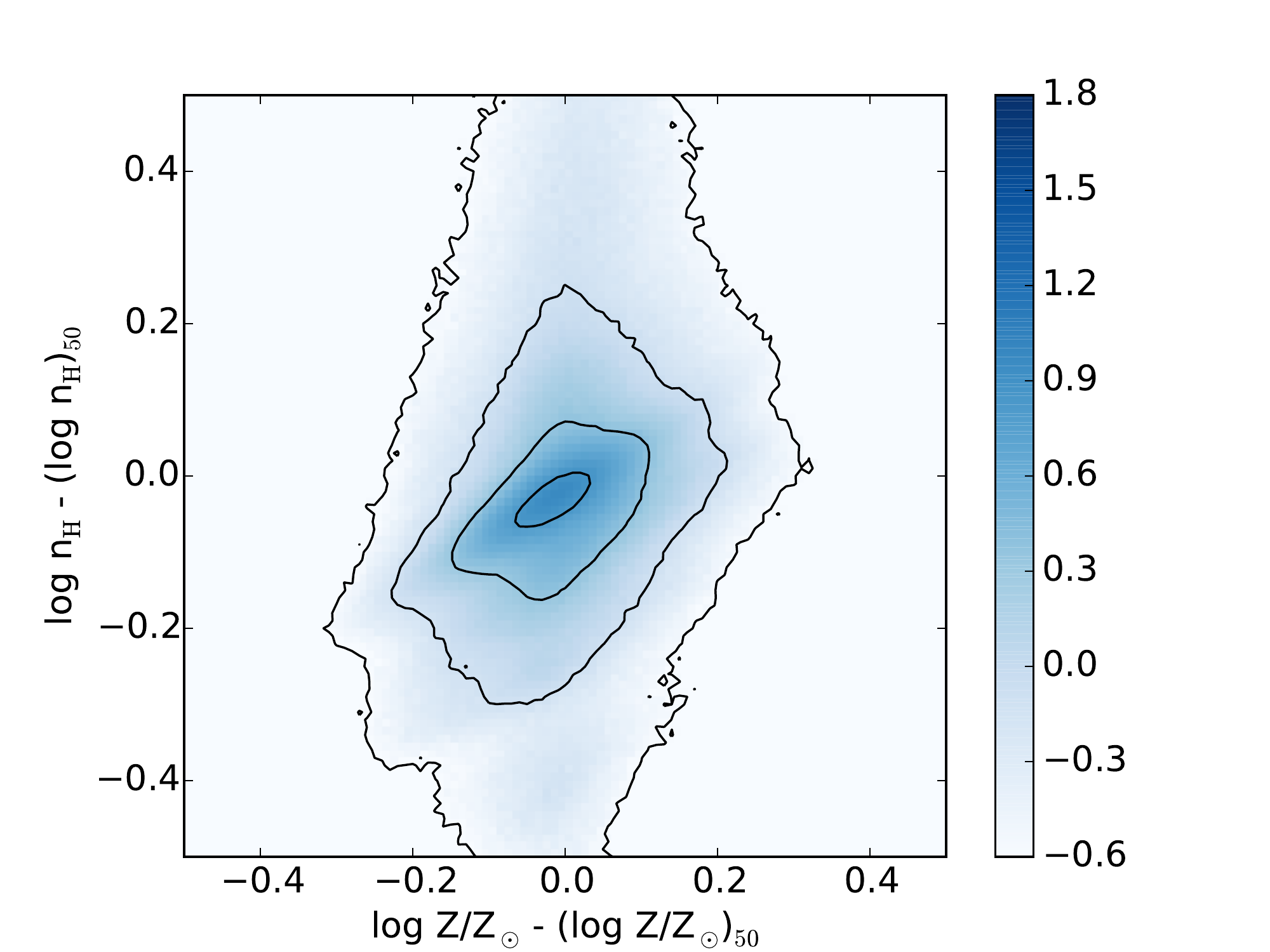}\\
\includegraphics[scale=0.44]{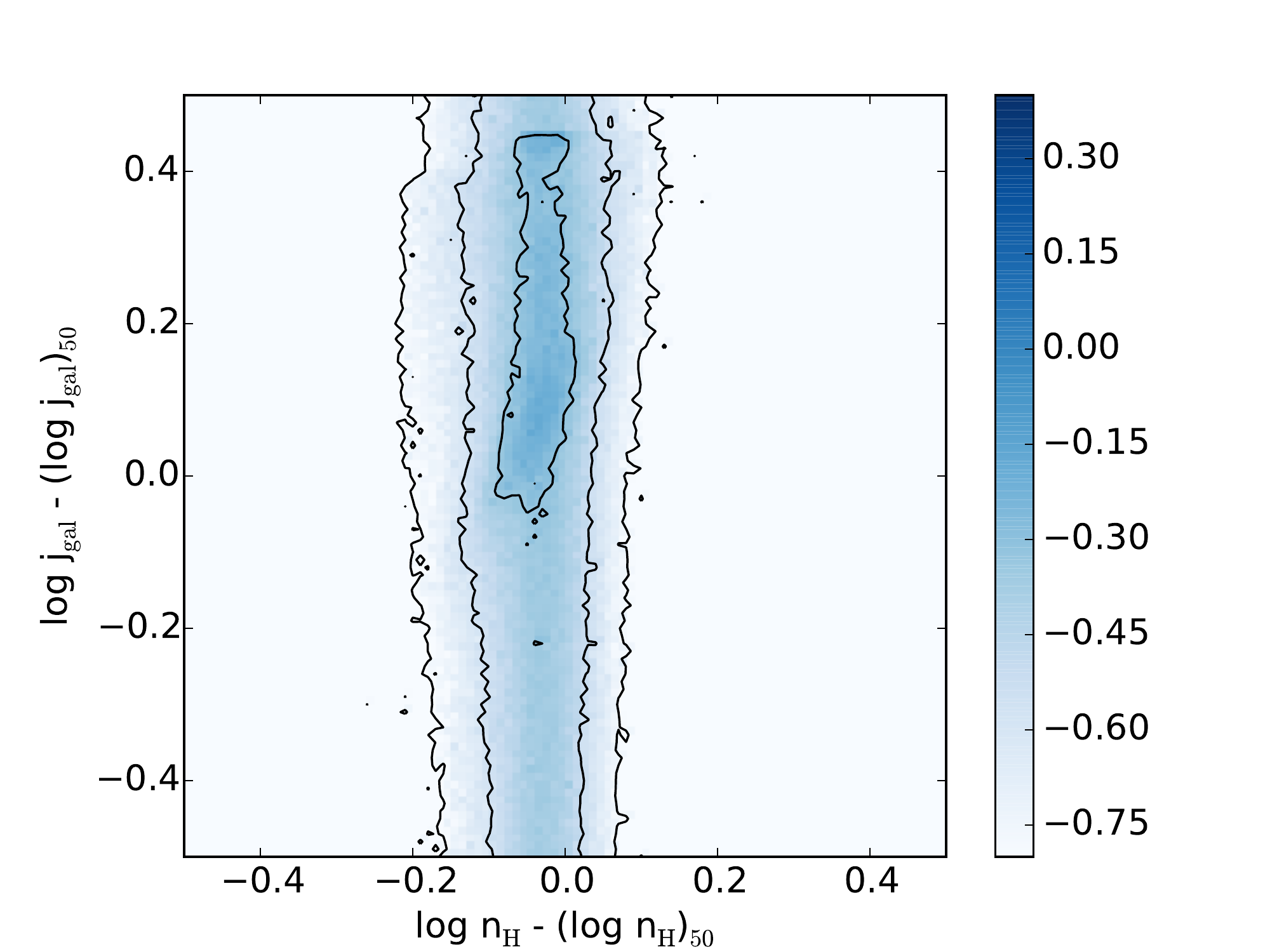}\\
\end{tabular}
\caption{Same as Figure \ref{fig:jointmin}, for the joint PDFs of metallicity and density (top)
and density and $j_{\rm gal}$ (bottom).  Due to the weak constraints on the contribution from local sources, 
the density PDF broadens.}\label{fig:jointsource}
\end{figure}

Before running the MCMC procedure on the data, in Figure \ref{fig:carbjj} we provide 
a visual example on how the ionisation stages of common elements are affected 
by the inclusion of additional sources of radiation. Specifically, the variation of ionisation for 
carbon in C$^{+}$, C$^{++}$, and C$^{+++}$ is shown as a function of $j_{\rm gal}$ and $j_{\rm qso}$, 
for constant redshift ($z = 2.5$), density ($\log n_{\rm H} = -2.5$), metallicity ($\log Z/Z_\odot=-2.5$), 
and column density ($\log N_{\rm HI} = 18$). As expected, increasing the contribution of local sources 
increases the ionisation parameter at fixed physical density, and a progressively 
larger fraction of carbon can be found in C$^{++}$ and C$^{+++}$. Furthermore, the addition of a
harder quasar SED for $\log j_{\rm qso} > -1.5$ induces ionisation of carbon in higher stages, 
most notably from C$^{++}$ to C$^{+++}$ \citep[cf][]{sim11}.

When using the source and source2 grid of models to infer the posterior PDFs for the
density and metallicity, we find that the metallicity of LLSs, as a population, 
is not particularly sensitive to variations in the radiation field. 
Comparing the posterior PDFs derived from the source and source2 models to the one 
inferred from the minimal model (Figure \ref{fig:pdfmet}), one can see that, 
especially for the HD-LLS sub-sample, the distributions retain a similar shape.
Furthermore, one can also see how the source and source2 models yield a virtually identical 
PDF for the metallicity of these LLSs. Conversely, substantial differences can be seen for the PDFs of 
the density (Figure \ref{fig:pdfden}). While the characteristic peak between 
$-3 \lesssim \log n_{\rm H} \lesssim -2$ is retained, appreciable discrepancies can be noted
when additional sources are included, particularly with an excess probability for 
$-2 \lesssim \log n_{\rm H} \lesssim -1$ compared to the results from the minimal model. 
This excess is a consequence of our treatment for the local radiation field, which can only 
boost the contribution from the UVB, thus skewing the density distribution to higher values.

The same trends emerge when inspecting the medians for the posterior PDFs for individual 
systems (not shown). Comparing the results of the minimal and source model,
a tight correlation is found, with a mean offset of $\sim 0.01$ dex and dispersion of $\sim 0.15$ dex.
Conversely, for the density, we find both a larger scatter and a 
systematic offset between the medians, with typical discrepancies 
of $0.33\pm 0.49$ dex. Notably, when we compare medians inferred assuming the \citet{haa12} UVB 
model to the ones inferred assuming the \citet{fau09} UVB model, we find a tight correlation 
for both metallicity ($0.02\pm0.11$ dex) and, although with larger scatter, densities 
($0.06\pm0.33$ dex). Thus, variations in the shape and amplitude of the UVB among different models 
are sub-dominant compared to uncertainties in a local radiation field.

The origin of the different behaviour in the density and metallicity PDFs is attributable to 
the degeneracy between density and intensity of the local radiation field. That is, 
the available data are able to quite precisely constrain the ionisation parameter, and thus the 
metallicity of LLSs. However, there is not enough information within most of these observations to
also constrain the shape and amplitude of the radiation field, 
as it is clear from the broad posterior PDFs for $j_{\rm gal}$ and $j_{\rm qso}$ (Figure \ref{fig:jqsogal}).
Such a broad PDF could in principle be the result of a superposition of many narrow
PDFs if LLSs had to be exposed to a diverse range of radiation fields. 
However, the shape of the mean joint PDF for the density and $j_{\rm gal}$ of individual 
systems (Figure \ref{fig:jointsource}) 
rather indicates that  $j_{\rm gal}$ (and $j_{\rm qso}$) are
mostly unconstrained\footnote{The joint PDF for the density and $j_{\rm qso}$ is not shown, but it has a similar shape.}, 
and thus we cannot precisely establish the importance of additional sources 
of radiation with this procedure. In response to a broad PDF for $j_{\rm gal}$ and $j_{\rm qso}$, 
the density PDF broadens (top panel of Figure \ref{fig:jointsource}), 
often with tails to higher $n_{\rm H}$ to compensate for additional sources of ionising 
radiation. 

We note that, despite the inclusion of two additional parameters, the 
source and source2 models do not appear to yield significantly better residuals, 
which remain comparable to what was found when using the minimal model. Again, we can attribute 
this result to the ambiguity in separating density and sources of radiation, which shifts the medians 
from the peaks of the PDFs. 
Finally, while we should refrain from strong conclusions 
based on the PDFs for $j_{\rm gal}$ and $j_{\rm qso}$ given the discussion above, 
it is worthwhile noting that high values of $j_{\rm gal}$ and, especially, of $j_{\rm qso}$ 
have low probability in this sample. Furthermore, there seems to be a preference for a low 
(perhaps non-zero) contribution from local sources, suggestive that the population of LLSs could 
be exposed to only modest levels of ambient radiation in excess to the UVB \citep[e.g.][]{sch06,pro13}.

\begin{table}
\caption{Free paramaters in the dust grid of models}\label{tab:dust} 
\centering
\begin{tabular}{c c c c}
\hline
\hline
Parameter   		        &  Min.     &	  Max.  &   Step     \\
            		        &	    &		&	     \\
\hline  
$\log Z/Z_\odot$                &     -4.0  &	   1.1	&     0.30   \\ %
$z$      	                &      0.0  &	   4.5	&     0.30   \\ %
$\log N_{\rm HI}$ [cm$^{-2}$]   &     17.0  &	  20.6	&     0.30   \\ %
$\log n_{\rm H}$ [cm$^{-3}$] 	&     -4.0  &	   0.2	&     0.30   \\ %
$F_*$        	                &     -1.5  &	   1.5	&     0.50   \\ %
\hline    
\end{tabular}
\flushleft{The columns of the table are: (1) the free parameter as described in the text; (2) the minimum allowed value; 
(3) the maximum allowed value; (4) the step adopted in the grid.}
\end{table}

\subsection{Dust depletion}\label{sec:dust}

Up to this point, we have considered idealised absorption systems with no dust and a gas phase 
abundance pattern equal to the one measured in the solar neighbourhood. Next we
examine, in simple terms, the impact that this assumption has on the metallicity 
determination for LLSs.  

Several authors have addressed the problem of characterising 
element by element depletion factors and their variation with physical properties, both in the 
nearby and in the distant Universe \citep[e.g.][]{sav96,pro07,jen09,raf12,dec13}.
These studies convincingly show that dust depletes elements onto grains in the denser 
astrophysical environments. Nevertheless, the complex astrophysics that regulates 
dust formation or the condensation and evaporation of elements onto and from grains 
hampers the formulation of a general theory for the depletion factors of all the elements of
interest, especially at high redshift and within the poorly-explored population of LLSs. 
        
For this reason, in the following, we construct a very simple grid of ``dust'' models
building on the empirical work of \citet{jen09}. Given the many unknowns, 
and the degeneracy between dust depletion and intrinsic deviations from the assumed 
solar abundance, we construct a model that relies on a single parameter that 
regulates both the dust-to-gas ratio at constant metallicity and the 
element-by-element depletion. Specifically, we expand the minimal model introducing ISM grains 
as specified by {\sc cloudy} with an abundance relative to solar composition 
defined by $Z_{\rm grn}/Z_\odot = \alpha_{\rm dtm} Z/Z_\odot$, where $Z/Z_\odot$ is the gas 
phase metallicity and $\alpha_{\rm dtm}$ is a free parameter that specifies the dust-to-metal ratio. 
It should be noted that the inclusion of grains in {\sc cloudy} affects both the gas thermal state
and the opacity. Next, we compute the depletion of relevant elements from the gas phase 
relative to their solar abundance following the fitting formulae provided by 
\citet[][their table 4 and equation 10]{jen09}, which we generalise to the case of systems with 
arbitrary metallicity. As described in detail in 
\citet{jen09}, while the absolute value of the gas phase abundance of each element is uncertain, 
it is possible to more accurately establish how rapidly each element is depleted onto dust grains 
by introducing a depletion strength factor, $F_*$, which varies from sightline to 
sightline.

\begin{figure}
\centering
\begin{tabular}{c}
\includegraphics[scale=0.42]{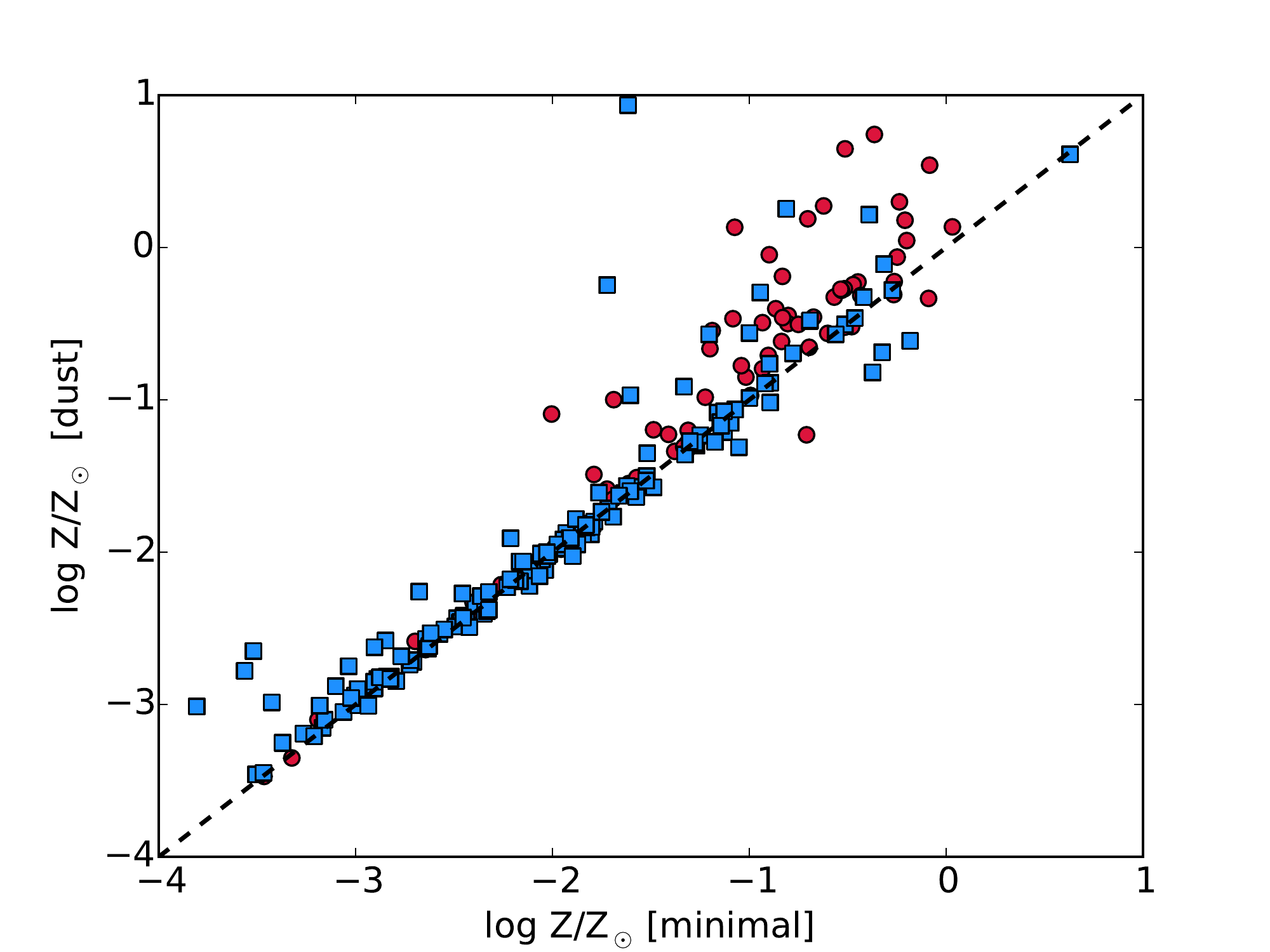}\\
\includegraphics[scale=0.42]{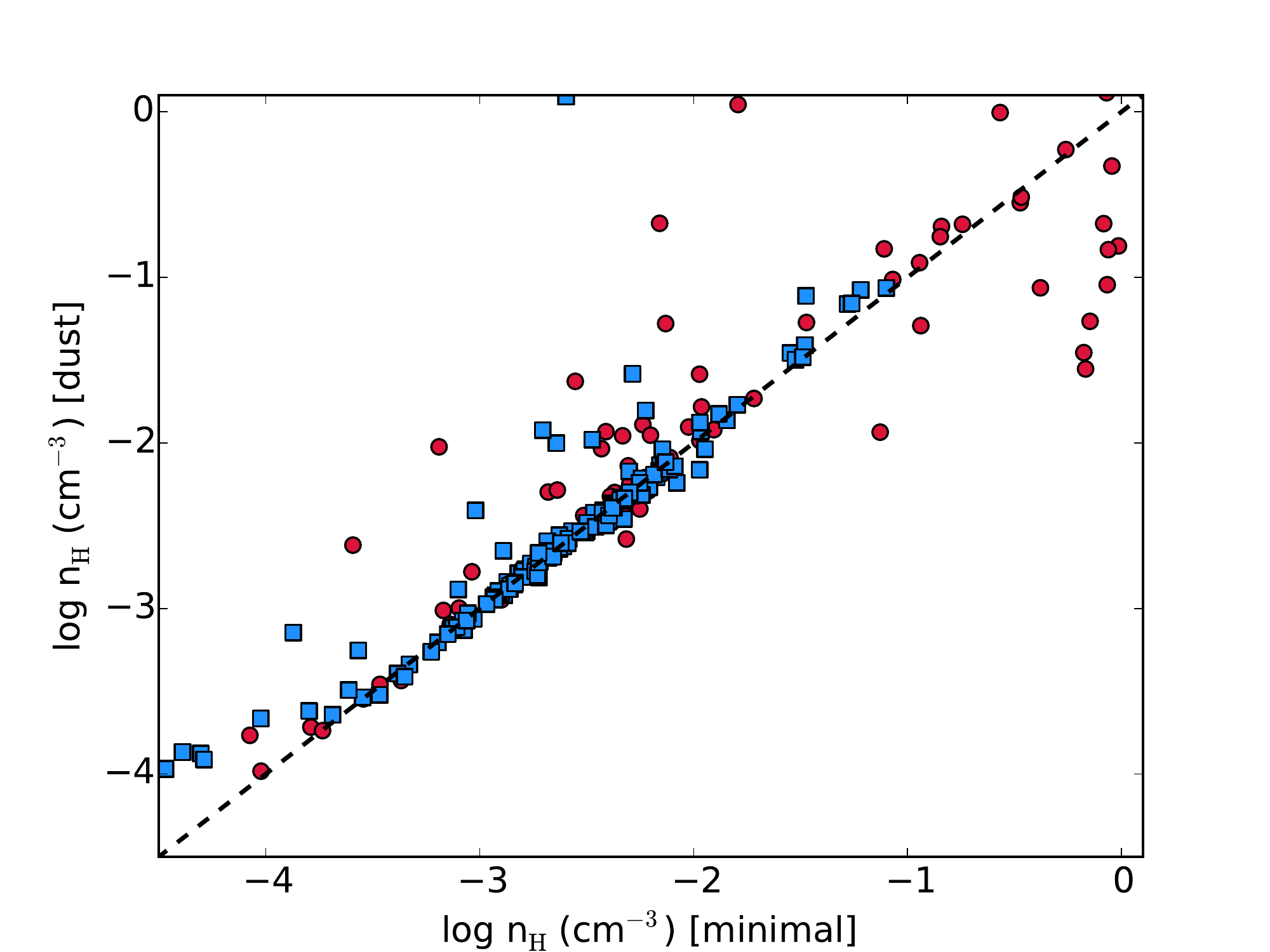}\\
\end{tabular}
\caption{Comparison of the median PDFs for metallicity (top) and density (bottom)
derived from the minimal and dust models in individual systems. Red circles represent LLSs from the literature, 
while the blue squares mark LLSs from the HD-LLS sample. The inferred quantities are tightly 
correlated, with increasing dispersion and systematic offsets towards high metallicity and lower 
redshifts.}\label{fig:cfrdust}
\end{figure}

\begin{figure*}
\centering
\begin{tabular}{ccc}
\includegraphics[scale=0.37]{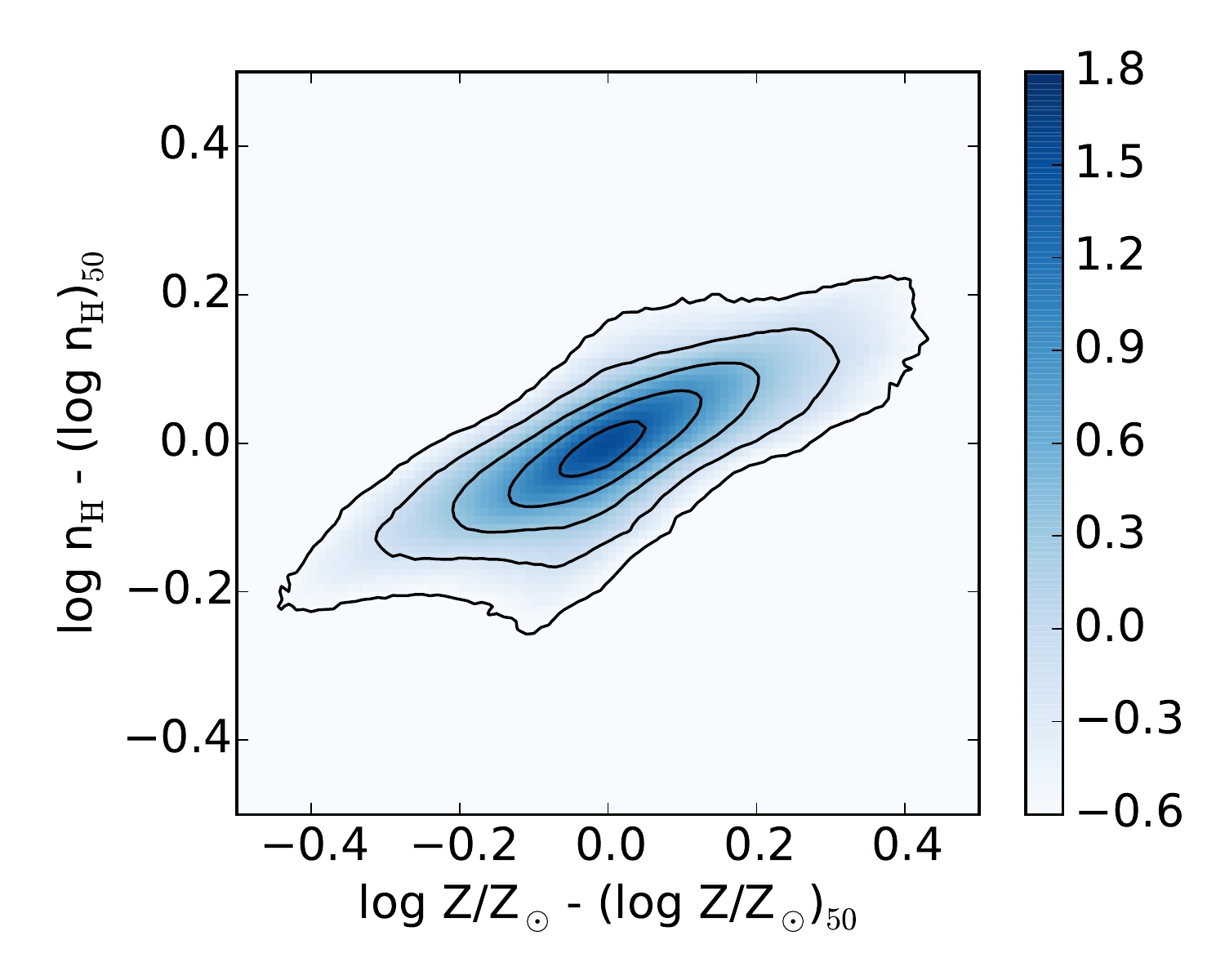}&
\includegraphics[scale=0.37]{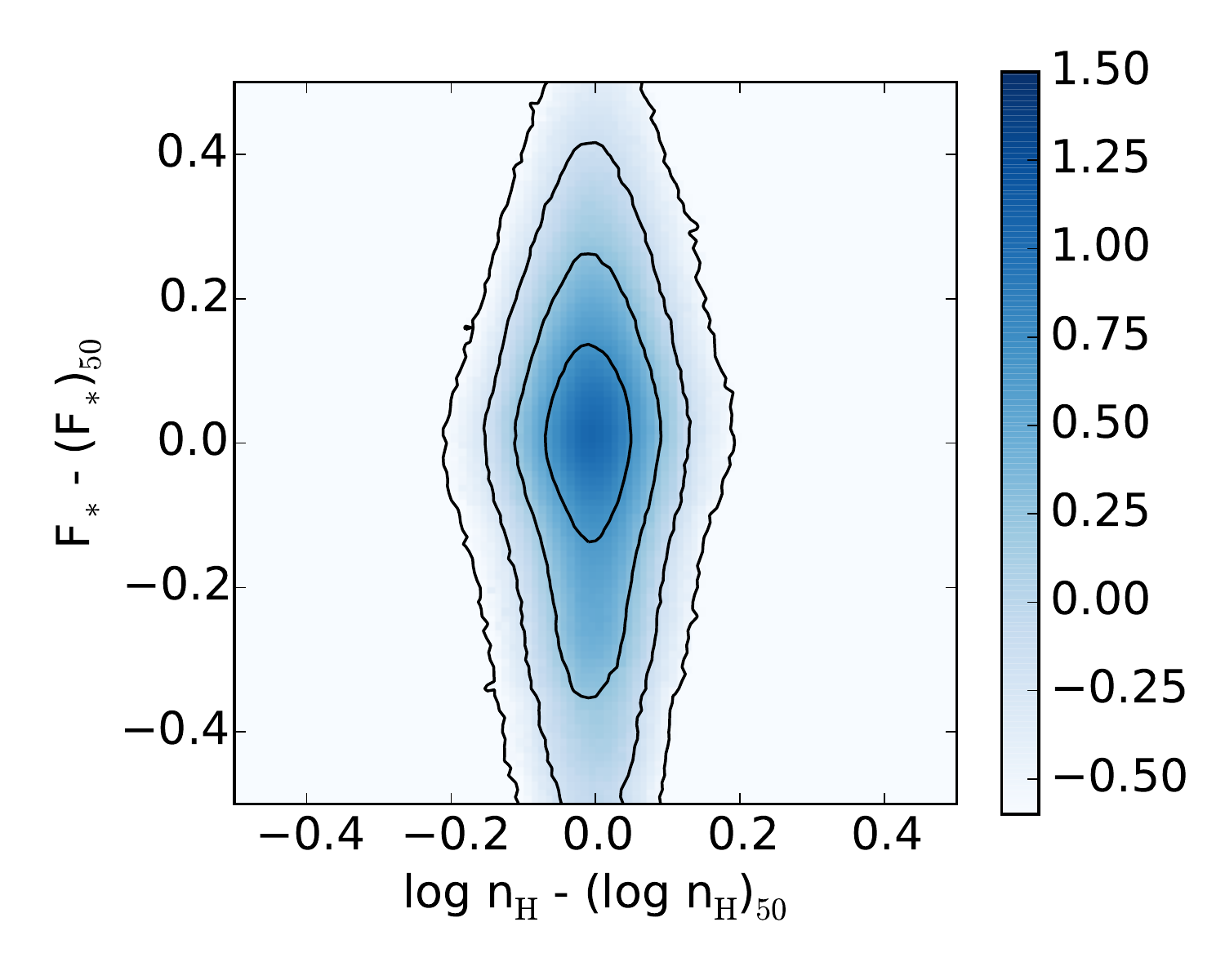}&
\includegraphics[scale=0.37]{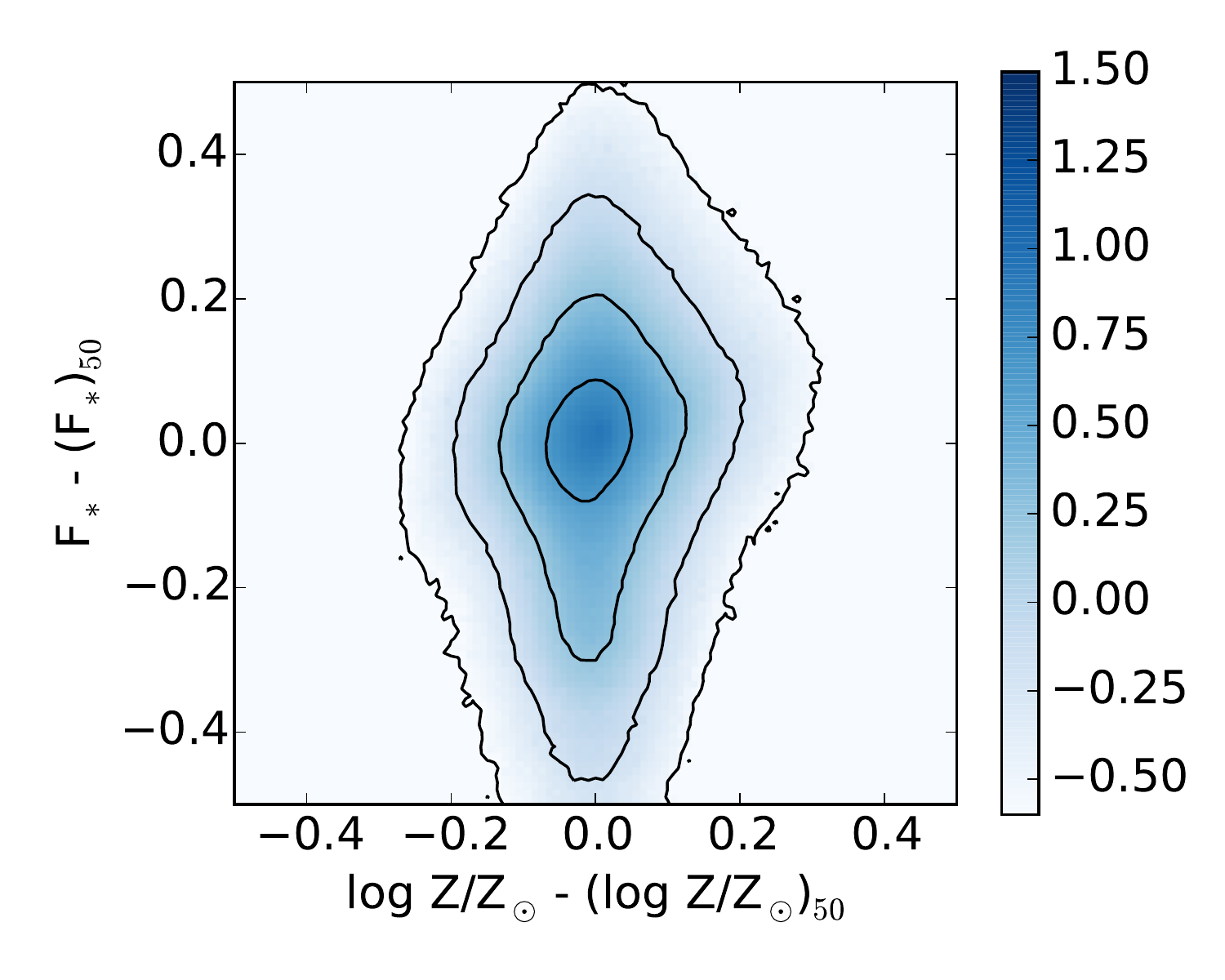}\\
\end{tabular}
\caption{Same as Figure \ref{fig:jointmin}, for the joint PDFs of metallicity and density (left)
and density and $F_*$ (centre), and metallicity and $F_*$ (right). Density and metallicity in the 
dust grid of models are well constrained, despite the broader PDFs.}\label{fig:jointdust}
\end{figure*}

Following the notation of \citet{jen09}, having defined the depletion from the gas phase of an 
element $X$ for a system with intrinsic abundance $\rm [X/H]_{\rm int}$
\begin{equation}
{\rm [X/H]}_{\rm dep} \equiv \log N_{\rm X} - \log N_{\rm H} - (\log N_{\rm X} - \log N_{\rm H})_{\rm int}\:,
\end{equation}
we can write the gas phase metallicity as
\begin{equation}\label{eq:dep}
{\rm [X/H]_{\rm gas}} = {\rm [X/H]}_{\rm int} - B_{\rm X} - A_{\rm X} (F_* - z_{\rm X})\:,
\end{equation}
where $B_{\rm X}$, $A_{\rm X}$, and $z_{\rm X}$ are locally-calibrated coefficients
taken from \citet{jen09} that define ${\rm [X/H]}_{\rm dep}$. 
Here, we further expand on the work of \citet{jen09} in two ways. Firstly, 
equation \ref{eq:dep} returns a non-zero level of depletion even for $F_* = 0$. 
For our purposes, however, we wish to recover a limiting case of no depletion for all elements, 
and we allow for values $F_* < 0$. We note, however, that the limiting case 
${\rm [X/H]}_{\rm dep}=0$ is recovered for different elements by means of different $F_*$.
Because of the use of a single $F_*$ parameter, the extrapolation of 
depletion factors for large negative values of $F_*$ could yield negative corrections for some elements, 
which we cap to 0. Similarly, there is no prior reason to impose the condition $F_* \le 1$,
which originates from empirical considerations in the analysis of \citet{jen09}.
Thus, in our work we allow for cases with $F_* > 1$. Indeed, 
although stronger depletion than observed in the local ISM is probably 
a rare occurrence for LLSs, it may be plausible for some of the super-solar 
systems. Secondly, we postulate a connection between  $\alpha_{\rm dtm}$ and $F_*$
such that $\alpha_{\rm dtm} = F_*$ for $F_* \ge 0$, and $\alpha_{\rm dtm} = 0$ otherwise.

We emphasise that this choice is rather arbitrary, as the exact scaling between 
the dust content and the depletion is unexplored for LLSs.
Nevertheless, this ansatz lets us specify a simple dust model that, 
with a single parameter $F_*$, provides an expansion of the minimal model with desirable 
characteristics. For $F_* > 0$, dust grains are included in the photoionization modeling 
with abundance proportional to $F_*$ and elements are depleted according to 
equation \ref{eq:dep}, also allowing for a variable 
metal-to-dust ratio. For $F_* = 0$, dust grains are not included, although 
residual deviations from the assumed solar abundance pattern are allowed. Finally, 
for $F_* < 0$, the behaviour of the minimal dust-free model is progressively recovered. 
As previously done for $j_{\rm qso}$ and $j_{\rm gal}$, we note that we 
do not attempt to assign a physical meaning to the precise values of $F_*$ (see below). 
A summary of the parameters included in the dust grid of models is 
provided in Table \ref{tab:dust}.

Besides the obvious effect of altering the relative ratios of ions when keeping constant all the other
parameters of the grid, the ionisation stages of the individual elements in the dust grid 
do not appreciably differ from the trends already discussed for the minimal models (Figure \ref{fig:carb}).
However, the inclusion of dust shapes the radiation transmitted through the cloud
and the gas thermal state. As a consequence, at the corner of the grid where the highest density, metallicity, 
and column densities are found, models with $F_* \gtrsim 0.5$ develop a molecular phase.
In turn, this leads to a reduction of the effective hydrogen column density in the neutral atomic phase 
when compared to a dust free model. Throughout our analysis, we account for this effect by considering 
the output $\log N_{\rm HI}$ in the computation of the priors. However, as we will show below, data prefer 
models with low dust content and thus the molecular phase is unimportant for our analysis.

\begin{figure}
\centering
\includegraphics[scale=0.63]{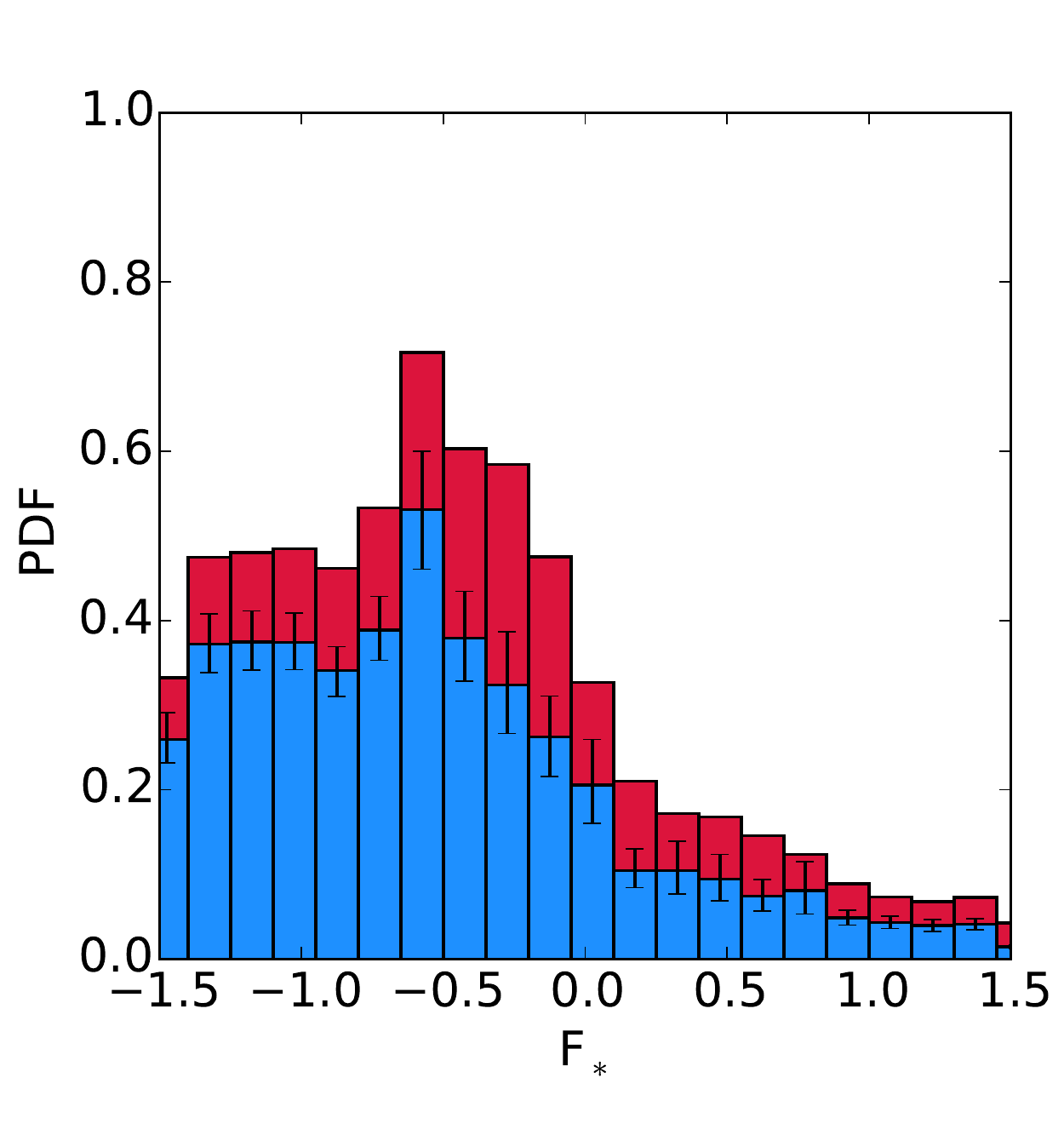}
\caption{Same as Figure \ref{fig:pdfden}, but for the posterior PDF of 
$F_*$ in the dust model. The majority of LLSs, especially at $z>2$, are characterised by a 
low dust content with $F_* \lesssim 0$.}\label{fig:pdffstar}
\end{figure}

When using the dust grid of models to infer the posterior PDF for density and metallicity, we recover
the distributions shown in Figure \ref{fig:pdfden} and Figure \ref{fig:pdfmet} (right panels).
The density distribution appears to be nearly insensitive to the inclusion of $F_*$ in the dust model, 
with the exception of the highest density tail for the literature sample. Conversely, 
the metallicity PDFs show some discrepancies, particularly for $\log Z/Z_\odot \gtrsim -1.5$
and for the literature sample dominated by low-redshift and high-column systems.  
To better understand the origin of this difference, we inspect the medians (Figure \ref{fig:cfrdust}) and
shapes (Figure \ref{fig:jointdust}) of the posterior PDFs for the density and metallicity, as well as the posterior 
PDF for $F_*$  (Figure \ref{fig:pdffstar}). 

For the HD-LLS sample, both the median density and metallicity 
appear to be tightly correlated, with a dispersion of $\sim 0.3$ dex. For the metallicity, 
most of this dispersion is driven either by data with $\log Z/Z_\odot > -1$ 
or $\log Z/Z_\odot \lesssim -3$. The lower-redshift higher-column density sample from the literature 
follows similar trends, with an even more pronounced dispersion in the metallicity for 
$\log Z/Z_\odot > -1.5$. Qualitatively, an increasing importance of dust towards 
lower redshifts is in line with the results from studies of \ion{Mg}{II} absorbers \citep[e.g.][]{men08}.
At higher metallicity, despite the small sample size, there is 
also evidence for a systematic offset, in the direction of having higher metallicity for the dust model.
This offset arises from a tentative  correlation between $F_*$ and $\log Z/Z_*$
(not shown), for which a higher intrinsic metallicity is required to model observations 
for the gas phase metal content in the presence of depletion.
Regarding the discrepancy at lower metallicity, instead, the deviant points are for LLSs with 
only one or two detected ions. Compared to the minimal model, it appears that values $F_* > -0.5$ 
provide a better fit for these detected ions, and in particular for the ubiquitous detection of 
\ion{C}{IV}. Lacking multiple ions in different ionisation stages, it is unclear whether this result 
is physical or whether it arises from a second-order degeneracy among parameters. 
As this discrepancy is seen in only 4 systems and does not impact our conclusions, we do not investigate 
it further.

Differently from the case of the source model, we note that the inclusion of an 
additional parameter in the grid does not introduce significant degeneracy. 
The shape of the mean joint PDFs (Figure \ref{fig:jointdust}) 
reveals that density, metallicity, and $F_*$ are generally well constrained.
Due to the simplicity of our dust model, however, it is not surprising that the 
posterior PDFs for $F_*$ are quite broad, and that the PDFs for the metallicity 
broaden compared to what seen in Figure \ref{fig:jointmin}. 
As before, this figure provides a qualitative way to 
disentangle the width in the PDF for individual systems  ($\sim 0.2-0.3$ dex) 
from the scatter within the population when examining the combined posterior PDFs 
for the entire sample.

Given that the shape of the PDFs for $F_*$ is generally well constrained, the 
posterior PDF for $F_*$ encodes some information on the dust properties of these LLSs, 
although in a model-dependent way. From Figure \ref{fig:pdffstar}, we see 
that most of the probability lies at $F_* < 0$, which suggests 
that LLSs typically reside in environments with low dust content. 
There is also evidence that $>50\%$ of the probability is contained between 
$-1 \lesssim F_* \lesssim 0$. This result stems almost entirely from the fact that, 
in our model, iron exhibits residual depletion even for $F_* < 0$ \citep{jen09}.
However, the physical interpretation of negative but small $F_*$ is complicated by the 
degeneracy between iron depletion onto dust grains and its unknown intrinsic abundances
relative to $\alpha-$elements \citep[cf][]{ber15}. Due to this ambiguity, we cannot make any inference on depletion 
based on the posterior PDF for $F_*$. We only note on empirical grounds that
a lower than solar iron abundance is preferred by the data given our simple model. Indeed, 
inspecting the residuals, we note that the inclusion of $F_*$ now ensures 
that the observed column densities are reproduced within 2 times the observational 
errors for $\sim 80\%$ of the LLSs. 

\begin{table}
\caption{Free paramaters in the CIE grid of models}\label{tab:cie} 
\centering
\begin{tabular}{c c c c}
\hline
\hline
Parameter   		        &  Min.     &	  Max.  &   Step    \\
            		        &	    &		&	    \\
\hline  
$\log Z/Z_\odot$                &     -4.0  &	   1.1	&     0.30   \\ %
$z$      	                &      0.0  &	   4.5	&     0.30   \\ %
$\log N_{\rm HI}$ [cm$^{-2}$]   &     17.0  &	  20.6	&     0.30   \\ %
$\log n_{\rm H}$ [cm$^{-3}$] 	&     -4.0  &	   0.2	&     0.30   \\ %
$\log T$ [K] 	                &      4.0  &	   6.1	&     0.30    \\ %
\hline  
\end{tabular}
\flushleft{The columns of the table are: (1) the free parameter as described in the text; (2) the minimum allowed value; 
(3) the maximum allowed value; (4) the step adopted in the grid.}
\end{table}

\begin{figure}
\centering
\includegraphics[scale=0.6]{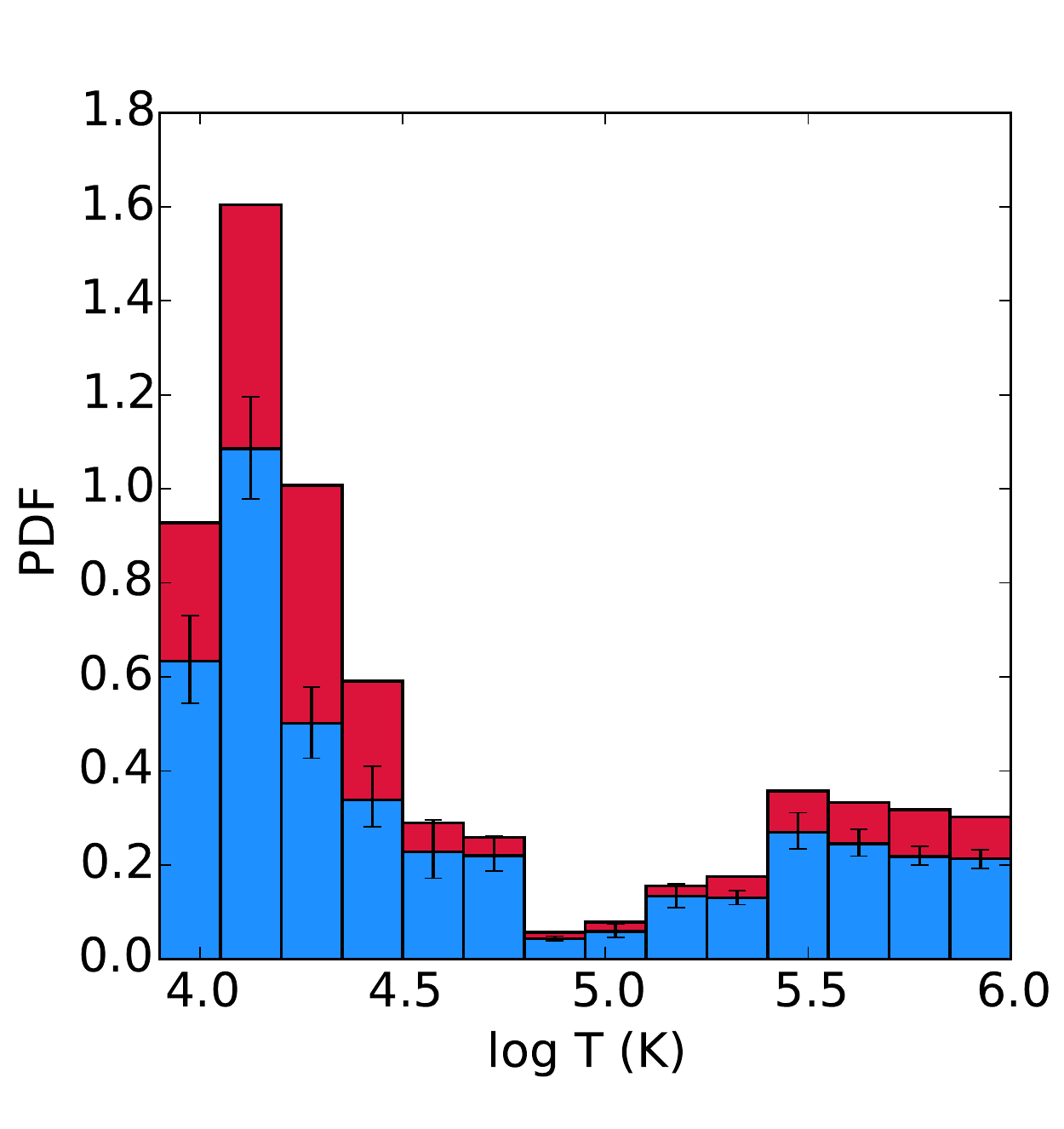}
\caption{Same as Figure \ref{fig:pdfden}, but for the posterior PDF of 
the temperature in the CIE model. The majority of LLSs are characterised 
by low temperatures, common for photoionised gas.}\label{fig:ciepdf}
\end{figure}

\subsection{Effects of collisions}

Up to this point, we have consider only models in which the gas is predominately photoionised. 
Indeed, for the majority of LLSs published in the literature, 
strong absorption from elements which are doubly or triply ionised 
\citep[e.g.][]{rib11,fum13,leh13,pro15} imposes a strong prior on the lack of significant collisional 
ionisation at temperatures $>10^5$ K. It is in fact well known that in collisional ionisation equilibrium (CIE) the 
most abundant elements can be found in neutral or singly/doubly ionised phase only for 
$T < 10^5$K \citep[e.g.][]{gna07,gna12}.
Additional constraints on the gas temperature at $< 10 ^5$ K in CIE 
originate from the fact that LLSs, although ionised, contain large amount of 
neutral hydrogen. Achieving \ion{H}{I} column densities as large as $10^{19}-10^{20}~\rm cm^{-2}$
for temperatures $\gg 10^5$~K would require extreme column densities ($>10^{24}~\rm cm^{-2}$) 
for the total atomic hydrogen.

To seek confirmation of this hypothesis in our own data, we construct a CIE grid of models as an extension 
of the minimal model, by setting the gas temperature to a constant value over the interval $10^4-10^6$ K. 
The parameters of this grid are summarised in Table \ref{tab:cie}. Differently from the previous calculations, 
we also impose a maximum column density for the total hydrogen of 
$\log N_{\rm H} = 10^{24}~\rm cm^{-2} \sim 8 \times 10^3 ~\rm M_\odot~pc^{-2}$
to avoid cases in which the desired \ion{H}{I} column density is achieved by means of implausibly large
columns of highly-ionised hydrogen. 

When using the CIE grid to model the data, in line with our expectations, we find that observations 
nearly exclusively prefer models with $T < 5\times 10^4 $ K (with $\sim 70\%$ of the probability), 
thus recovering the limiting case of gas 
that is photoionised (Figure \ref{fig:ciepdf}). The tail at $T > 10^5$ K arises from either a tail in the 
temperature PDF for systems with large residuals (i.e. for which there is no satisfactory model in the grid) or for 
low metallicity systems that are dominated by non detections. In the latter case, the temperature PDFs are 
sometimes double peaked. High temperature models with very low column densities for most of the ions, particularly 
singly or double ionised, represent an acceptable solution even if limits or values for lines from
triply ionised elements (i.e. \ion{C}{IV} and \ion{Si}{IV}) are overpredicted. Overall, however, 
the accepted models from the CIE grid yield worse residuals compared to what found using photoionisation models.

While this analysis favours photoionisation over collisional ionisation at high temperature, 
we do not exclude the existence of a second gas phase where collisional ionisation is important or even dominant. 
Indeed, other conditions than those found in photoionised gas are likely required to reproduce the strong 
\ion{O}{VI} absorption seen in many LLSs \citep[e.g.][]{leh14}.

\begin{figure*}
\centering
\includegraphics[scale=0.4]{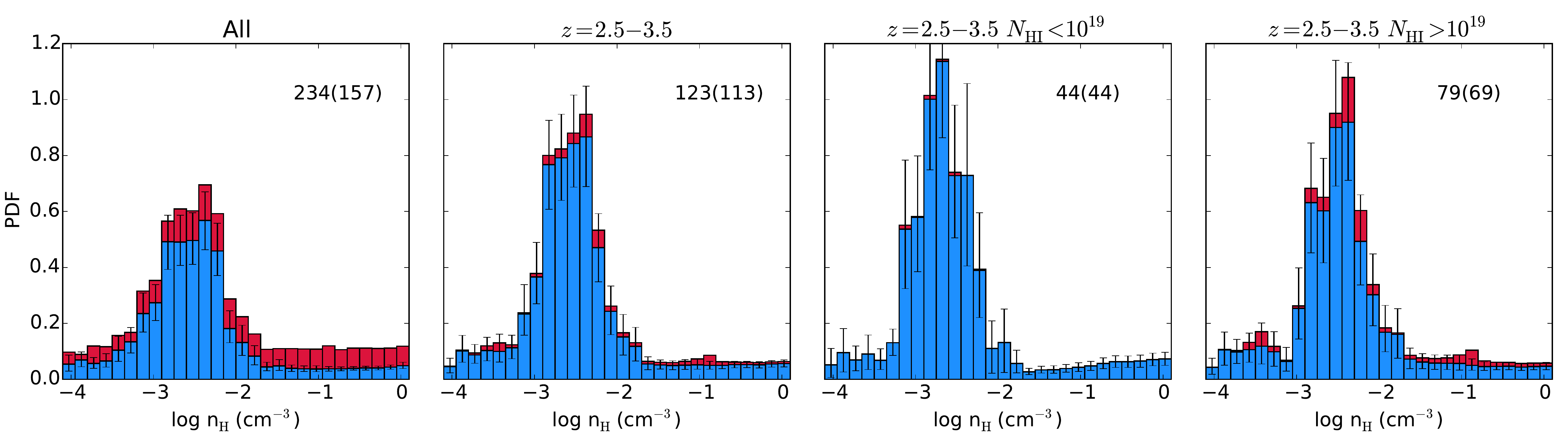}
\caption{The posterior PDF for the density of all LLSs included in this study (red)
and for a subset from the HD-LLS survey (blue), both of which are normalised to the total 
sample size. Error bars indicate the 10th and 90th 
percentile from bootstrapping. From left to right, panels show the PDFs for different cuts 
in redshift and column density, as labelled. The number of systems included in each bin 
is shown, with those from the statistical sample in parenthesis. Despite uncertainties in the 
ionisation corrections, the majority of LLSs is characterised by densities 
between $10^{-3.5}-10^{-2}~\rm cm^{-3}$.}\label{fig:llsdens}
\end{figure*}

\begin{figure*}
\centering
\includegraphics[scale=0.4]{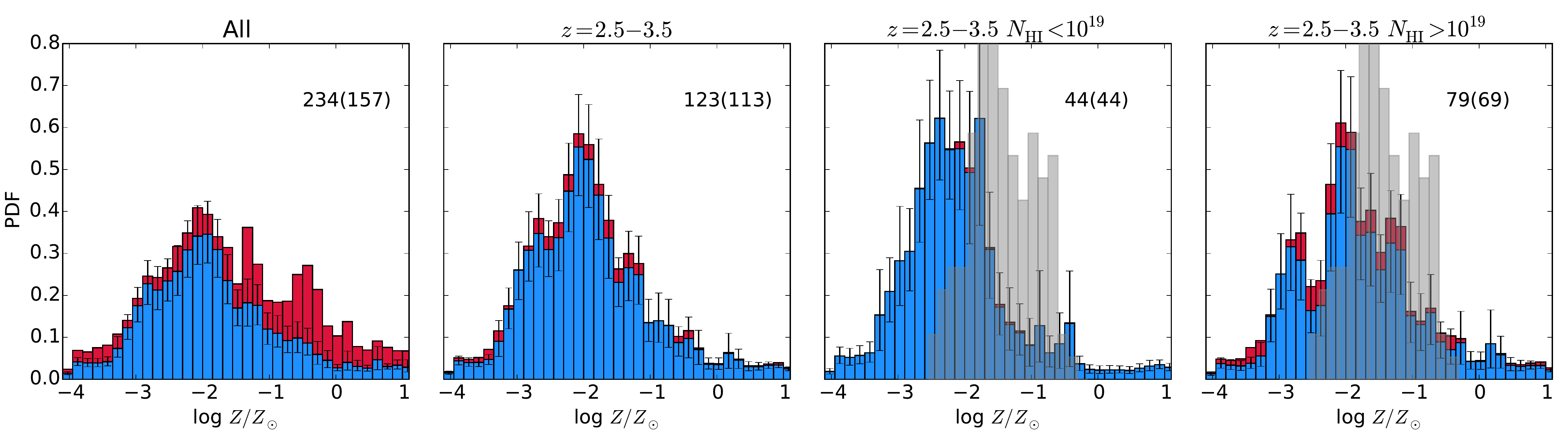}
\caption{Same as Figure \ref{fig:llsdens}, but for the PDFs of the metallicity.
The metallicity distribution of $z\sim 2.5-3.5$ DLAs is shown in grey for comparison 
in the rightmost panels. DLAs are on average more enriched than LLSs, with 
the lower column density LLSs being the most metal poor.}\label{fig:llsmet}
\end{figure*}

\section{Physical properties of LLS\lowercase{s} and astrophysical implications}\label{sec:metal}

In the previous section, we have used Bayesian techniques
combined with grids of ionisation models
to infer the posterior PDFs for the metallicity and density of a sample of 234 LLSs between 
redshift $z\sim 0 - 4.4$, including a homogeneous subset of 157
LLSs from the HD-LLS survey, which form the main statistical sample 
for this analysis. Through different radiative transfer calculations, 
we have assessed whether the analysis of multiple ions in individual 
systems yields robust inference on the physical properties of LLSs.
Under the basic assumptions of a single phase medium in ionisation
equilibrium, we have shown that the inferred PDFs for the LLS
metallicity is well converged, while the inferred PDFs for the 
hydrogen density are less robust, being more sensitive to the 
assumptions made on the shape and intensity of the 
radiation field that illuminates the clouds. 

Next, we  discuss the astrophysical implications of our analysis, 
focusing on the physical properties of LLSs across cosmic epochs
and their connection to accretion and feedback processes. Throughout the remainder of this 
paper, we  use the results derived assuming the dust grid of models, which 
provide a satisfactory description of the data\footnote{Additional tests
based on mock data and the dust grid of models 
can be found in Appendix \ref{sec:valid}.}, as we have shown in Section \ref{sec:dust}.

\begin{figure}
\centering
\begin{tabular}{c}
\includegraphics[scale=0.43]{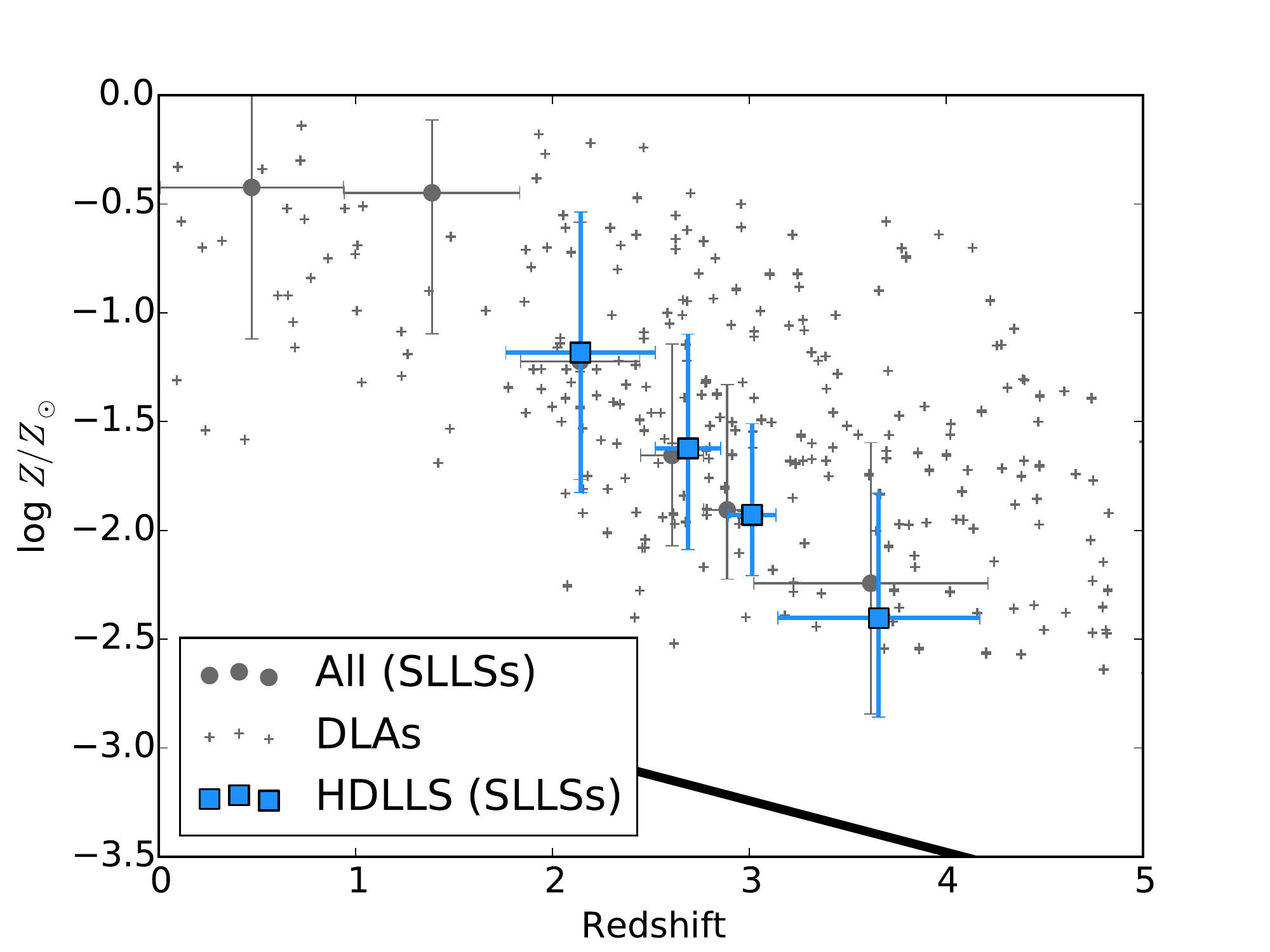}\\
\includegraphics[scale=0.43]{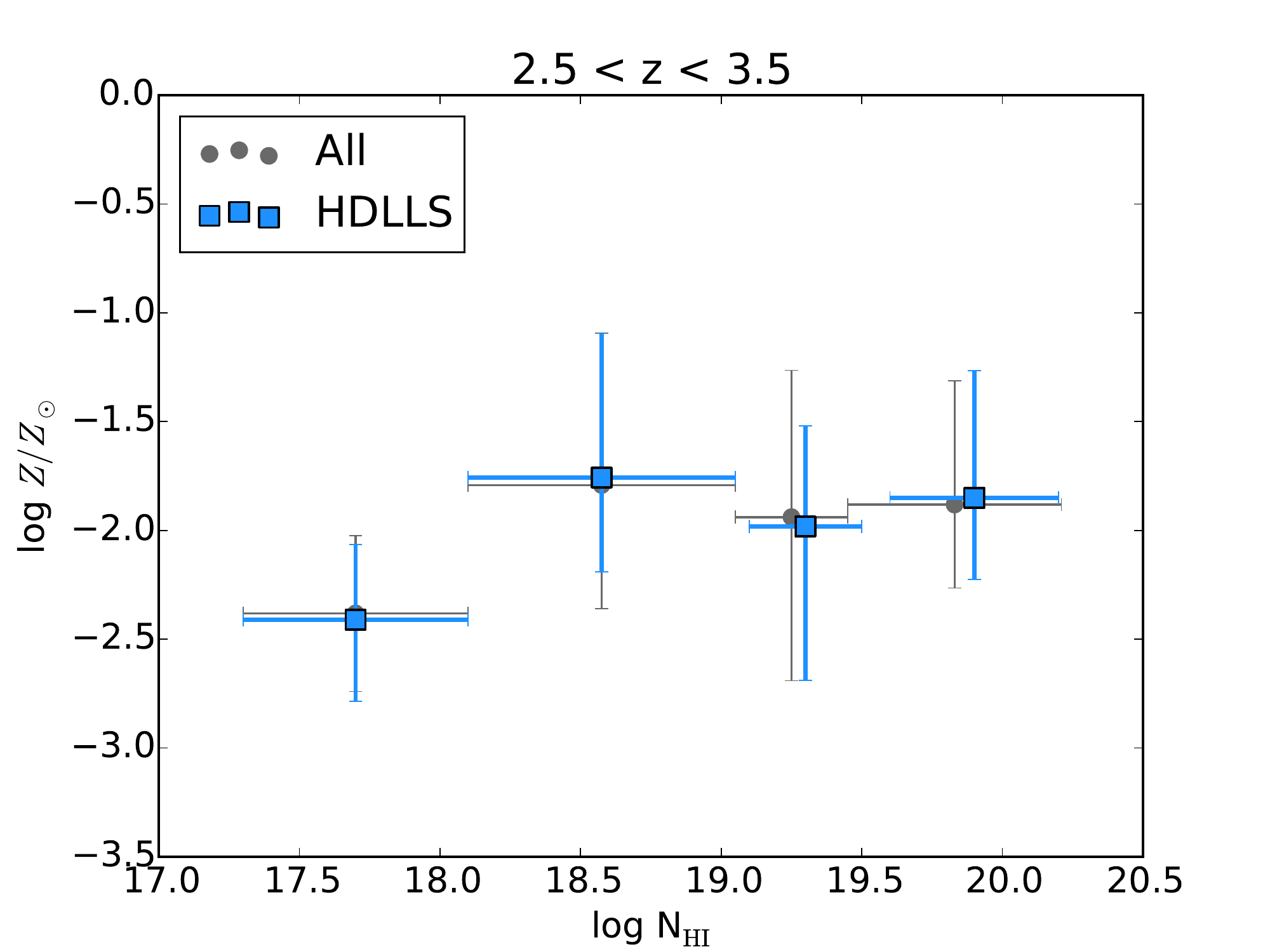}\\
\end{tabular}
\caption{{\it Top.} Redshift evolution of the metallicity for all the LLSs with $\log N_{\rm HI} \ge 19$ 
in our sample (grey circles), and for the subset from the HD-LLS survey (blue squares). 
Data points and their associated error bars represent the median and the 25th/75th percentiles of 
the composite posterior PDFs in redshift bins that are chosen to contain at least 
25 LLSs each. Metallicity measurements for individual DLAs from the literature are also 
superimposed (crosses). The black line marks the median IGM metallicity \citep{sim11}. 
A clear trend with redshift is visible. {\it Bottom.} Dependency on the
metallicity as a function of column density for all the LLSs at $z \sim 2.5-3.5$ (grey circles) 
and the HD-LLS subset (blue squares), showing tentative evidence that lowest column density systems
are the most metal poor.}\label{fig:llsmetcoldep}
\end{figure}

\subsection{Redshift evolution and column density dependence}\label{sec:colred}

Our best estimate for the posterior PDFs for the density and metallicity inferred from both the 
total sample and the HD-LLS sub-sample using the preferred dust model 
are shown in Figure \ref{fig:llsdens} and \ref{fig:llsmet}.
The left-most panel in these figures show the distributions from the entire sample. As was clearly shown 
in Figure \ref{fig:sample}, however, the full sample spans a wide range of redshifts and \ion{H}{I}
column density, but not uniformly. Thus, the reconstructed PDFs carry the imprint of the
selection functions for our data. To gain insight into the underlying physical properties
of high redshift LLSs, we therefore restrict our analysis to the HD-LLS sample in the 
redshift interval $z\sim 2.5-3.5$ (second panel in Figure \ref{fig:llsdens} and \ref{fig:llsmet}), 
where the data more homogeneously sample the full range of column densities. 
Given the large sample size, we can further subdivide the sample in two intervals of 
column densities (third and fourth panels), with a cut at $\log N_{\rm HI} = 19$ for comparisons with 
previous literature on SLLSs (see Sect. \ref{sec:intro}).

Considering the density PDF first, we find that $\sim 80\%$ of the probability 
is contained within $10^{-3.5} \le n_{\rm H} \le 10^{-2}~\rm cm^{-3}$, with a hint of 
higher density for the SLLS subset. However, due to both sample variance and the substantial uncertainties in the
density determination for individual systems, we regard this difference as marginal 
given current data. Again, we remark that while the density for the entire population can be 
constrained, at least loosely, our study confirms a fundamental limitation in the use of photoionisation 
models to infer the density of individual LLSs. More significantly, we conclude that any 
inference on the sizes of the absorbing clouds relying on comparisons between the column density and the gas 
physical density are affected by significant systematic uncertainties, typically in the direction of higher 
density and smaller sizes for clouds that are illuminated by a local radiation field.

Focusing on the metallicity next, LLSs at $z \sim 2.5-3.5$ are characterised by a unimodal distribution
with a peak around $\log Z/Z_\odot \sim -2$ and broad tails towards both high and low metallicity. 
Thus, the shape of the metallicity distribution of high-redshift LLSs does not exhibit the bimodality that has been 
reported at lower redshift by \citet{leh13}. We note, however, that these authors included many systems with 
$\log N_{\rm HI} < 17.2$ in their analysis, and therefore do not strictly consider only optically-thick LLSs. 

From the analysis of the metallicity PDF, we also conclude that LLSs at $z\sim 2.5-3.5$ are metal poor, 
with $\sim 70\%$ ($\sim 85\%$) of the probability at $\log Z/Z_\odot \le -1.5$ 
($\log Z/Z_\odot \le -1$). These findings extend with a $\sim 10$ times larger sample 
some of the results from previous studies based on only a handful of systems \citep{ste90,pro99,fum13,coo15}.
Despite the low metal content, the incidence of very metal poor LLSs ($\log Z/Z_\odot \le -3$) 
is only of the order of $\sim 10\%$, implying that metals are already widespread in moderate column 
density systems at these redshifts. The handful of metal free gas clouds currently known 
are thus rare outliers at $z<3.6$ \citep{fum11pri,pro15,coo15}.

When split in bins of column density, a difference between SLLS and the lower column density LLSs 
can be seen. For LLSs with $\log N_{\rm HI} < 19$, $\sim 60\%$  of the probability is 
at $\log Z/Z_\odot \le -2$ while, for SLLSs with $\log N_{\rm HI} \ge 19$, only 
$\sim 40\%$ of the probability is below $\log Z/Z_\odot \le -2$ (see also Figure \ref{fig:llsmetcoldep}).
Finally, compared to the coeval population of DLAs with $\log N_{\rm HI} \ge 20.3$ 
from the samples of \citet{raf12} and \citet{nee13}, we see that LLSs
are significantly less enriched than DLAs \citep[see also][]{pro15,coo15}.
Even when restricting to systems with $\log N_{\rm HI} > 19$, 
DLAs are still more enriched than, or at least have comparable metallicity to, SLLSs. 
This result is in contrast with claims according to which SLLSs are more metal rich 
than DLAs \citep[e.g.][]{kul07,bat12,som13,som15}. We note however that those claims rely on small samples
at $z>2$, and are largely based on extrapolations of trends apparent at lower redshift. According to our 
homogeneous analysis, either the median metallicity of SLLSs crosses the one for DLAs below $z \sim 2$, 
or a reassessment of the data from the literature may be necessary 
\citep[see][for additional discussion on the topic]{des09}. 
Furthermore, as previously noted by other authors \citep{pro06,per06b,fum11pri},
while the metallicity of DLAs is well described by a Gaussian with mean $\log Z/Z_\odot \sim -1.5$
and width  $\log Z/Z_\odot \sim 0.6$ \citep{raf12}, LLS metallicity spans a range up to 
$> 4$ dex.

Additional insight into the redshift evolution of the metal enrichment and its dependence on 
the \ion{H}{I} column density can be gained by further studying the median properties of 
this sample in bins of redshift and column density (Figure \ref{fig:llsmetcoldep}). As a function of redshift, 
LLSs with $\log N_{\rm HI} > 19$ appear to evolve rapidly with time, 
with a median $\log Z/Z_\odot \sim -2.4$ at $z\sim 3.6$
reaching a median $\log Z/Z_\odot \sim -1.2$ at $z\sim 2.1$. Also, the metallicity of LLSs 
with $\log N_{\rm HI} > 19$ appears to evolve more rapidly in comparison to DLAs. 
In agreement with previous studies \citep[e.g.][]{per07},
we observe a ten-fold increase in the median metallicity of SLLSs over $\sim 1.5$ 
Gyr of cosmic evolution, although with $\sim 1$ dex scatter about the median relation.
This evolution also continues at lower redshift, although we caution that a better 
understanding of the selection function for the systems from the literature is required 
before extrapolating this trend to $z<2$ using this sample. 
Such an evolution points towards a very rapid enrichment of the absorbing gas
below $z<4$ and/or the disappearance of the most metal-poor LLSs which could be ionised
or accreted onto galaxies at later times.

Finally, when examining the dependence on column density, the
metallicity appears to be only a very weak function of $\log N_{\rm HI}$, although with 
an apparent drop of $\sim 0.5$ dex for the lowest column density bin. Thus, the difference between 
lower column density LLSs and SLLSs seen in Figure \ref{fig:llsmet} appears to be driven mostly by 
systems at the lowest column density. 
This trend suggests a smooth transition between the population of LLSs and the IGM.
Thus, while a boundary at $\log N_{\rm HI} = 17.2$ is physically motivated given the
ionisation condition of the gas, the separation between optically-thick absorbers 
that reside in biased environments from those in the IGM may be less clear cut
already at $\log N_{\rm HI} \lesssim 18$ (see Sect \ref{sec:cgm}). 
Studies based on even larger samples should confirm this conclusion, which remains tentative.
  
\begin{figure}
\centering
\begin{tabular}{c}
\includegraphics[scale=0.43]{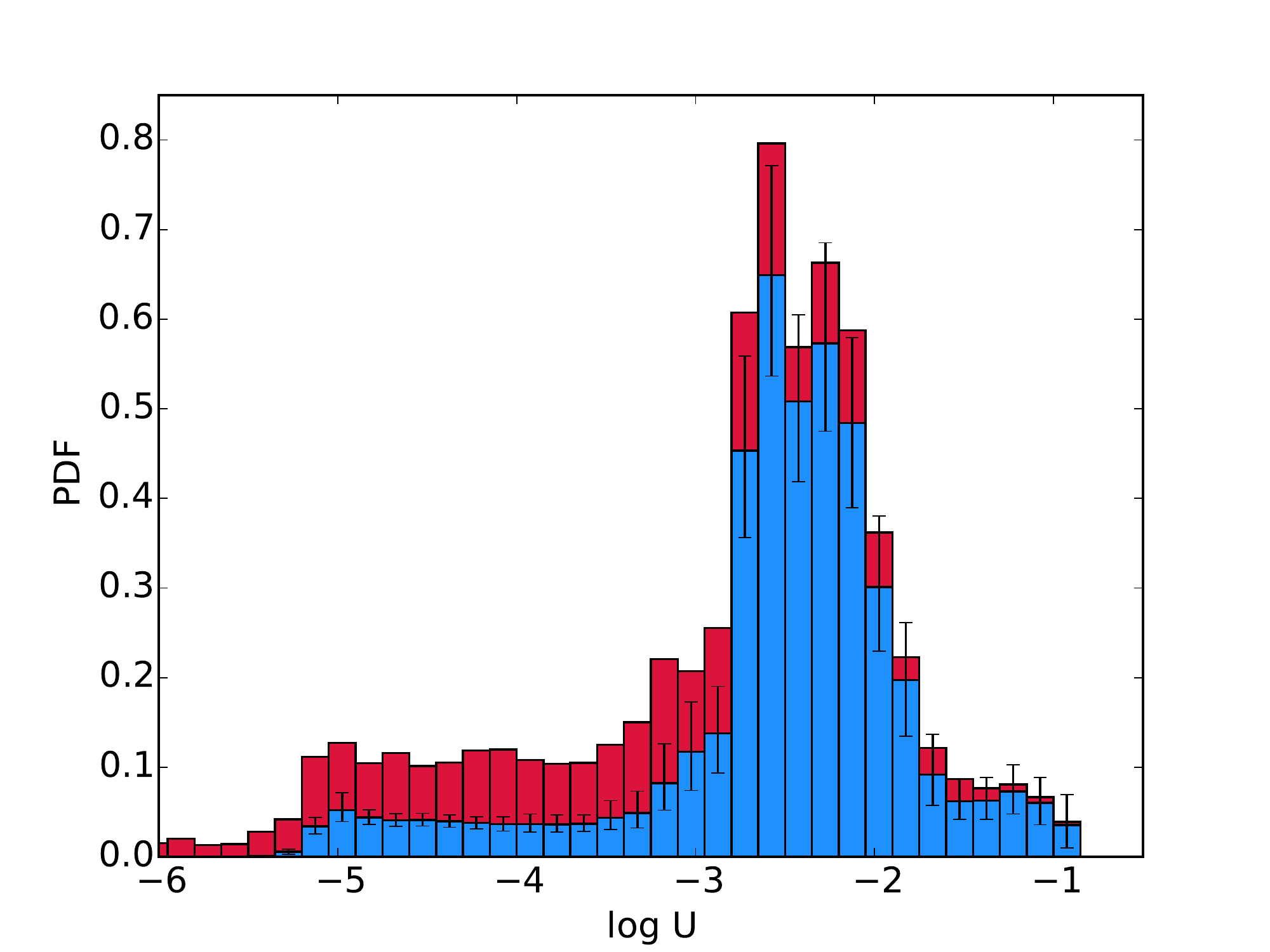}\\
\includegraphics[scale=0.43]{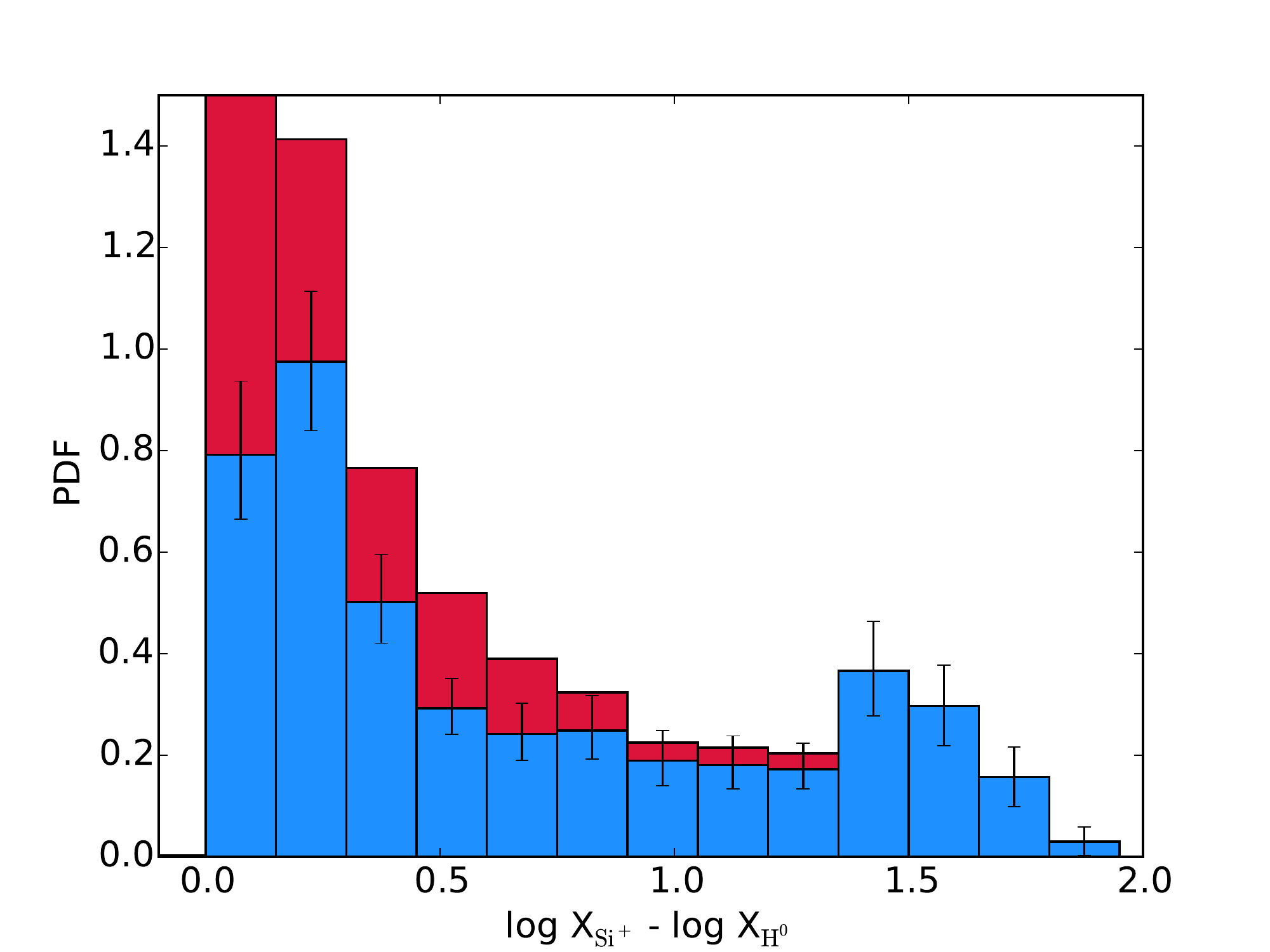}\\
\end{tabular}
\caption{{\it Top.} The PDF of the ionisation parameter for all the LLSs included in this study
(red) and for the HD-LLS sub-sample (blue), computed from bootstrap resampling.
{\it Bottom.} ICs for \ion{Si}{II} relative to \ion{H}{I}, computed as the logarithmic difference of 
the Si$^{+}$ and H$^{0}$ fraction. The near totality of LLSs are highly ionised and differential 
ICs cannot be neglected when computing the metallicity.}\label{fig:ionprop}
\end{figure}

\begin{figure}
\centering
\begin{tabular}{c}
\includegraphics[scale=0.43]{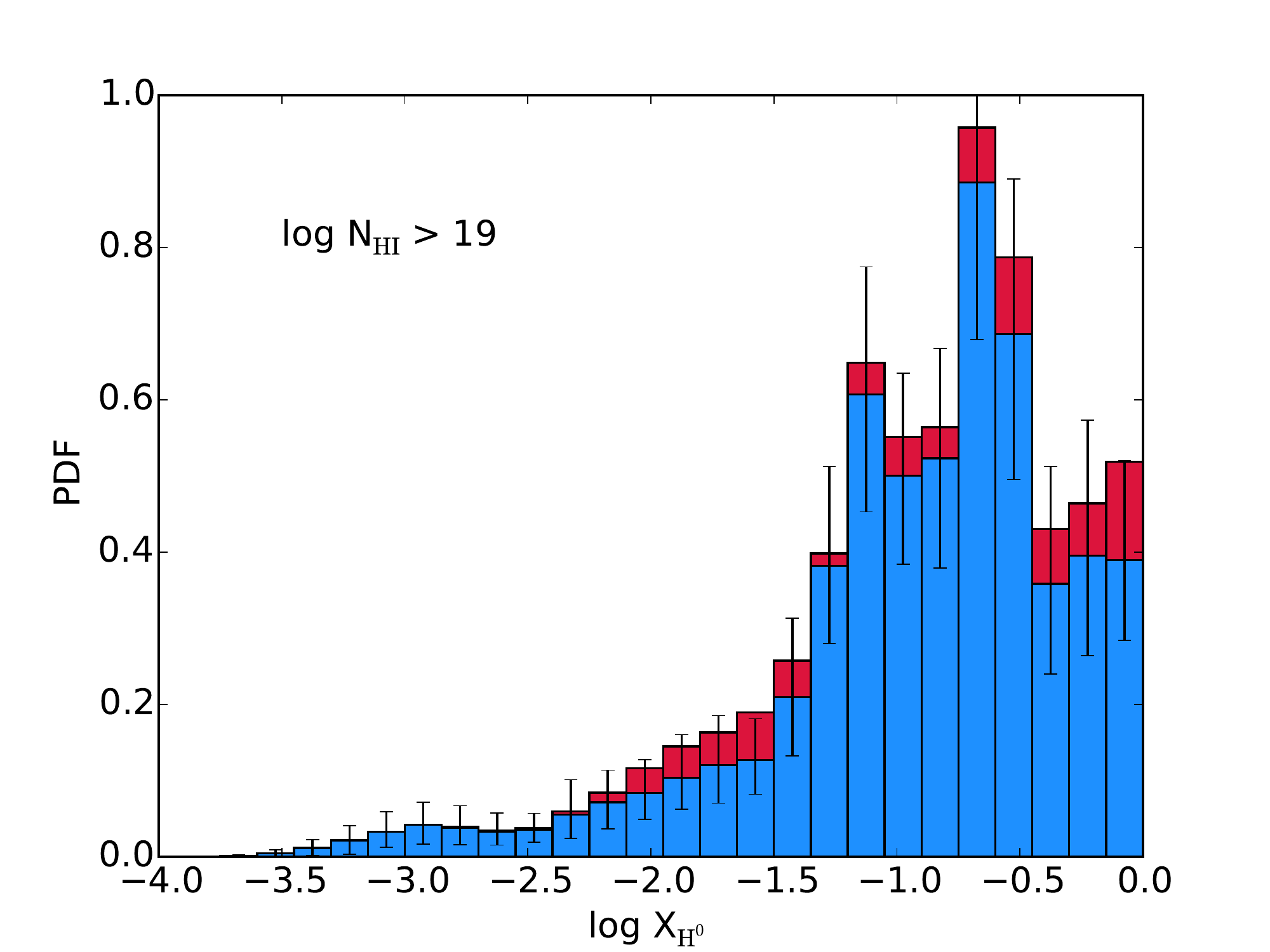}\\
\includegraphics[scale=0.43]{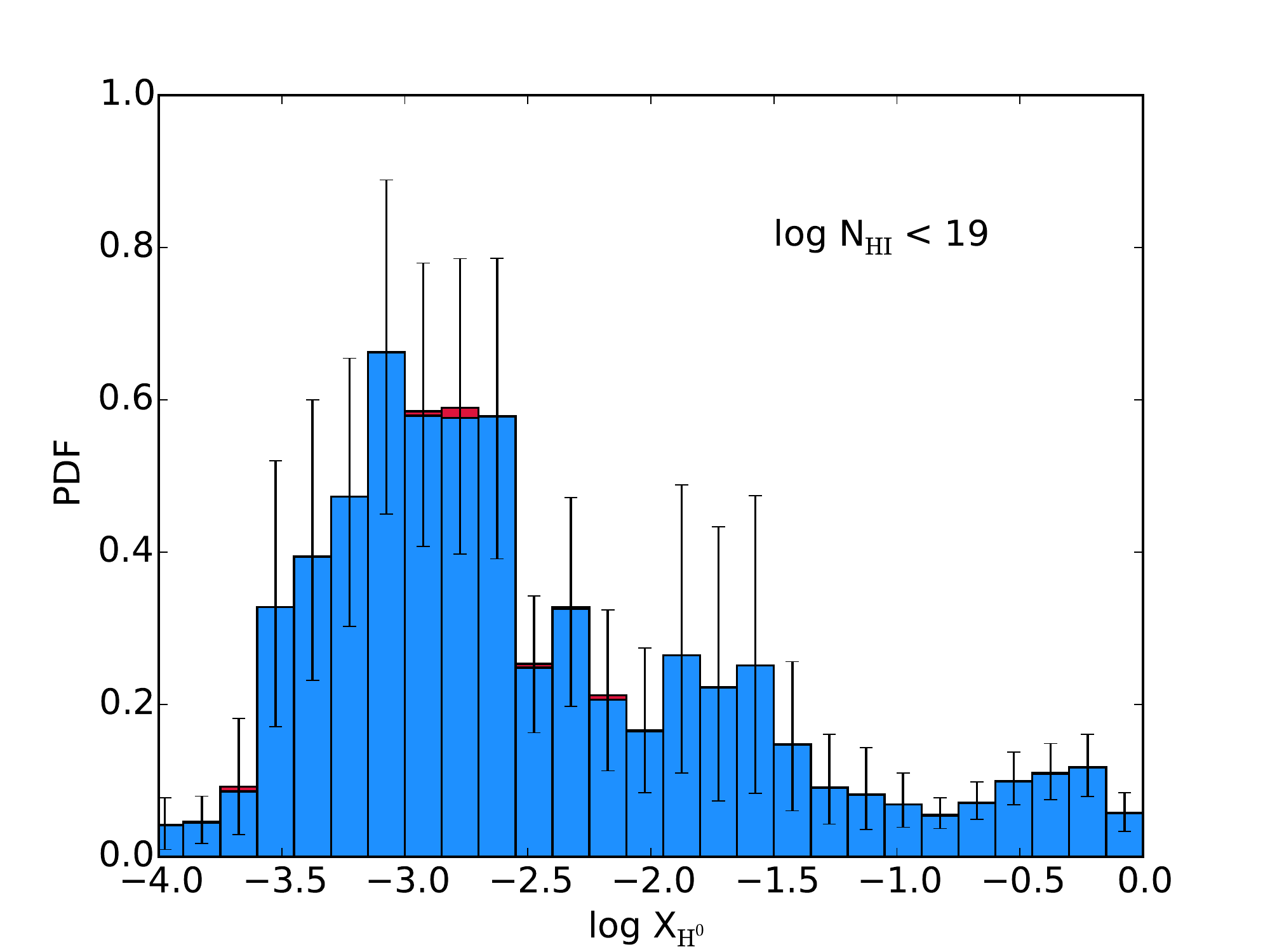}\\
\end{tabular}
\caption{Same as Figure \ref{fig:pdfden}, but showing the PDF of the hydrogen neutral fraction
between $z = 2.5-3.5$ for the SLLS sub-sample (top) and the sub-sample of LLSs with $\log N_{\rm HI} < 19$
(bottom).}\label{fig:hifrac}
\end{figure}

\subsection{Ionisation corrections}\label{sec:ioncor}

With a model for the ionisation properties of LLSs, we now discuss the importance of
ICs in deriving the metallicity. To this end, we compute the PDF of the ionisation 
parameter, $U$, and of the ICs that are needed for inferring the gas metallicity using \ion{Si}{II} 
column densities. Both these quantities are shown in Figure \ref{fig:ionprop}. 
For completeness, we define the ionisation parameter as the dimensionless 
ratio of the ionising photon flux to hydrogen density, $U = \phi/(n_{\rm H} c)$, with 
$c$ the speed of light. The IC for \ion{Si}{II} relative to \ion{H}{I} is 
instead defined as the logarithmic difference of the fraction of silicon in the first ionisation stage and the fraction of 
atomic hydrogen in the neutral phase.
To reconstruct the PDF for $U$ in the entire sample, we first compute PDFs for individual systems 
combining the density PDF and the ionising photon flux from the UVB at the relevant redshifts, and 
then we coadd the individual PDFs following the procedures we have adopted for the density and metallicity 
case. For the ICs, instead, we extract the ionisation fraction of Si$^{+}$ and H$^{0}$
at each position of the parameter space that has been sampled by the MCMC. 

As seen  in Figure \ref{fig:ionprop}, the PDF for the ionisation parameter shows a clear peak 
between $-3 \lesssim \log U \lesssim -2$, as commonly found for the analysis of individual 
LLSs \citep{pro99,fum11pri,coo15}. This means that, within the HD-LLS sample, there is a $>80\%$ probability 
of finding a LLS that is significantly ionised, with $\log U \ge -3$, including a $\sim 15\%$ probability 
of finding a very high ionisation for  $\log U \ge -2$. Thus, the majority of the LLSs at $z\gtrsim 2$, 
including SLLSs, are highly ionised, a conclusion that is in line with more 
empirical assessments based on ion ratios \citep{pro15}. At lower redshifts, where the 
UVB photoionisation rate declines, and for the highest column densities that are more common 
in the literature sample, the ionisation parameter progressively decreases below $\log U \lesssim -3$, as 
visible from a long tail in the PDF.

Inspecting the PDF of the ICs for \ion{Si}{II} as a metallicity tracer, we note that these are generally small 
($\lesssim 0.5$ dex) for most of the systems. However, the large fraction of ICs above $\sim 0.2$
already for SLLSs, including a long tail that extends beyond $\sim 1$ for most of the LLSs, 
clearly shows that differential ionisation effects among elements cannot be neglected when computing 
the metallicity of optically-thick absorbers for all \ion{H}{I} column densities (at least at redshifts $z\sim 2-3$). 

Finally, since the \ion{C}{III} ($\lambda$977) and \ion{Si}{III} ($\lambda$1206) 
transitions lie within the Ly$\alpha$ forest, we do not have clean measurements for these ions, 
especially at high redshifts \citep[see][]{pro15}.  We can however use the results of the ionisation models
to verify the hypothesis that most of the carbon  (and silicon) in LLSs are doubly-ionised 
(cf. Figure \ref{fig:carb}). By computing the fraction of carbon in the singly, doubly, 
and triply ionisation stages along the chains, we find $\log X_{\rm C^{++}} \gtrsim -0.4$ for 
the majority of the SLLSs and $\log X_{\rm C^{++}} \gtrsim -0.2$
for LLSs with $\log N_{\rm HI} < 19$. Thus, the majority of the carbon is indeed predicted to be
observed as \ion{C}{III}. Similar conclusions hold for silicon. We emphasise that, in turn, this result implies that our 
estimates for the metallicity hinge on ions that trace only a fraction of the total mass in metals.

\begin{figure*}
\centering
\begin{tabular}{c}
\includegraphics[scale=0.6]{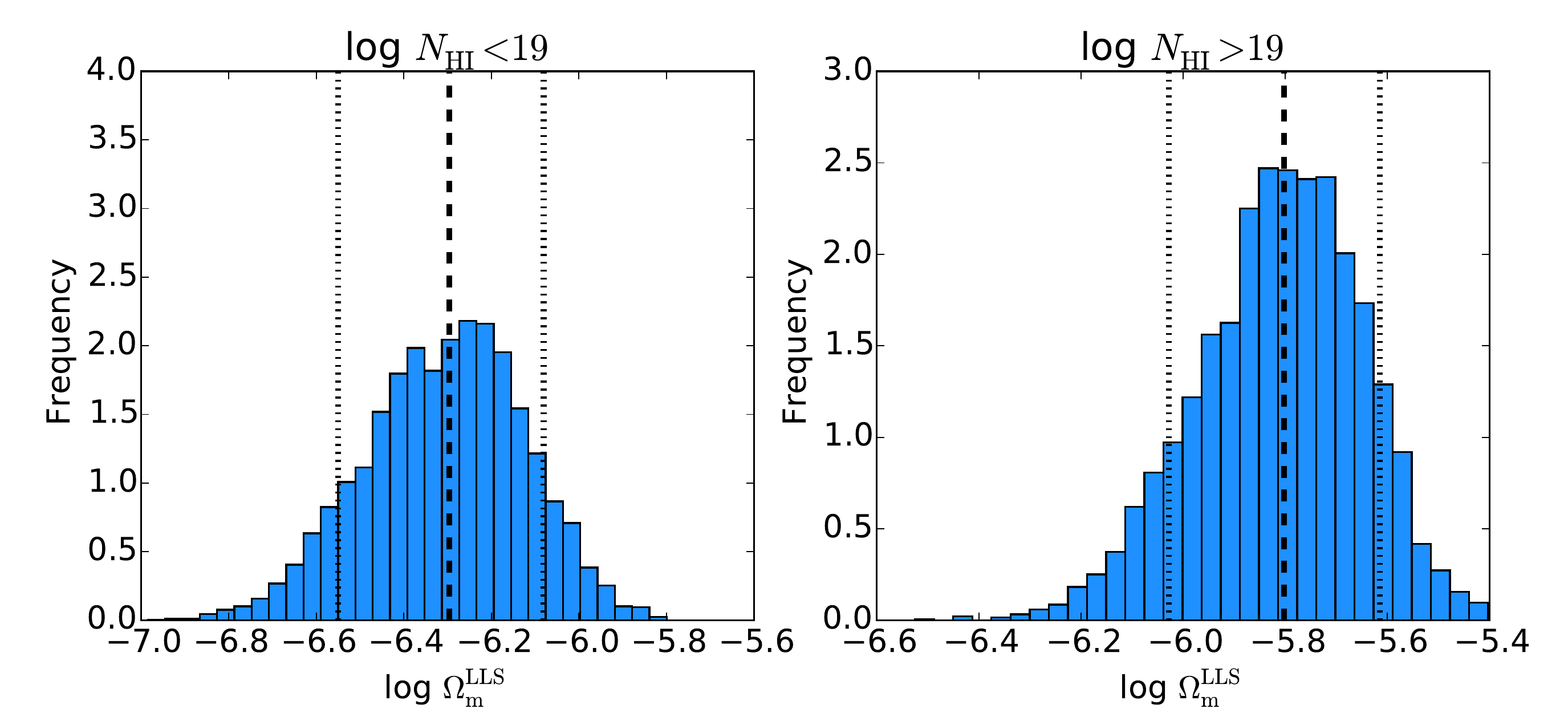}
\end{tabular}
\caption{Frequency distributions for the cosmic metal content of optically-thick 
absorbers in the HD-LLS sample at $z = 2.5-3.5$. The different contribution of SLLSs (right) and LLSs with 
$\log N_{\rm HI} < 19$ (left) is shown. Once ionisation corrections are accounted for,
LLSs contribute to $\sim 15\%$ of the total metal budget at these redshifts, 
with large uncertainties.}\label{fig:omegam}
\end{figure*}

\subsection{The cosmic metal budget}\label{sec:budget}

With an estimate for the metallicity PDF in a large sample of HI-selected LLSs 
between $z\sim 2.5 - 3.5$, we revisit the question of what fraction of the metals ever 
produced in stars is locked in optically-thick absorbers, also comparing to
DLAs and the Ly$\alpha$ forest.  Calculations for the metal budget in DLAs and LLSs can be 
found in the literature \citep[e.g.][]{pet06,kul07,bou07,pro06,per07,raf14,leh14}, but we 
are now able to fill the gap between $\log N_{\rm HI} \sim 17-19$, a range of column density 
that remained largely unexplored by previous studies. 

Following the notation in \citet{pro06}, the metal mass density of LLSs, 
$\Omega_{\rm m}^{\rm LLS}$, is defined as
\begin{equation}\label{eq:omega}
\Omega_{\rm m}^{\rm LLS} = \frac{1.3 m_{\rm p} H_0}{c \rho_{\rm crit}} \int_{N_{\rm low}}^{N_{\rm high}} 
\frac{1}{X_{\rm H^0}} N_{\rm HI} f(N_{\rm HI}) Z_{\rm LLS} dN_{\rm HI}\:, 
\end{equation}
with $X_{\rm H^0}$ the atomic hydrogen neutral fraction, $f(N_{\rm HI})$ the column density distribution 
function, $Z_{\rm LLS}$ the mass in metals in the LLS gas, $m_{\rm p}$ the proton mass and 
$\rho_{\rm crit}$ the critical density at redshift $z=0$. 
The factor $1.3$ accounts for helium. 
From equation \ref{eq:omega}, the first 
ingredient for computing $\Omega_{\rm m}^{\rm LLS}$ is the metallicity distribution, which is 
available from the previous section. 
Next, for the column density distribution, we assume the spline 
function tabulated in \citet{pro14}, which we evaluate at $z = 3$.

For the final ingredient of this calculation, we need to characterise the hydrogen neutral 
fraction in our sample, which we compute following the same procedures used 
to derive the PDF for the \ion{Si}{II} ICs, i.e. by extracting the hydrogen neutral fractions 
along the chains that sample the grid parameter space. In Figure \ref{fig:hifrac}, we show the 
PDFs for SLLSs and for LLSs with $\log N_{\rm HI} < 19$, restricted to the redshift interval 
$z = 2.5-3.5$. In line with our previous discussion on ionisation, SLLSs are generally 
characterised  by neutral fractions $X_{\rm H^0} \gtrsim 0.1$, while LLSs are much more ionised, 
typically with $X_{\rm H^0} \lesssim 0.01$. As the neutral fraction is a function of $N_{\rm HI}$, 
we refine our estimate for $X_{\rm H^0}$ with a simple functional form 
$\log X_{\rm H^0} = \alpha \log N_{\rm HI} + \beta$, which is found to be a good 
description for our data. We adopt linear regression to compute the coefficients 
$\alpha$ and $\beta$ for our statistical sample at $z = 2.5-3.5$,
also accounting for the width of the $X_{\rm H^0}$ PDFs for individual systems. The best-fit values
are $\alpha = 0.99 \pm 0.06$ and $\beta = -20.3 \pm 1.2$, implying that the hydrogen 
neutral fraction changes by one order of magnitude for every decade of \ion{H}{I} and that 
the transition between neutral and ionised hydrogen is crossed at the SLLS/DLA boundary. 
This dependence also implies that the total hydrogen column density 
in LLSs is comparable to the one in SLLSs.

To compute $\Omega_{\rm m}^{\rm LLS}$, we evaluate equation \ref{eq:omega} 
in bins of column density, chosen to contain $\gtrsim 15$ systems each. 
In each bin, we integrate the total hydrogen column density using our 
fitting formula for $\log X_{\rm H^0}$ and weighting by the column density distribution 
function. We then multiply this integral value by the mean metallicity, in linear space,
computed for the LLSs within that bin. To account for sample variance, we repeat this calculation
for 5000 samples drawn from our parent distribution allowing for repetitions. 
The resulting distributions, split for LLSs and SLLSs, are shown in Figure \ref{fig:omegam}.

For the assumed $f(N_{\rm HI})$ and cosmology, LLSs with $\log N_{\rm HI} < 19$ and SLLSs 
respectively contain $\Omega_{\rm HI} \sim 1.0 \times 10^{-5}$ and 
$\Omega_{\rm HI} \sim 2.2 \times 10^{-4}$, which is a factor 
$\sim 5-100$ less than the amount of neutral hydrogen locked in the higher column density DLAs, 
for which $\Omega_{\rm HI} \sim 1.0 \times 10^{-3}$. The difference among these populations 
arises mainly from the intrinsic \ion{H}{I} column density. However, after accounting for 
ionisation corrections and folding in the metallicity distributions, LLSs with 
$\log N_{\rm HI} < 19$ contain 
$\Omega_{\rm m}^{\rm LLS} = 5.1 \times 10^{-7}$ with a 10th/90th percentile interval of
$(2.8,8.3)\times 10^{-7}$. SLLSs, instead, have a median 
$\Omega_{\rm m}^{\rm LLS} = 1.6 \times 10^{-6}$ with a 10th/90th percentile interval of 
$(0.9,2.4)\times 10^{-6}$. Jointly, these optically-thick absorbers contain a total mass 
in metals of $\Omega_{\rm m}^{\rm LLS} \sim 2.1 \times 10^{-6}$.

According to our estimate, SLLSs contribute about a factor $\sim 3$ more than LLSs to the total 
cosmic metal budget, which is a consequence of having a few very metal rich SLLSs in our sample
with no counterparts at $\log N_{\rm HI} < 19$.
We emphasise however that, while the extremes of the distribution do not significantly affect the 
median metallicity discussed in previous sections, the estimate of $\Omega_{\rm m}$ depends on the mean 
(linear) metallicity, which is much more sensitive to the tail of the metallicity PDF. Thus, solar and super-solar
systems, although a small fraction of the entire sample, contribute significantly to $\Omega_{\rm m}$ \citep{pro06}.
To illustrate this point in quantitative terms, we perform a simple idealised experiment, by adding 
three fake LLSs with $\log N_{\rm HI} = 17.5, 18.0, 18.5$ and metallicity $\log Z/Z_\odot = 0.4$ to the sample 
of 44 LLSs in our statistical sample. With these fake systems, the median metallicity of the ensemble increases 
by only $\sim 0.1$ dex. Conversely, the estimate for the total metal content is significantly perturbed by the 
inclusion of systems with $\sim 100$ times more metals than the mean population, shifting
$\Omega_{\rm m}^{\rm LLS}$ to $\sim 3.7 \times 10^{-6}$.
This example is clearly based on an arbitrary number of fake systems, but it highlights how sensitive 
$\Omega_{\rm m}^{\rm LLS}$ is to extremes in the metallicity distribution. 

For a similar reason, the uncertainty in the metallicity for individual systems, which is generally computed in 
logarithmic space, bias the estimate of $\Omega_{\rm m}^{\rm LLS}$ in one direction. To test this effect, we perturb 
the median metallicity of individual systems with errors drawn from a Gaussian with width 0.2 dex 
(e.g. Figure \ref{fig:jointdust}), and we repeat the measurements of $\Omega_{\rm m}^{\rm LLS}$
for 5000 realisations. Due to the use of the arithmetic mean, realisations with positive errors weight more 
than those with negative errors, skewing the distribution for $\Omega_{\rm m}^{\rm LLS}$ towards 
higher values, by a factor $\gtrsim 2-3$. Thus, on the top of the uncertainty in the sample variance 
captured by Figure \ref{fig:omegam}, a factor of $\gtrsim 2$ systematic uncertainties may affect 
this measurement. 

With this uncertainty in mind, we compare our estimate to literature values. Our new measurement 
for $\Omega_{\rm m}^{\rm LLS}$ for SLLSs is comparable to the value reported by \citet{pro06}, who quote 
$\Omega_{\rm m}^{\rm LLS} \sim (2-5) \times 10^{-6}$ for SLLSs at $z\sim 2$, with their range being 
dependent on the choice for ionisation correction. Similar considerations apply for the estimate in \citet{per07}.
For a more detailed comparison, especially for SLLSs, we should consider not only the shape 
of the metallicity distributions adopted by different authors, but also the rapid metallicity 
evolution shown in Figure \ref{fig:llsmetcoldep}. In our estimate, we have restricted to the redshift 
range $z\sim 2.5-3.5$ in an attempt to reduce the redshift dependence while retaining a sufficiently
large sample size. As $\Omega_{\rm HI}$ only mildly evolves with redshift \citep{per05}, however, 
a steep evolution in the metallicity for SLLSs likely weights our measurement towards lower redshifts. 
Accounting for further evolution to $z \sim 2$, our determination is likely to be in even better agreement with 
the estimate of \citet{pro06} and \citet{per07}. 

Compared to the $z \sim 3$ DLA population, instead, LLSs as a whole account for 
$\sim 3$ times the metals in DLAs, according to the recent value of $\Omega_{\rm m}^{\rm DLA} \sim 6.2 \times 10^{-7}$ 
reported by \citet{raf14}, who revise previous estimates downward by a factor of $\sim 3$. 
The contribution from the Ly$\alpha$ forest is instead estimated at 
$\Omega_{\rm m}^{\rm Ly\alpha} \sim 4.6 \times 10^{-6}$, that is a factor of $\sim 2$ higher than 
for LLSs, albeit with substantial uncertainties \citep{sch03,bou07,sim11}.
Altogether, hydrogen absorbers at $z\sim 2.5-3.5$ contain a total amount of metals 
$\Omega_{\rm m}^{\rm ALS} \sim 7.3 \times 10^{-6}$. Given a conservative estimate for 
the metals produced by Lyman break galaxies (LBGs), 
$\Omega_{\rm m}^{\rm LBG} \sim 1.6 \times 10^{-5}$ \citep{raf14}, we find that 
LLSs account for $\sim 15\%$ of the metals ever produced by UV-selected galaxies, 
with the total population of HI absorbers accounting for $\sim 45\%$ \citep[cf][]{bou07}. 
From this analysis, we conclude that LLSs are significant repository of 
metals at $z\sim 3$. It should be noted that in our estimate we are considering only metals that are 
locked in the main cool gas phase, which gives rise to the bulk of the \ion{H}{I} absorption. 
A second (likely warmer) more ionised phase (e.g. traced by \ion{O}{VI}) may account for an 
even greater fraction of metals \citep{leh14}.

Clearly, this rough budget should be interpreted with caution, as several uncertainties still 
hamper a precise determination on many quantities that are relevant in this calculation, 
including the metal yields, the faint end of the galaxy luminosity function, the contribution 
from an obscured galaxy population, and ionisation corrections as a function of column density. 
Furthermore, we emphasise again that, as we have discussed at length, the LLS population is characterised 
by an intrinsic scatter in the metallicity that, in turn, is reflected in the scatter for 
$\Omega_{\rm m}^{\rm LLS}$ in Figure \ref{fig:omegam}. More significantly,
the cosmic metal budget is sensitive to the high-end of the metallicity distribution, 
and thus our estimate based on this sample is likely not to have converged \citep{pro06}.
Refined measurements will require much larger samples or physically motivated models for 
the metallicity distribution that enters equation \ref{eq:omega}.

\subsection{Optically-thick gas and the circumgalactic medium}\label{sec:cgm}

We conclude with a discussion of the implications of our analysis for studies of 
the nature of optically-thick absorption line systems and for studies of
the physical properties of the CGM at high redshift.

As noted above, the exact shape of the density distribution for LLSs is not well constrained because 
of a degeneracy with the radiation field. With this caveat in mind, our analysis has 
nevertheless provided indications that the bulk of LLSs have densities 
in the range $n_{\rm H} \sim 10^{-3.5}-10^{-2}~\rm cm^{-3}$.
Thus, for a mean cosmic density of $\sim 1.2\times10^{-5}~\rm cm^{-3}$ at $z\sim 3$, our work 
places LLSs in overdense structures, with contrast densities $\delta \sim 30-800$, which are 
comparable to or greater than the virial densities for most of the systems. 
In turn, these observations lead to the natural 
inference that a large fraction of LLSs are associated to galaxy halos, adding to the many pieces of empirical evidence 
that suggest a connection between LLSs and the CGM, such as the redshift evolution 
of the number of LLSs per unit redshift \citep[e.g.][]{sar89,fum13}. 

If indeed the link between LLSs and the CGM is established at $z \sim 2-3$ by more direct measurements, 
such as direct imaging of the galaxies giving rise to LLSs or a measurement of the LLS 
bias \citep{fum14}, the physical properties of large samples of LLSs will become 
some of the most constraining observables for models of galaxy formation. 
Hydrodynamic simulations consistently predict a connection between LLSs and the halo of galaxies.
More specifically, different simulations agree in predicting that the elusive gas accretion 
onto halos should be manifest in the form of LLSs absorbers \citep[e.g.][]{fau11,fum11sim,van12,fum14}.
However, up until now, most of these predictions have been tested with simple metrics,
currently available in observations, such as the covering fraction of optically-thick gas around LBGs 
or quasars \citep{rud12,pro13}. Unfortunately, the outcome of these tests have been weakened
by the small sample sizes or by the fact that observations probe halos that are not always 
representative of the entire galaxy population of interest. 
Our analysis offers a way to alleviate some of these limitations,
by providing new additional metrics that can be used to test predictions of 
hydrodynamic simulations and, in turn, improve our understanding of the properties of cold gas 
accretion and feedback around galaxies. 

A detailed comparison between our observations and simulations is beyond the scope of this work, 
but we provide some qualitative considerations of this type of analysis.
As an example, one could compare the range of physical densities and hydrogen neutral fractions 
inferred from data to the predictions of detailed radiative transfer calculations for the denser 
components of the CGM in hydrodynamic simulations \citep{fau10,fum11sim}.
A cursory look to published predictions suggests general agreement between these quantities.

Moreover, our measurement of the metal distribution of LLSs opens a new window to quantitatively 
constrain feedback models, and the interaction between metal-poor cold gas accretion and the 
ejection of metal-rich gas in galactic winds. Indeed, it is clearly emerging that different 
implementations of stellar feedback alter the observable properties of the CGM 
\citep[e.g.][]{hum13,lia15}. Fundamental questions on the reliability of cosmological simulations 
to predict the transport and mixing of metals on small scales remain, but 
a quantitative comparison between the observed
metallicity PDFs and the results of hydrodynamic simulations will be 
a simple yet powerful diagnostic of the efficiency of metal ejection and mixing in the halos 
of high redshift galaxies.

While we cannot derive firm conclusions without a proper statistical analysis
on large samples of simulated halos, it is interesting to note that 
simulations often predict drastic discrepancies for the metal distributions in the CGM. 
Specifically, if the majority of the LLSs indeed 
arise in the cold   gas near to galaxies, some of the simulations that implement efficient 
feedback to overcome rapid gas cooling and excessive star formation may fail to reproduce 
the metallicity PDF \citep[see e.g.][who predict metallicity $\log Z/Z_\odot \sim -1$ for LLSs]{she13}. 
Puzzlingly, models with weak feedback implementations that 
fail instead to reproduce the correct fraction of baryons in stars \citep{fum11sim},
may more closely reproduce the metallicity PDF for LLSs with a peak around $\log Z/Z_\odot \sim -2$. 

Furthermore, the low metal content of the LLS population at $z\sim 2.5-3.5$ poses an additional 
constraint in the interpretation of the strong equivalent widths in low ionisation metal transition 
near to galaxies \citep{ste10}. Models that are often invoked to explain these observations
assert that these lines arise from far-reaching metal-enriched outflows inside the halos. 
However, a better understanding of the gas 
kinematics and of the metal distribution in the wind cold phase becomes critical to assess 
the relative contribution of column densities and Doppler parameters in shaping the strong (saturated) 
metal transitions, while keeping the metallicity of LLSs below $\log Z/Z_\odot \sim -1.5$, 
in line with our observations.

Finally, we remark that, from a theoretical point of view, these possible tensions 
motivate ongoing and future efforts to study in detail the coexistence of metal-poor inflows 
and metal-enriched outflows around high redshift galaxies.
Similarly, from the observational point of view, establishing 
to what extent LLSs and the halo of galaxies are connected is becoming a necessary task, so as to fully exploit 
the many diagnostics available in LLSs for studies of accretion and feedback. 

\section{Summary and Conclusions}\label{sec:summary}

We have presented a detailed study of the physical properties of a sample of 157 optically-thick 
absorption line systems with redshifts $z\sim 1.8-4.4$. This sample has been 
selected from the HD-LLS survey by \citet{pro15}, which provides us with a representative 
population of high-redshift LLSs. We have further expanded this statistical sample with 
77 additional systems from the literature, with redshifts down to $z \sim 0$.

To infer the chemical composition and the physical state of the absorbing gas, we have
computed ionisation models by means of radiative transfer calculations at equilibrium
and for a single gas phase. These calculations are the input of a Bayesian 
formalism that exploits Markov Chain Monte Carlo techniques to derive the posterior probability 
distribution function for quantities of interest, such as the gas-phase metallicity, 
the physical density, the temperature, and the dust content of the absorbing 
gas. To explore the dependency of our results on the assumptions of the 
input ionisation models, we have computed five different grids of models, varying the 
shape and intensity of the UVB, and the contribution of local sources, of
dust and metal depletion onto grains, and of collisional ionisation. 

Through comparisons of the PDFs inferred under different model assumptions, we have shown 
that simple photoionisation models provide a good description of the general LLS population, 
and that the physical properties of LLSs are not extremely sensitive to the 
assumed ionisation corrections. However, the predictions for individual systems, and
particularly for their density and size, are more prone to systematic effects attributable 
to ionisation corrections.

Our findings on the physical properties of our statistical sample of LLSs between $z\sim 2.5-3.5$ 
can be summarised in the following way.

\begin{itemize}

\item[--] LLSs arise from photoionised gas, with temperatures $T < 5\times 10^4$ K,
and ionisation parameters $U \sim 10^{-3}-10^{-2}$. Thus, ionisation corrections for hydrogen 
and metal lines are critical and cannot be neglected when inferring the metallicity of the 
absorbing gas.

\item[--] LLSs have typical densities between $n_{\rm H}\sim 10^{-3.5}-10^{-2}~\rm cm^{-3}$.
However, the detailed shape of the density PDF is sensitive to the amplitude of the ionisation 
radiation field because of a well-known degeneracy between density and radiation. Indeed,
in our analysis, we have found that observations of the most common ions cannot be generally used to 
robustly constrain the properties of the ionising radiation field, or the size 
of the absorbing gas.

\item[--] The population of $z\sim 2.5-3.5$ LLSs is metal poor, with a peak at
$\log Z/Z_\odot \sim -2$, which is below that observed in higher 
column-density DLAs with a mean $\log Z/Z_\odot \sim -1.5$. Further, LLSs 
appear to contain only modest amounts of dust. The inferred metallicity distribution 
is very broad, extending over 4 orders of magnitude.
The probability of finding a metallicity $\log Z/Z_\odot \le -1.5$ is $\sim 70\%$, 
but the probability of finding very metal poor systems with $\log Z/Z_\odot \le -3$ is modest, 
being $\sim 10\%$. The metal content of SLLSs with 
$\log N_{\rm HI} \ge 19$ rapidly evolves with redshift, with a ten-fold increase between $z\sim 2.1$ 
and $z\sim 3.6$. We have also reported tentative evidence that the lower column density LLSs 
with $\log N_{\rm HI} \lesssim 18.5$ are the least enriched, suggesting a smooth transition with 
the IGM. 

\item[--] After accounting for ionisation corrections, LLSs with 
$\log N_{\rm HI} < 19$ and SLLSs with  $\log N_{\rm HI} \ge 19$ 
jointly contain a total mass in metals of $\Omega_{\rm m}^{\rm LLS} \sim 2.1 \times 10^{-6}$, 
which is $\sim 3$ time more than the amount of metals locked in DLAs. 
Compared to the metals produced by UV-selected galaxies, LLSs account for 
$\sim 15\%$ of all the metals, although systematic uncertainties as large as a factor 
$\gtrsim 2$ affect this estimate. Moreover, $\Omega_{\rm m}^{\rm LLS}$ is likely not converged
in our sample, as rare systems in the tail of the metallicity PDF may contribute significantly 
to the cosmic metal budget.  
\end{itemize}

Despite our efforts to quantify the extent to which 
the inferred physical properties of LLSs are robust against systematic uncertainties 
in the input ionisation models, our analysis has not 
exhausted all possible scenarios, such as the presence of 
multiple gas phases, a non-constant density profile \citep[e.g.][]{pet92,asc10},
or non-equilibrium effects \citep[e.g.][]{opp13}. Future work should expand on the 
formalism we have developed to gain an even deeper understanding of
the behaviour of ionisation corrections in more realistic astrophysical
environments. 

By characterising the physical properties of LLSs at $z>2$ in a large statistical sample, our 
work has provided new clues for the origin of LLSs and new empirical constraints for theories
of cold accretion and feedback. Given that a significant fraction of 
LLSs have densities comparable to or higher than the virial 
densities, the statistical properties of samples of hundreds of 
optically-thick absorbers can now be compared quantitatively to the predictions of 
numerical simulations for the CGM, providing new
metrics to constrain the efficiency of metal ejection and the mixing between 
enriched gas in the outflows and metal-poor inflows. A simple qualitative 
comparison already reveals a possible tension. Indeed, successful galaxy formation models 
that eject baryons from galaxies to avoid an overproduction of stars appear to 
enrich LLSs in the halo above what is suggested by 
our observations. Conversely, models with weak feedback that overpredict 
the fraction of baryons locked in stars may better reproduce the metallicity PDF
of LLSs. Future work is now needed to characterise
and address this possible discrepancy between theory and observations,
and to explore the role of mixing between inflows and outflows in shaping 
the observed metallicity distribution. 

\section*{Acknowledgements}
We thank N. Lehner, C. Howk, and M. Rafelski for useful discussions
on LLSs, and M. Fossati for valuable suggestions on the MCMC analysis. 
M.F. acknowledges support by the Science and 
Technology Facilities Council [grant number ST/L00075X/1]. 
This work used the DiRAC Data Centric system at Durham University, operated by 
the Institute for Computational 
Cosmology on behalf of the STFC DiRAC HPC Facility (www.dirac.ac.uk). This equipment was funded by BIS 
National E-infrastructure capital grant ST/K00042X/1, STFC capital grant ST/H008519/1, and 
STFC DiRAC Operations grant ST/K003267/1 and Durham University. DiRAC is part of the National 
E-Infrastructure. 
Some of data used in this work were obtained at the W.M. Keck Observatory, 
which is operated as a scientific partnership among the California Institute of Technology, 
the University of California and the National Aeronautics and Space Administration. 
The Observatory was made possible by the generous financial support of the W.M. Keck Foundation.
The authors wish to recognize and acknowledge the very significant cultural role and reverence that the 
summit of Maunakea has always had within the indigenous Hawaiian community.  We are most 
fortunate to have the opportunity to conduct observations from this mountain.
Some of the data presented in this work were obtained from the Keck Observatory Database of Ionized 
Absorbers toward Quasars (KODIAQ), which was funded through NASA ADAP grant NNX10AE84G 
\citep{ome15}.  For access to the data used in this paper and analysis codes, please contact the authors or visit \url{http://www.michelefumagalli.com/codes.html}.

\begin{figure*}
\centering
\begin{tabular}{ccc}
\includegraphics[scale=0.28]{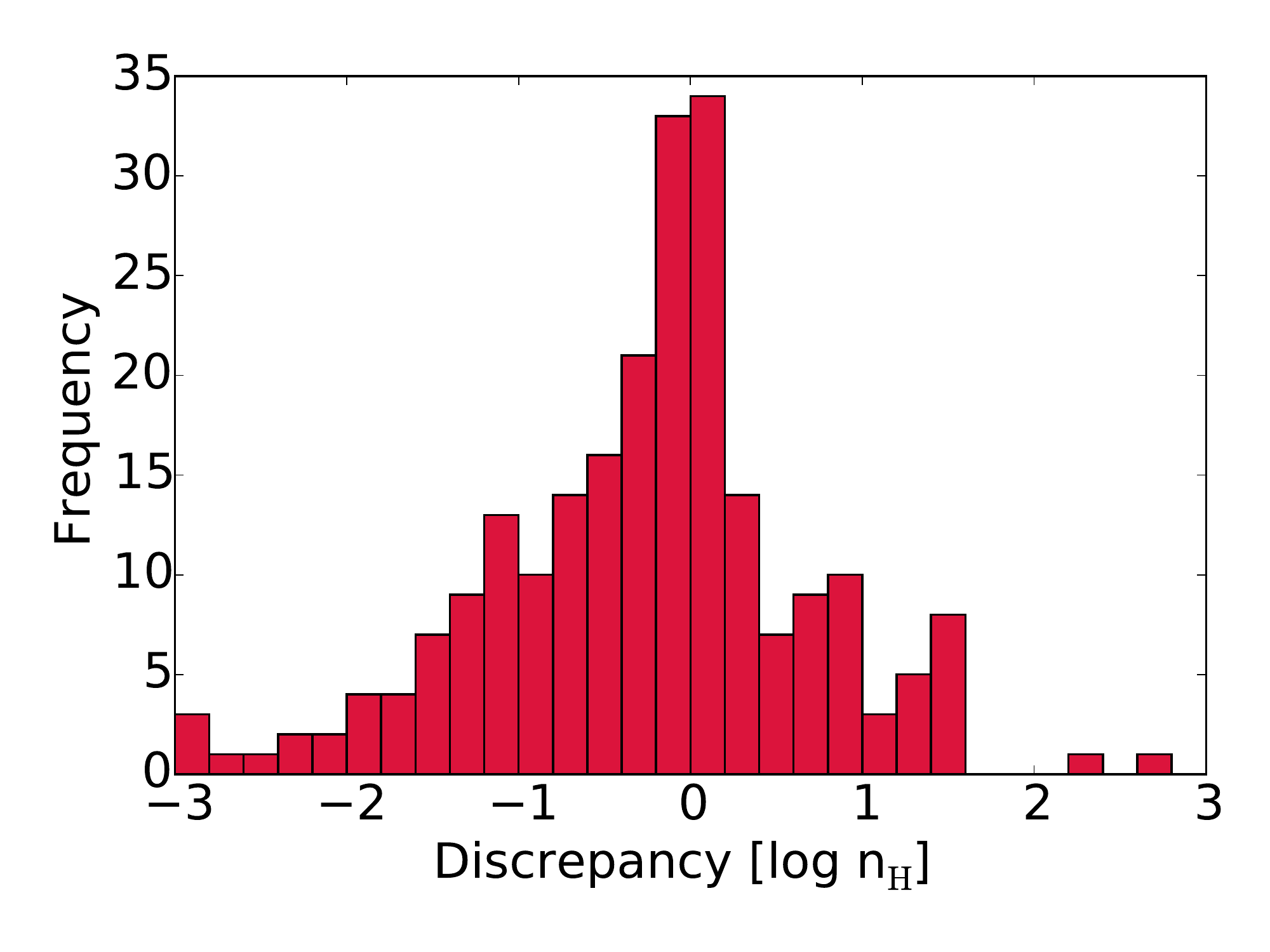}&
\includegraphics[scale=0.28]{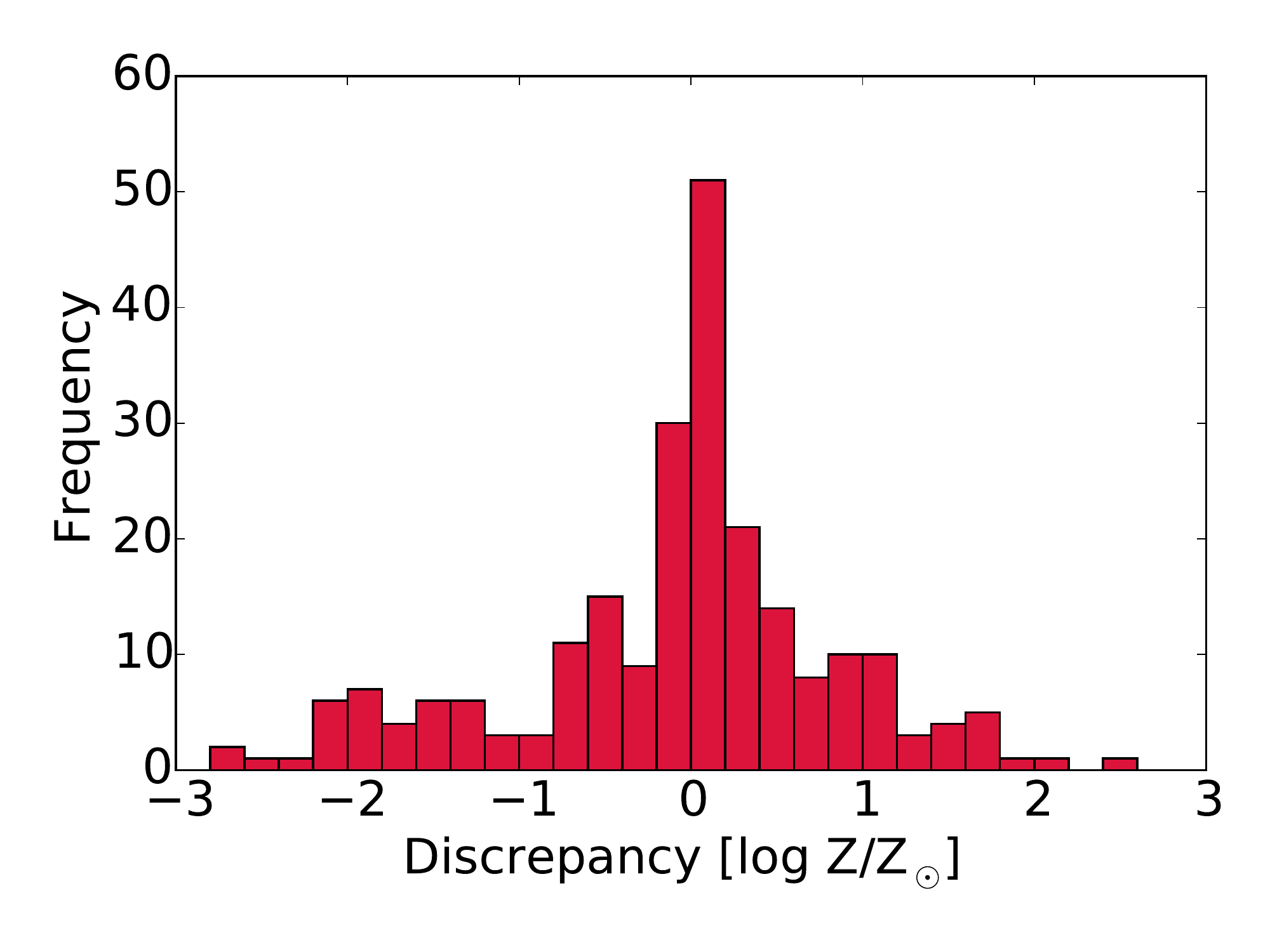}&
\includegraphics[scale=0.28]{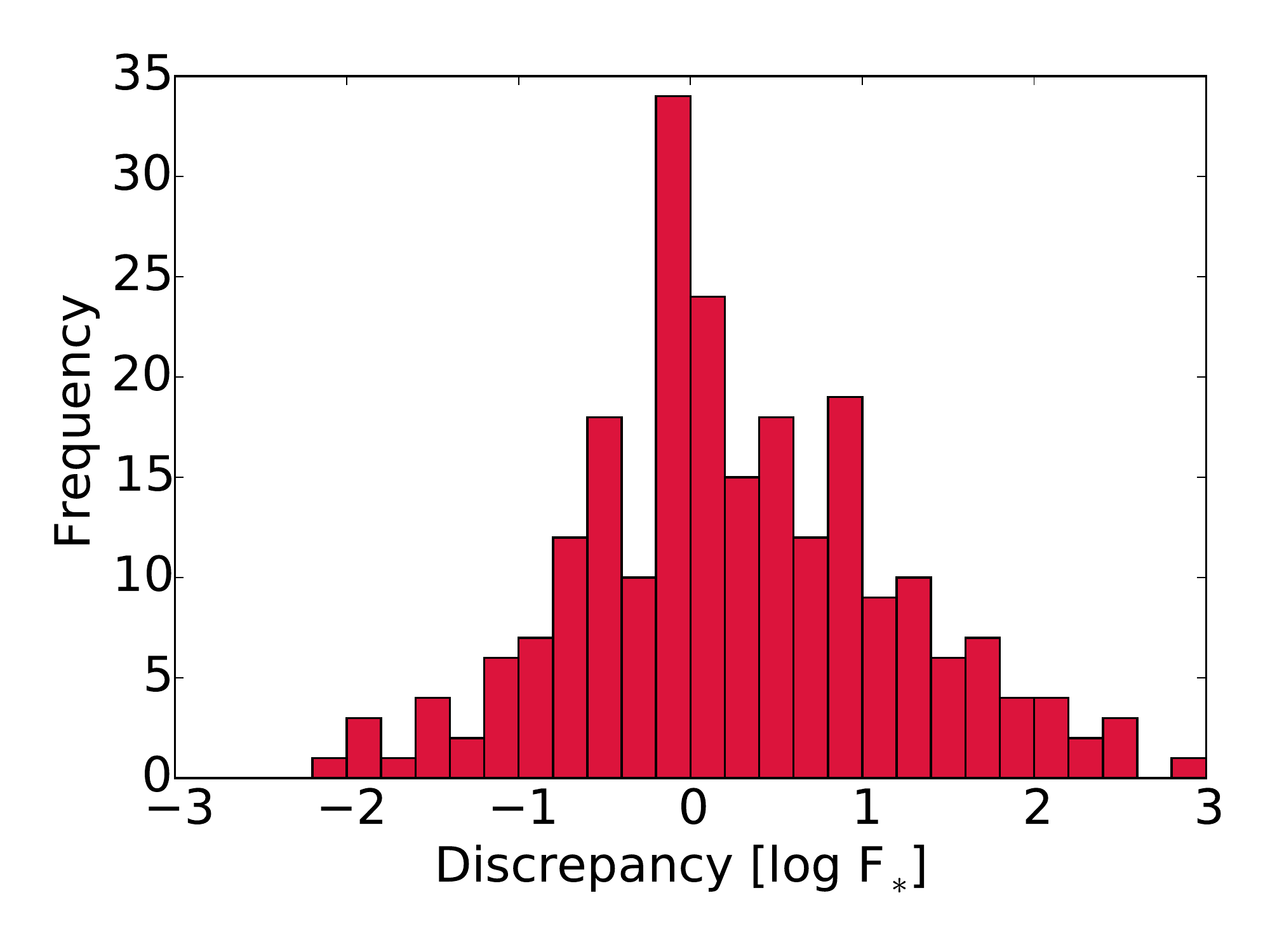}\\
\end{tabular}
\caption{The discrepancy between the input and recovered 
density (left), metallicity (centre), and $F_*$ (right) for a mock sample of LLSs
that is matched to our observed sample. 
This discrepancy is quantified as the median of the posterior PDF and the input value,
normalised to the PDF width from the 25th and 75th percentiles. For the majority of the 
mock LLSs, the input values are well within the first and third quartile of the
posterior PDF.}\label{fig:mockreal}
\end{figure*}

\begin{figure*}
\centering
\begin{tabular}{ccc}
\includegraphics[scale=0.28]{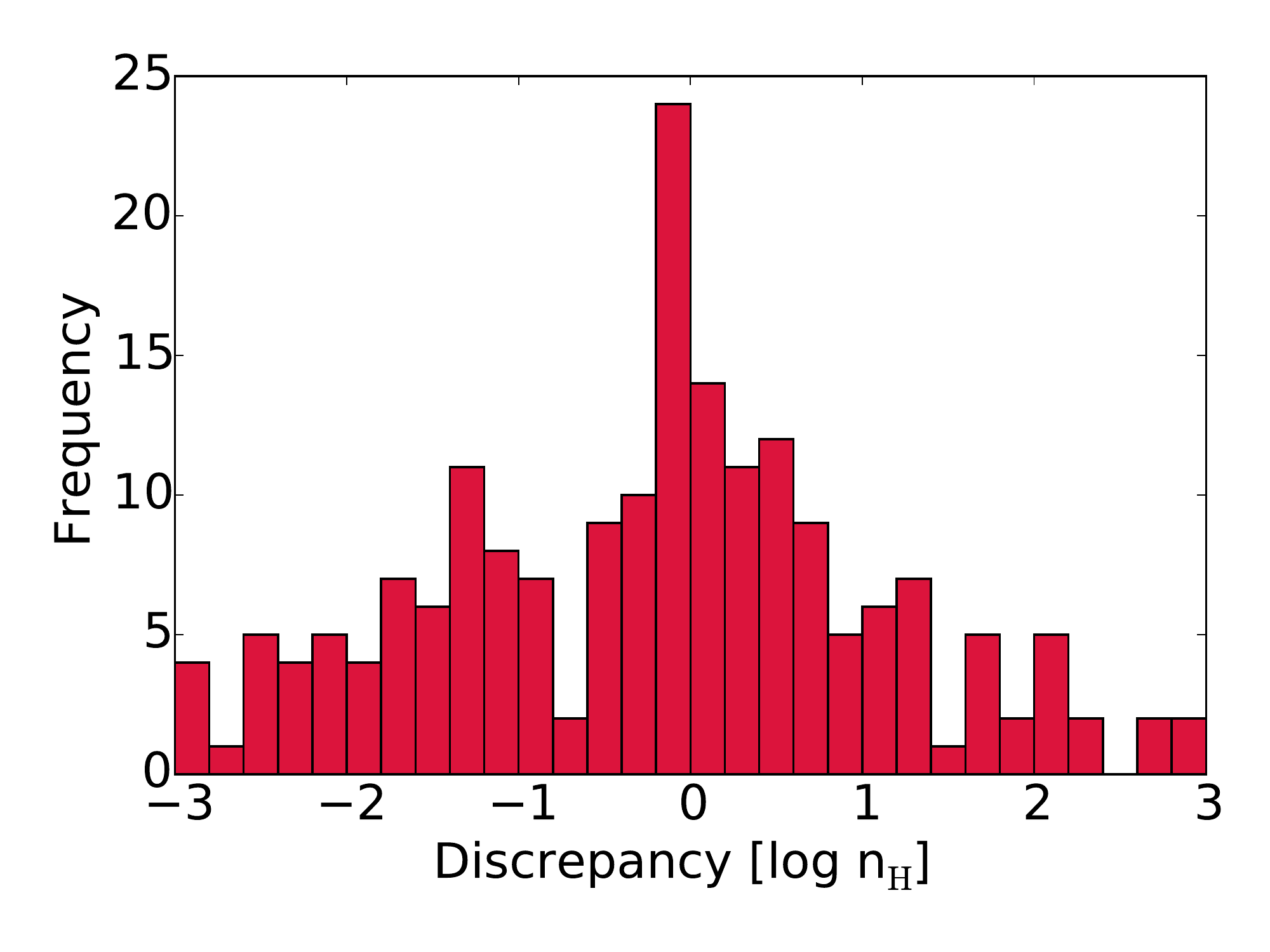}&
\includegraphics[scale=0.28]{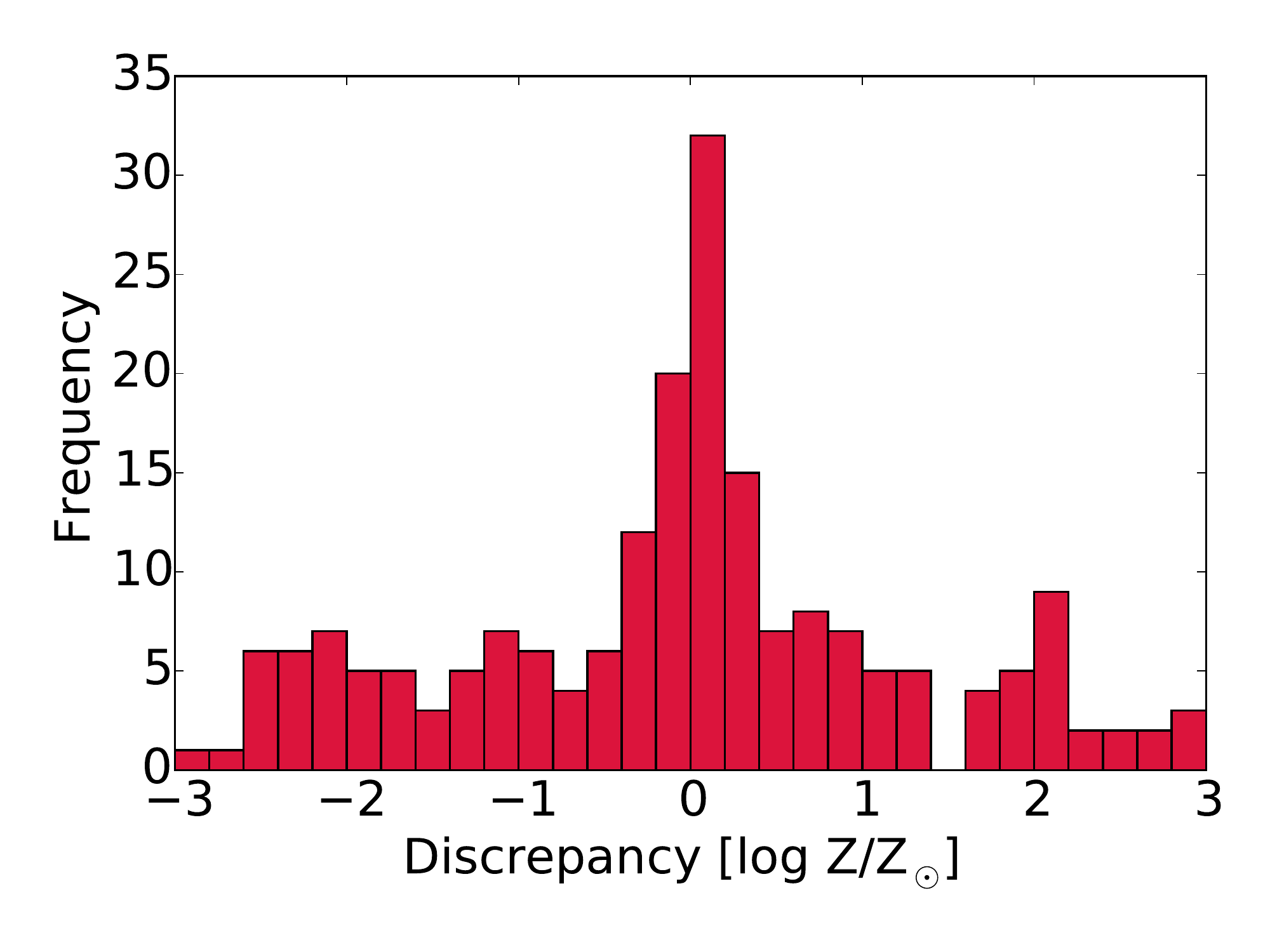}&
\includegraphics[scale=0.28]{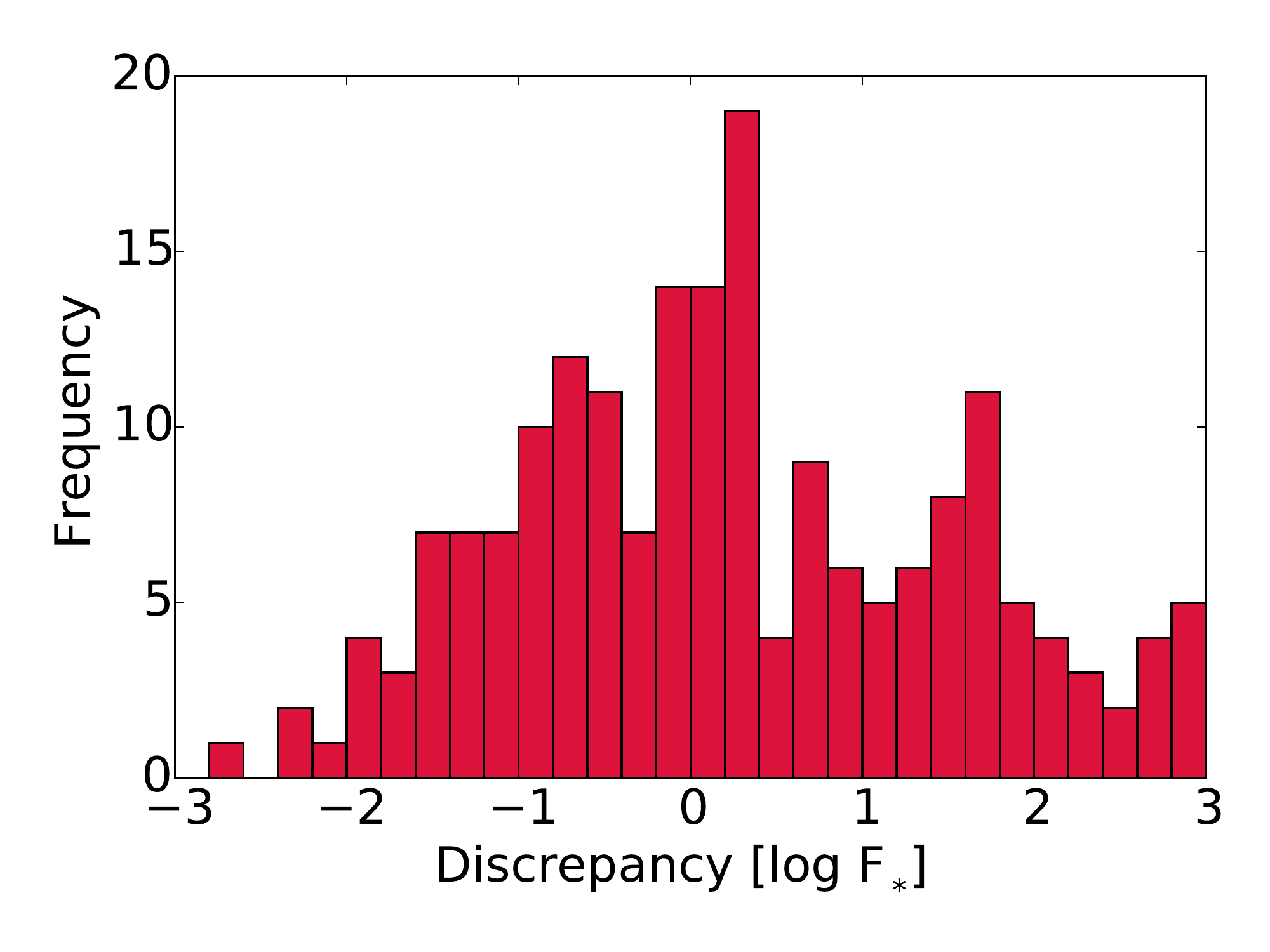}\\
\end{tabular}
\caption{Same as Figure \ref{fig:mockreal}, but including systematic errors on the 
ion column densities.}\label{fig:mocksys}
\end{figure*}

\begin{figure*}
\centering
\begin{tabular}{ccc}
\includegraphics[scale=0.28]{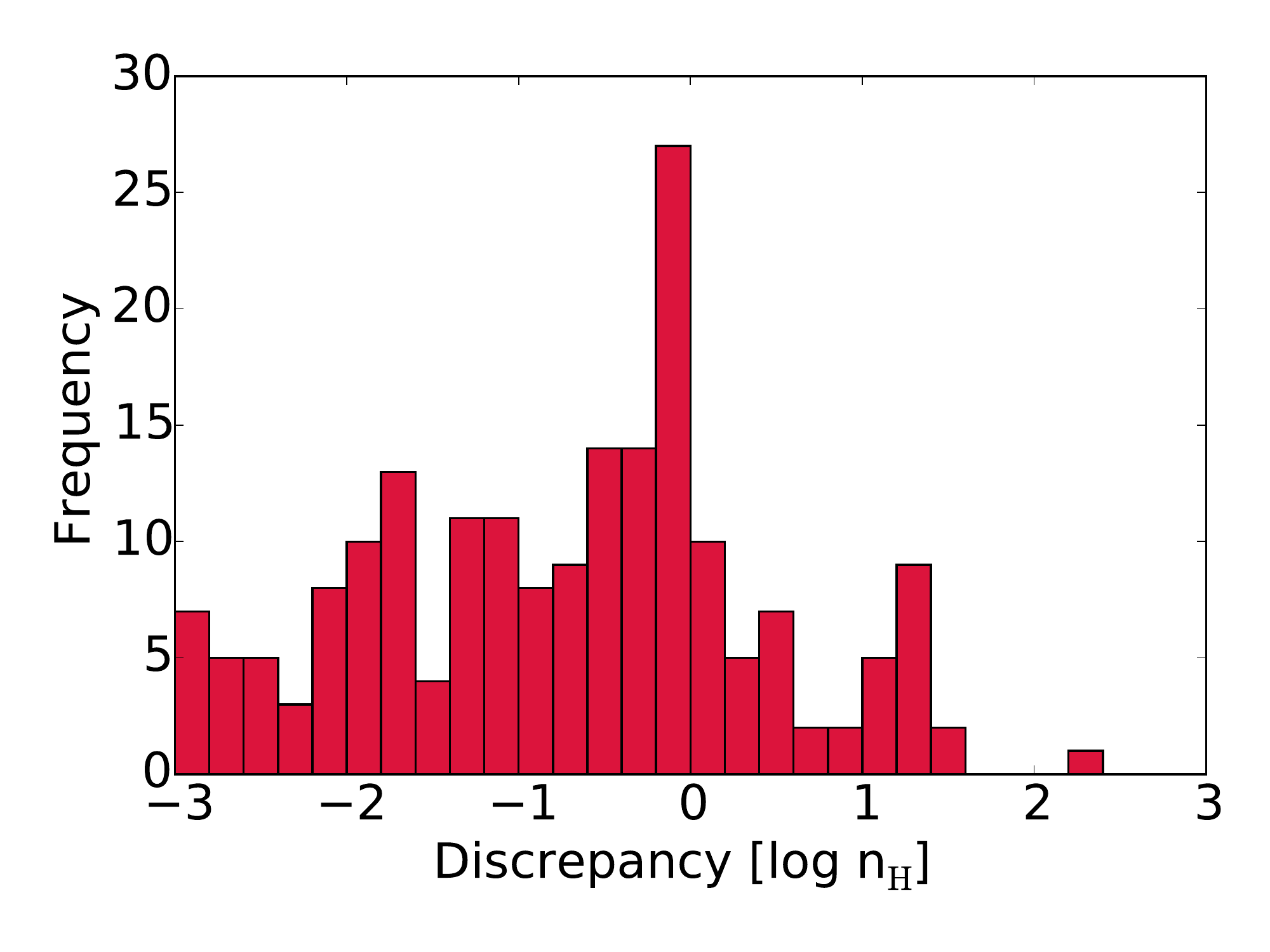}&
\includegraphics[scale=0.28]{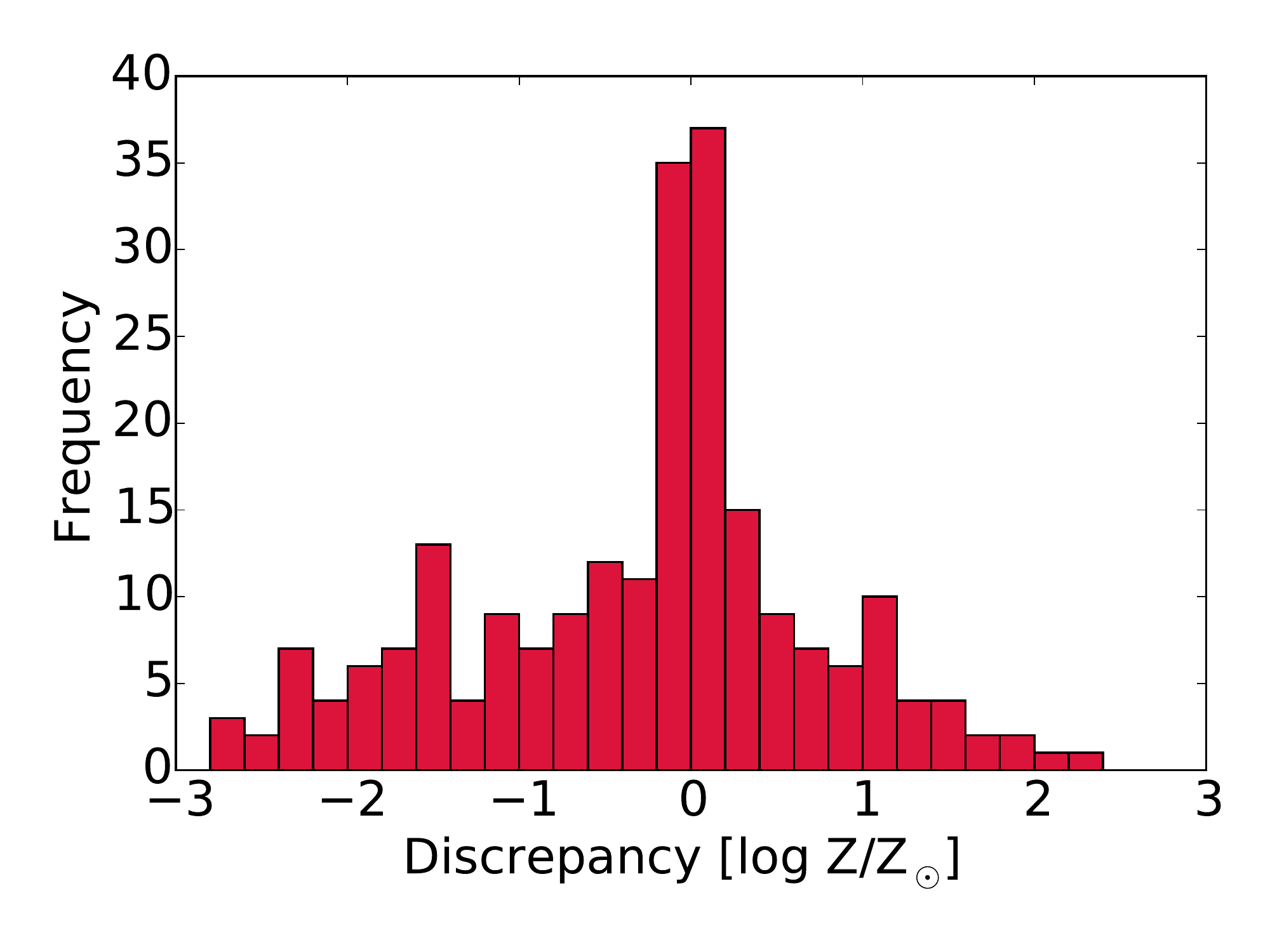}&
\includegraphics[scale=0.28]{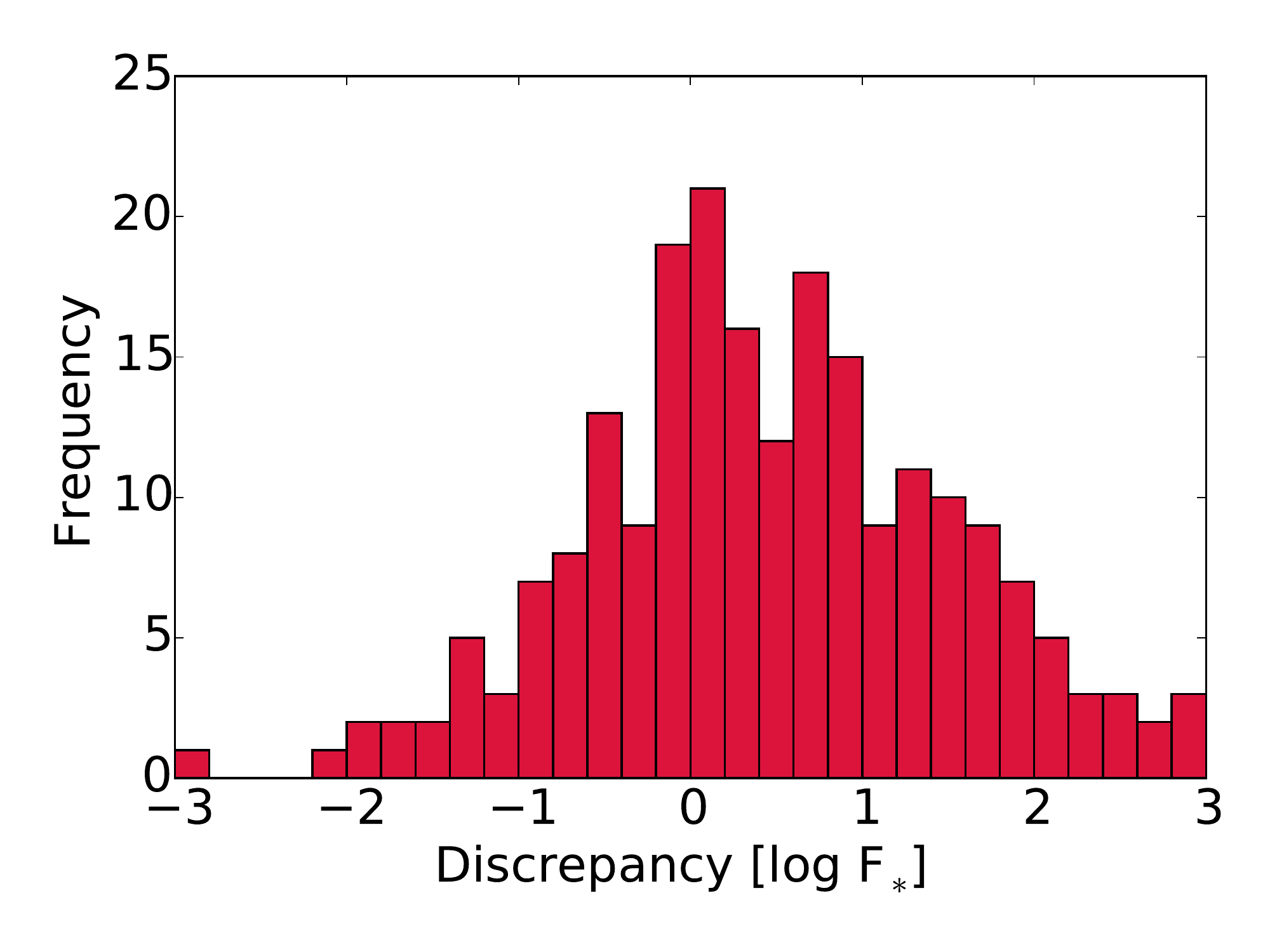}\\
\end{tabular}
\caption{Same as Figure \ref{fig:mockreal}, but including additional contribution to the
\ion{C}{IV} column density from a second, more ionised, gas phase.}\label{fig:mockciv}
\end{figure*}

\appendix
\section{Mock tests and method validation}\label{sec:valid}

To validate the analysis method and to characterise the performances of the MCMC code in reconstructing 
the underlying PDFs, we generate sets of mock data that we then process with the same analysis
tools used throughout this paper.

As a trivial test to validate the code, we generate a mock set of 234 LLSs with the redshift
and density distribution of the observed sample, and with a Gaussian density distribution centred 
at $\log n_{\rm H} = -2.5$ with a width of 0.3 dex, and a metallicity distribution centred at 
$\log Z/Z_\odot = -2$ with a width of 1 dex. Using the minimal grid, we then generate an idealised mock 
catalogue of column densities, where each ion is measured with an uncertainty of 0.05 dex. We then perform the 
MCMC analysis on this dataset, successfully recovering the input density and metallicity PDFs
for the full sample, as well as a one-to-one relation between the input values
and the medians of the metallicity and density PDFs for individual systems.

To better understand the performance of the MCMC analysis on more realistic data, we construct a new set of mocks, 
using the dust grid of models with $F_*$ values distributed as a Gaussian centred at $F_*=0$ and width 0.6 dex. 
For this set, we match the data quality to the observed one, including for each simulated 
LLS only the observed ions in the corresponding real systems, 
and preserving the associated uncertainties and upper/lower limits.
Thus, differently from the previous case, this mock sample is characterised by a variety of data quality, 
including systems with only a handful of observed ions, larger errors on the column densities, or 
a large number of lower/upper limits. After performing the MCMC analysis, we find a tight correlation between the 
input quantities and the medians of the posterior PDFs, although with a scatter and the presence of outliers. 
As a metric of the ability of the MCMC analysis to recover the input data, we compute for each quantity
the deviation between the median of the reconstructed PDF and the input value, which we normalise to the
a characteristic width of the posterior PDF using the 25th and 75th percentiles.

The histogram of the discrepancies for these mock LLSs is shown in Figure \ref{fig:mockreal}.
From this test, we conclude that the posterior PDFs reconstructed with our analysis technique
contain the input value within the first and third quartile of the distribution for 
the majority of the systems, with the near totality being contained by twice the PDF width
for individual LLSs. Furthermore, the discrepancies are well centred at zero.
However, among the outliers, there is a hint of a small preference for larger $F_*$ and smaller $n_{\rm H}$
which we attribute in part to the degeneracy between parameters in the dust model, and to the 
skewness of the posterior PDFs for the LLSs with least constraining data (e.g. all upper limits).

We can also test how robust is the MCMC method to: i) the presence of systematic errors in the ion column 
densities that are unaccounted for by error bars; ii) a possible excess of column density in 
high ionisation species (e.g. \ion{C}{IV}) from a second phase along the line of sight. For the first test, 
we add a systematic offset to the ion column densities in the mock sample by drawing from a Gaussian 
centred at zero and with a width of 0.15. This means that approximately one in three ions for each LLS
have their column density offset by $>0.15$ dex, thus beyond the typical error bars on the column density.
After performing the MCMC analysis, despite the addition of systematic errors, we see that the input values
are still recovered without large systematic errors (Figure \ref{fig:mocksys}), although,
unsurprisingly, the number of outliers increases, particularly between 1-2 times the width of the posterior PDFs.
For the second test, we boost the \ion{C}{IV} column density by a random amount in the interval 
$(0,0.5)$ dex. Figure \ref{fig:mockciv} shows the resulting discrepancy from the analysis of these
new mock data. On average, the input PDFs distribution are recovered without significant biases, 
although there are a handful of individual cases in which the input and output values 
are in worse agreement, with a hint of systematic effect, which is to be expected given the 
one-sided perturbation in the data. 

In summary, these tests confirm that our MCMC procedure is performing well for the case of idealised 
mock data. Moreover, despite discrepancies for individual systems (especially the LLSs with 
the least constraining data), our analysis appears robust in recovering the underlying 
PDFs for the LLS population, even in presence of systematic errors on the column densities and a 
possible second, more ionised, gas phase.

\label{lastpage}

\end{document}